\documentclass[12pt]{article}
\usepackage{amsmath,graphicx,amssymb,amsthm,amsfonts}
\usepackage{epsfig}
\usepackage{amssymb}
\usepackage{lscape}
\usepackage{cleveref}
\usepackage{subfig}
\usepackage{setspace}

\begin{document}

\title{A model-based assessment of the cost-benefit balance and the plea
bargain in criminality -- A qualitative case study of the Covid-19 epidemic
shedding light on the \textquotedblleft car wash
operation\textquotedblright\ in Brazil\\
\bigskip }
\author{Hyun Mo Yang{$^a$}\thanks{%
Corresponding author: tel. + 55 19 3521-6031} , Ariana Campos Yang{$^b$},
Silvia Martorano Raimundo{$^c$} \bigskip \bigskip \\
{$^{a,c}$}Department of Applied Mathematics \\
State University of Campinas \\
Pra\c{c}a S\'{e}rgio Buarque de Holanda, 651; CEP: 13083-859 \\
Campinas, SP, Brazil \\
{$^b$}Division of Allergy and Immunology \\
General Hospital of the Medicine School of University of S\~{a}o Paulo \\
Av. Dr. Eneas Carvalho de Aguiar, 255, CEP: 05403-000 \\
S\~{a}o Paulo, SP, Brazil \\
\\
E-mails: {$^a$}hyunyang@ime.unicamp.br, {$^b$}arianacy@gmail.com, \\
{$^c$}silviamr@unicamp.br \\
\\
\\
Running head: Evaluating the cost-benefit balance \\
and the plea bargain in criminality}
\date{ }
\maketitle

\begin{abstract}
\textbf{Objectives}: We developed a simple mathematical model to describe
criminality and the justice system composed of the police investigation and
court trial. The model assessed two features of organized crime -- the
cost-benefit analysis done by the crime-susceptible to commit a crime and
the whistleblowing of the law offenders.

\textbf{Methods}: The model was formulated considering the mass action law
commonly used in the disease propagation modelings, which can shed light on
the model's analysis. The crime-susceptible individuals analyze two opposing
\textquotedblleft forces\textquotedblright\ -- committing crime influenced
by the law offenders not caught by police neither imprisonment by the court
trial (benefit of enjoying the corruption incoming), and the refraction to
commit crime influenced by those caught by police or condemned by a court
(cost of incarceration). Moreover, we assessed the dilemma for those
captured by police investigation to participate in the rewarding
whistleblowing program.

\textbf{Results}: The model was applied to analyze the \textquotedblleft car
wash operation\textquotedblright\ against corruption in Brazil. The model
analysis showed that the cost-benefit analysis of crime-susceptible
individuals whether the act of bribery is worth or not determined the basic
crime reproduction number (threshold); however, the rewarding whistleblowing
policies improved the combat to corruption arising a sub-threshold. Some
adopted mechanisms to control the Covid-19 pandemic shed light on
understanding the \textquotedblleft car wash operation\textquotedblright and
threatens to the fight against corruption.

\textbf{Conclusion}: Appropriate coverage of corruption by media,
enhancement of laws against white-collar crimes, well-functioning police
investigation and court trial, and the rewarding whistleblowing policies
inhibited and decreased the corruption.

\vspace{0.5cm}

\end{abstract}

\section*{Declarations}

\textbf{Funding}: This research received no specific grant from any funding
agency, commercial or not-for-profit sectors.

\noindent \textbf{Conflicts of interest/Competing interests}: The authors
have no conflicts of interest/competing interests to declare for this study.

\noindent \textbf{Availability of data and material}: Not applicable.

\noindent \textbf{Code availability}: We will provide under request.

\noindent \textbf{Authors' contributions}: Hyun Mo Yang: Conceptualization,
Methodology, Formal analysis, Writing - Original draft preparation,
Supervision. Ariana Campos Yang: Validation, Investigation. Silvia Martorano
Raimundo: Simulation, Visualization. The first draft of the manuscript was
written by Hyun Mo Yang and all authors commented on previous versions of
the manuscript. All authors read and approved the final manuscript.

\noindent \textbf{Ethics approval}: Not applicable.

\noindent \textbf{Consent to participate}: Not applicable.

\noindent \textbf{Consent for publication}: Not applicable.

\noindent \textbf{MSC codes}: 92B05 and 37N25.

\section{Introduction}

Silva \cite{silva} showed that \textquotedblleft corruption is not simply a
kind of crime, rather, it is an ordinary economic activity that arises in
some institutional environments\textquotedblright . He applied his idea to
describe Brazil's corruption theoretically by adopting some economic models
to understand corruption better. Kleemans and Poot \cite{kleemans} analyzed
quantitative and qualitative information of 979 criminal careers involved in
organized crime and white-collar crime based on the concept of social
opportunity structure. Hence, it is appropriate a cost-benefit analysis
(CBA) in criminology.

Abrams \cite{abrams} analyzed a cost-benefit approach to incarceration:
\textquotedblleft An excessive rate of incarceration cost the taxpayers
large amounts of money; however,\ too little imprisonment harm society
through costs to victims and even non-victims who must increase precautions
to avoid crime\textquotedblright . Dossetor \cite{dossetor} stated that
\textquotedblleft CBA is an analytical tool that compares the total costs of
an intervention or program against its total expected benefits. The
substantial costs of crime and the limited resources available for crime
prevention programs provide a compelling argument for a systematic approach
for allocating scarce public resources among competing programs or policies
on the basis of CBA\textquotedblright . Roman and Farrell \cite{roman}
stated that \textquotedblleft CBA can be focused on different levels, from
the evaluation of philosophies and perspectives, to assessment of
strategies, policies, tactics, specific activities, or the manner in which
combinations of these are applied in specific circumstances. However, since
crime prevention efforts typically need to be tailored to specific crime
types and contexts, the theoretical spectrum of applications of crime
prevention, and hence of the CBA required, could be
infinite\textquotedblright .

Brown \cite{brown} explored \textquotedblleft the prospects for integrating
criminal law into the widespread trend elsewhere in the executive branch of
using CBA to improve criminal justice policy making and enforcement
practice\textquotedblright . The whistleblowing policy is one of those
mechanisms to improve the justice system. Buccirossi et al. \cite{buccirossi}
analyzed \textquotedblleft the interaction between rewards for
whistleblowers, sanctions against fraudulent reporting, judicial errors and
standards of proof in the court case on a whistleblower's allegations and
the possible follow-up for fraudulent allegations\textquotedblright . They
warned that \textquotedblleft when the risk of retaliation is severe, larger
rewards are needed. The precision of the legal system must be sufficiently
high, hence these programs are not viable in weak institution environments,
where protection is imperfect and court precision low, or where sanctions
against false reporting are mild\textquotedblright .

The CBA quantifies the costs and benefits of different policies, with costs
and benefits being monetized in terms of local currency (dollars, for
instance). The quantitative models were based on criminal network \cite%
{easton} and principal-agent model \cite{groenen}. Instead of applying this
classical definition, we model the perception of the crime-susceptible
individuals being influenced by the inefficiency of the justice system
(police investigation and court trial). These individuals analyze the cost
(imprisonment) and benefit (escaping the justice system and using the
products of robbery or bribe) before participating in crime organizations or
in corruption groups to commit a crime. The model is formulated similarly to
the epidemiological modelings by using ordinary differential equations.
Hence, we discuss criminality (corruption in Brazil) compared with the
current coronavirus disease 2019 (Covid-19) declared pandemic by the WHO in
March 2020.

We apply our model to explore qualitatively \textquotedblleft an
investigation of an isolated instance of corruption within a Brazilian oil
company expanded into an immense anticorruption operation known as Opera\c{c}%
\~{a}o Lava Jato (`Operation Car Wash'). This investigative operation has
penetrated deep within Brazil's government and corporate elite to root out
systemic state-sanctioned corruption. Its criminal cases also appear to be
instating new legal norms for how corruption cases are handled in Brazil,
giving citizens hope that Lava Jato's impact will be felt far into the
future\textquotedblright\ \cite{moro}.\ Medeiros and Silveira \cite{medeiros}
provided a comprehensive review of the media coverage on the `Operation Car
Wash' performed by digital editions of Folha de S\~{a}o Paulo (newspaper)
and Veja Magazine.

The paper is structured as follows. In section 2, a general model is
formulated to describe organized crime. In section 3, the model is applied
to corruption. Section 4 presents a discussion of white-collar crime
compared with the control of the Covid-19 epidemic, and conclusions are
given in section 5.

\section{Material and methods}

Raimundo et al. \cite{raimundo} described the criminal contagion from inside
Brazil's prison system to outside susceptible individuals with criminal
propensity. In the model, they assumed that the behavioral contagion rate
was proportional to the product between the numbers of crime-susceptible and
incarcerated individuals, which is known in the disease epidemics modelings
as mass action law \cite{anderson}. They analyzed the crime-inducing
parameters to help policy-makers design crime control strategies to decrease
the number of inmates.

However, our approach here is the cost-benefit involved in committing a
crime (organized crime and corruption) and the whistleblowing policies. The
objective of the law offenders is not to be caught by the justice system;
hence the main goal is the incarceration of criminals (mainly white-collar
crime) to inhibit the practice of corruption. In the modeling, the justice
system is composed of police investigation and court trial. To decide to
participate in criminal activities, crime-susceptible individuals analyze
the possibility of escaping from the justice system. We model the law
offending rate as the balance between the benefit of usufruct the product of
corruption (not being caught) and the cost of being condemned and
incarcerated. The whistleblowing policies target those caught by police
investigation by offering rewards to help police investigation and court
trial. The model is formulated based on the flowchart shown in Figure \ref%
{Fig_diagram}.

\begin{figure}[h]
\vspace{-0.5cm} \centering \includegraphics[scale=0.70]{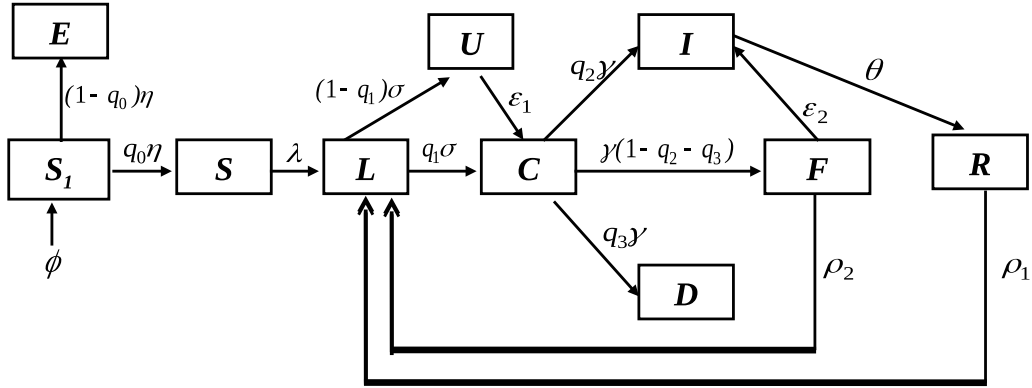} \vspace{%
-5cm}
\caption{The flowchart of the crime model. The arrow representing natural
mortality rate $\protect\mu $ is omitted in all compartments. }
\label{Fig_diagram}
\end{figure}

We describe the model variables (compartments) and parameters and the
hypotheses to formulate a model to describe the justice system acting
against criminal activities.

The model considers a naive population ($S_{1}^{\prime }$) divided into
crime-protected ($E^{\prime }$) and crime-susceptible subpopulations. The
latter subpopulation is divided into eight compartments: crime-susceptibles (%
$S^{\prime }$), new law-offenders ($L^{\prime }$), individuals caught ($%
C^{\prime }$) and uncaught ($U^{\prime }$) by police investigation, those
collaborating with justice system ($D^{\prime }$) under whistleblowing
program, individuals waiting for court trial in freedom ($F^{\prime }$) and
those sentenced by judge and incarcerated ($I^{\prime }$), and inmates
released from prison ($R^{\prime }$). The sum of all individuals in these
ten compartments is the size of the population denoted by $N$. For the
population's vital dynamic parameters, we denote the per-capita natality
(birth) and mortality rates by $\phi $ and $\mu $. We assume the absence of
additional mortality during the justice system's actions.

We describe the cost-benefit analysis done by crime-susceptible individuals.
We assume that an individual is influenced to commit crime by those uncaught
by police and waiting for court trial in freedom (benefit). On the other
hand, the crime-susceptible individual is inhibited from committing a crime
by those caught by police investigation and condemned by court trial (cost).
Hence, the flow from class $S^{\prime }$ to class $L^{\prime }$ at rate $%
\lambda $ obeys the mass action law, with the per-capita crime incidence or
the force of law-offending $\lambda $ being defined by%
\begin{equation}
\lambda =\displaystyle\frac{\beta _{1}U^{\prime }+\beta _{2}F^{\prime }}{%
N+\kappa _{1}C^{\prime }+\kappa _{2}I^{\prime }},  \label{lambda}
\end{equation}%
where $\beta _{1}$ and $\beta _{2}$ are the per-capita crime-influencing
rates (dimension $years^{-1}$), and $\kappa _{1}$ and $\kappa _{2}$ are the
per-capita inhibition coefficients (dimensionless). The name for the rate $%
\lambda $ is borrowed from epidemiology, where $\lambda $ is known as the
force of infection \cite{anderson}. Once police investigations catch law
offenders, they can participate in the whistleblowing program collaborating
with the justice system receiving rewards. Those individuals help police
investigation with police-collaboration rate $\epsilon _{1}$ and court trial
with judge-collaboration rate $\epsilon _{2}$ (both in $years^{-1}$),
defined by%
\begin{equation}
\begin{array}{ccccc}
\epsilon _{1}=\varepsilon _{10}+\varepsilon _{1}\frac{D^{\prime }}{N} &  & 
\mathrm{and} &  & \epsilon _{2}=\varepsilon _{20}+\varepsilon _{2}\frac{%
D^{\prime }}{N},%
\end{array}
\label{epsilon}
\end{equation}%
where $\varepsilon _{10}$ and $\varepsilon _{20}$ are voluntary (self)
collaboration rates, and $\varepsilon _{1}$ and $\varepsilon _{2}$ are
collaborator-dependent rates (those adhering to the whistleblowing program).
The collaborations of individuals in compartment $D$ result in the arrest of
uncaught individuals (flow from $U$ to $C$ by $\epsilon _{1}$) and the
incarceration of individuals waiting for court trial in freedom (flow from $%
F $ to $I$ by $\epsilon _{2}$).

We have two approaches to describe the avoidance of crime by a population.
One of them considers a fraction $1-q_{0}$ flowing instantaneously from $%
S_{1}^{\prime }$ to $E^{\prime }$ described by the Dirac delta function $%
\delta \left( t\right) $, where $q_{0}$ is the crime-susceptibility
proportion \cite{yang3}. Although this approach is appropriate in a
short-run dynamic, in the long-run, the number of individuals in the class $%
E^{\prime }$ goes to zero if additional pulses of inflow are not allowed. A
second approach is described by the naive individuals ($S_{1}^{\prime }$)
entering into the crime-protected class ($E^{\prime }$) at rate $\left(
1-q_{0}\right) \eta $ or to the crime-susceptible class ($S^{\prime }$) at
rate $q_{0}\eta $, where $\eta $ is the crime-prevention rate (dimension $%
years^{-1}$), which is influenced by many factors such as justice education
and social equity \cite{dossetor}. In this case, a fraction $q_{0}$ does not
adhere to the crime-prevention policies. We adopt the second approach
because the populational law-related behavior and justice system changes are
essentially a long-term dynamic. Yang et al. \cite{yang4} presented a severe
acute respiratory syndrome coronavirus 2 (SARS-CoV-2) transmission model
using this approach.

The law-offending dynamic begins with the flow from the crime-susceptible
class $S^{\prime }$ to the new law-offenders class $L^{\prime }$ at a rate $%
\lambda $ given by equation (\ref{lambda}). After an average period of
cover-up $1/\sigma $ in class $L^{\prime }$, where $\sigma $ is the uncover
rate, offenders enter into the caught or discovered by police class $%
C^{\prime }$ (with probability $q_{1}$) or uncaught or remaining covered-up
class $U^{\prime }$ (with probability $1-q_{1}$). Possibly uncaught
individuals can be incriminated by police investigation due to the
whistleblowers and enter into class $C^{\prime }$ at a rate $\epsilon _{1}$
given by equation (\ref{epsilon}). After an average period of court trial $%
1/\gamma $ in class $C^{\prime }$, where $\gamma $ is the court sentencing
(incarcerating) rate, individuals caught by police are incarcerated (class $%
I^{\prime }$, with probability $q_{2}$), or collaborate with justice (class $%
D^{\prime }$, with probability $q_{3}$) or delay the court sentence in
freedom (class $F^{\prime }$, with probability $1-q_{2}-q_{3}$). However,
individuals in the latter class are incarcerated at rate $\epsilon _{2}$,
equation (\ref{epsilon}), due to the whistleblowers in class $D^{\prime }$.
After an average period $1/\theta $ in prison (class $I^{\prime }$), where $%
\theta $ is the releasing rate, inmates are released and enter into the
class $R^{\prime }$. After average periods $1/\rho _{1}$ and $1/\rho _{2}$
in classes $R^{\prime }$ and $F^{\prime }$, where $\rho _{1}$ and $\rho _{2}$
are the relapsing rates of released and waiting for court trial in freedom
individuals, they relapse and commit a crime and re-enter into class $%
L^{\prime }$. Dimension of all rates is $years^{-1}$. Table \ref{Tab_param}
summarizes the model parameters.

\begin{table}[h]
\caption{Summary of the model parameters and respective mean values
(dimension is $years^{-1}$, except $q_{0}$, $q_{1}$, $q_{2}$, $q_{3}$, $%
\protect\kappa _{1}$ and $\protect\kappa _{1}$\ are dimensionless). The
non-linear parameters are varied ($^{\ast }$).}
\label{Tab_param}\centering       
\begin{tabular}{llll}
\hline
Symbol & Meaning &  & Value \\ \hline
$\phi $ & per-capita birth rate &  & $1/78$ \\ 
$\mu $ & per-capita mortality rate &  & $1/78$ \\ 
$\eta $ & crime-protection rate &  & $1/15$ \\ 
$q_{0}$ & proportion of crime-prevention failure &  & $0.01$ \\ 
$q_{1}$ & proportion of law offenders caught by police &  & $0.2$ \\ 
$q_{2}$ & proportion of individuals condemned by court trial &  & $0.2$ \\ 
$q_{3}$ & proportion of collaborators with justice system &  & $0.01$ \\ 
$\kappa _{1}$ & inhibition of criminality by those caught by police &  & $%
^{\ast }$ \\ 
$\kappa _{2}$ & inhibition of criminality by those condemned by court &  & $%
^{\ast }$ \\ 
$\beta _{1}$ & crime-influence rate by uncaught individuals &  & $^{\ast }$
\\ 
$\beta _{2}$ & crime-influence rate by individuals waiting court trial in
freedom &  & $^{\ast }$ \\ 
$\varepsilon _{10}$ & natural collaboration rate related to uncaught
individuals &  & $0.001$ \\ 
$\varepsilon _{20}$ & natural collaboration rate related to individuals
waiting trial &  & $0.001$ \\ 
$\varepsilon _{1}$ & $D$-dependent collaboration rate to uncaught individuals
&  & $^{\ast }$ \\ 
$\varepsilon _{2}$ & $D$-dependent collaboration rate to individuals waiting
trial &  & $^{\ast }$ \\ 
$\sigma $ & per-capita uncover rate &  & $1/1.5$ \\ 
$\gamma $ & per-capita court setencing (incarcerating) rate &  & $1/3$ \\ 
$\theta $ & per-capita releasing rate &  & $1/4$ \\ 
$\rho _{1}$ & per-capita relapsing rate of those released from prison &  & $%
1/100$ \\ 
$\rho _{2}$ & per-capita relapsing rate of those waiting court trial in
freedom &  & $1/50$ \\ \hline
\end{tabular}%
\end{table}

The population's vital dynamic disregarding criminality is described by%
\begin{equation*}
\frac{d}{dt}N=\left( \phi -\mu \right) N,
\end{equation*}%
where $N=S_{1}^{\prime }+E^{\prime }+S^{\prime }+L^{\prime }+U^{\prime
}+C^{\prime }+D^{\prime }+F^{\prime }+I^{\prime }+R^{\prime }$. Notice that $%
N$ will remain unchanged if deaths and births are equal ($\phi =\mu $,
resulting in $dN/dt=0$, and $N$ is constant in all-time). Instead of the
numbers of individuals in each compartment, the model can be formulated
using the fractions defined by%
\begin{equation*}
\begin{array}{ccc}
X=\displaystyle\frac{X^{\prime }}{N}, & \mathrm{where} & 
X=S_{1},E,S,L,U,C,D,F,I,R;%
\end{array}%
\end{equation*}%
hence, we suppress the prime ($^{\prime }$) for the fractions. However, the
constant population size results in%
\begin{equation}
\frac{1}{N}\frac{d}{dt}X^{\prime }=\frac{d}{dt}\frac{X^{\prime }}{N}=\frac{d%
}{dt}X,  \label{derfrac}
\end{equation}%
and equation (\ref{lambda}) for the force of law-offending becomes%
\begin{equation}
\lambda =\displaystyle\frac{\beta _{1}U+\beta _{2}F}{1+\kappa _{1}C+\kappa
_{2}I}  \label{lambda1}
\end{equation}%
in terms of the fractions. Table \ref{Tab_var} summarizes the compartments
(variables) of the model in terms of the fractions.

\begin{table}[h]
\caption{Summary of the model compartments (variables) given in fractions.}
\label{Tab_var}\centering                                                    
\begin{tabular}{lll}
\hline
Symbol &  & Meaning \\ \hline
$S_{1}$ &  & naive individuals \\ 
$E$ &  & crime-protected individuals \\ 
$S$ &  & crime-susceptible individuals \\ 
$L$ &  & law-offender individuals \\ 
$U$ &  & individuals uncaught by police investigation \\ 
$C$ &  & individuals caught by police investigation \\ 
$D$ &  & individuals colaborating with justice system \\ 
$F$ &  & individuals waiting for court trial \\ 
$I$ &  & individuals condemned by court trial \\ 
$R$ &  & individuals released from prison \\ \hline
\end{tabular}%
\end{table}

Based on the above assumptions and definitions, and considering equation (%
\ref{derfrac}) for the fractions, the justice system acting against criminal
activities model is described by the non-linear differential equations%
\begin{equation}
\left\{ 
\begin{array}{lll}
\displaystyle\frac{d}{dt}S_{1} & = & \mu -\left( \mu +\eta \right) S_{1} \\ 
\displaystyle\frac{d}{dt}E & = & \left( 1-q_{0}\right) \eta S_{1}-\mu E \\ 
\displaystyle\frac{d}{dt}S & = & q_{0}\eta S_{1}-\frac{\beta _{1}U+\beta
_{2}F}{1+\kappa _{1}C+\kappa _{2}I}S-\mu S \\ 
\displaystyle\frac{d}{dt}L & = & \frac{\beta _{1}U+\beta _{2}F}{1+\kappa
_{1}C+\kappa _{2}I}S-\left( \mu +\sigma \right) L+\rho _{1}R+\rho _{2}F \\ 
\displaystyle\frac{d}{dt}U & = & \left( 1-q_{1}\right) \sigma L-\left(
\varepsilon _{10}+\varepsilon _{1}D\right) U-\mu U \\ 
\displaystyle\frac{d}{dt}C & = & q_{1}\sigma L+\left( \varepsilon
_{10}+\varepsilon _{1}D\right) U-\left( \mu +\gamma \right) C \\ 
\displaystyle\frac{d}{dt}F & = & \left( 1-q_{2}-q_{3}\right) \gamma C-\left(
\varepsilon _{20}+\varepsilon _{2}D\right) F-\left( \mu +\rho _{2}\right) F
\\ 
\displaystyle\frac{d}{dt}I & = & q_{2}\gamma C+\left( \varepsilon
_{20}+\varepsilon _{2}D\right) F-\left( \mu +\theta \right) I \\ 
\displaystyle\frac{d}{dt}D & = & q_{3}\gamma C-\mu D \\ 
\displaystyle\frac{d}{dt}R & = & \theta I-\left( \mu +\rho _{1}\right) R,%
\end{array}%
\right.  \label{system}
\end{equation}%
where we assumed $\phi =\mu $ in the equation for $S_{1}$ (we substituted $%
\phi $ by $\mu $). The sum of all equations is zero resulting in a constant
population, for this reason the sum of all fractions is $%
S_{1}+E+S+L+U+C+D+F+I+R=1$. Notice that the equations for $S_{1}$ and $E$
can be decoupled from the system of equations.

The initial conditions supplied to the system of equations (\ref{system}) are%
\begin{equation}
\left\{ 
\begin{array}{l}
S_{1}(0)=S_{1}^{0},E(0)=E^{0},S(0)=S^{0}-f,L(0)=f,U(0)=0, \\ 
C(0)=0,F(0)=0,I(0)=0,D(0)=0,R(0)=0,%
\end{array}%
\right.  \label{init}
\end{equation}%
where $f$ is a small fraction of new law offenders, and $S_{1}^{0}$, $E^{0}$%
, and $S^{0}$, with $S_{1}^{0}+E^{0}+S^{0}=1$, are given in Appendix \ref%
{p_trivial}. These conditions describe the beginning of the corruption in a
country, which can be modified. Notice that when $f=1/N$, one new offender
is introduced in a completely crime-free population.

The system of equations (\ref{system}) in terms of the fractions has
equilibrium points ($N$ is constant, with $\phi =\mu $). The trivial ($P^{0}$%
) and non-trivial ($P^{\ast }$) equilibrium points are presented in
Appendices \ref{p_trivial} and \ref{p_n_trivial}. We apply the model to the
white-collar (corruption) crime, briefly describing the meaning of the
equilibrium points.

\subsection{Corruption-free society -- Trivial equilibrium point $P^{0}$}

The trivial equilibrium $P^{0}$ is locally stable if $R_{g}<1$ (the
corruption is eradicated or eliminated), with the gross crime reproduction
number $R_{g}$ given by equation (\ref{Rg}) in Appendix \ref{local}. In a
special case (see Appendix \ref{part1}), we showed that the trivial
equilibrium point $P^{0}$ is unique and stable when $R_{ef}\leq R^{c}$, with
the sub-threshold $R^{c}<1$. Appendix \ref{interp} presents the
interpretation of each term of $R_{g}=R_{0}+Q$, the basic ($R_{0}$) and
additional ($Q$) crime reproduction numbers. Notice that whenever the basic
crime reproduction number is $R_{0}=0$, the additional number $Q$ must also
be zero because $R_{0}$ act at the beginning of the dynamic, and $Q$ affects
later \cite{yang2}.

In epidemiological modelings, the reproduction number $R_{g}$ measures the
transmissibility of a parasite. In other words, $R_{g}$ is the average
number of secondary infections originated from a primary infection
introduced in a completely susceptible population \cite{anderson}. In this
particular criminality modeling, $R_{g}$ is the average number of
co-optations to commit a crime in an entirely crime-free population. Hence,
the higher $R_{g}$, the easier is the cooptation of crime-susceptible
individuals to offend the law. Our main goal is decreasing the gross crime
reproduction number $R_{g}$ aiming at the reduction of corruption. In other
words, the justice system's effectiveness difficult the cooptation to
criminality avoiding systemic corruption.

From equation (\ref{Rpart}) in Appendix \ref{local}, the gross crime
reproduction number $R_{g}$ can be diminished by increasing the
effectiveness of the justice system described by the parameters $q_{1}$, $%
q_{2}$, and $q_{3}$. Notice that the collaborator-dependent rates $%
\varepsilon _{1}$ and $\varepsilon _{2}$ do not affect on $R_{0}$ and $Q$;
still, the justice system can handle these parameters to increase the police
efficiency to catch offenders and incarcerate them by court trial. The
parameters $\kappa _{1}$ and $\kappa _{2}$ also do not appear in $R_{0}$ and 
$Q$; but affect the criminality's level. These contributions are evaluated
by the non-trivial equilibrium point $P^{\ast }$.

\subsection{Corruption prevailing society -- Non-trivial equilibrium point $%
P^{\ast }$}

The non-trivial equilibrium point $P^{\ast }$, given by equation (\ref{Pstar}%
) in Appendix \ref{p_n_trivial}, means an inefficient justice system; hence,
the objective is improving the investigation and more rigid laws to catch
and incarcerate law offenders. In other words, if $R_{g}$ is not reduced
below unity, entrance in the criminal activities must be inhibited by using,
for instance, handcuffs, coercive conduction, and pre-trial detentions
(increasing the inhibition coefficients $\kappa _{1}$ and $\kappa _{2}$).
Notice that when the per-capita inhibition coefficients $\kappa _{1}$ and $%
\kappa _{2}$ are zero, we have the classical epidemiological model (see \cite%
{yang8} for a non-bilinear incidence modeling), and for $\kappa
_{1}\rightarrow \infty $ and $\kappa _{2}\rightarrow \infty $, we have $%
C^{\ast }\rightarrow 0$ and $S^{\ast }\rightarrow S^{0}$ implying that the
corruption was controlled (remember that we have $R_{g}>1$, hence not
eradicated). Another crime-prevention control is decreasing $U$ and $F$
through plea bargain by whistleblowing policies (justice-collaboration rates 
$\varepsilon _{1}$ and $\varepsilon _{2}$) resulting in the increased $C$
and $I$ (see equation (\ref{lambda1}) for the force of law-offending
depending on these four classes). Hence, the whistleblowing program and
corruption-avoiding measures increase the justice system's effectiveness 
\cite{bechara}.

In the Covid-19 epidemic modelings, the non-trivial equilibrium point must
be decreased to achieve the trivial equilibrium by implementing controls,
such as the quarantine, adoption of individual (sanitization of hands and
use of face masks) and collective (social distancing) protective measures,
and vaccine \cite{yang6}. All these protective measures decrease individuals
harboring SARS-CoV-2. However, for criminality, when corruption prevails,
described by the non-trivial equilibrium point $P^{\ast }$, the justice
system can be improved by increasing the corruptors being caught ($C^{\ast }$%
) and incarcerated ($I^{\ast }$).

The equilibrium point $C^{\ast }$ is the positive solution of $5^{th}$
degree polynomial $Pol_{5}(C)$ given by equation (\ref{equil2}) in Appendix %
\ref{p_n_trivial}. To better understand the dynamic behavior of the model
described by equation (\ref{system}), we analyze two particular cases. When $%
\beta _{1}=0$, $\kappa _{2}=0$ and $\rho _{1}=0$, $Pol_{5}(C)$ in equation (%
\ref{equil2}) becomes a $4^{th}$ degree polynomial $Pol_{4}(C)$ given by
equation (\ref{equil3}). This polynomial may have up to two equilibrium
points named $P_{+}^{\ast }$ (using big solution $C_{+}^{\ast }$) and $%
P_{-}^{\ast }$ (using small solution $C_{-}^{\ast }$) when $R_{g}<1$ (see
Appendix \ref{part1}). The local stability of these points is assessed
numerically (see Appendix \ref{stab}). The second case deals with $%
\varepsilon _{1}=\varepsilon _{2}=0$, when we have a unique non-trivial
equilibrium point $P^{\ast }$ for $R_{g}>1$, with $C^{\ast }$ given by
equation (\ref{Csol}) (see Appendix \ref{part2}). In this case, forward
bifurcation occurs at $R_{g}=1$ and, additionally, the trivial equilibrium
point $P^{0}$ is globally stable.

Notice that, when $\varepsilon _{1}>0$ and $\varepsilon _{2}>0$, but for the
first particular case ($\beta _{1}=0$, $\kappa _{2}=0$ and $\rho _{1}=0$),
we have only one positive solution $C^{\ast }$ for $R_{g}>1$, but zero or
two positive solutions ($C_{+}^{\ast }$ and $C_{-}^{\ast }$) for $R_{g}<1$.
In other words, the forward bifurcation will occur if only zero solution is
found at $R_{g}=1$, but backward bifurcation will occur if a positive
solution is found at $R_{g}=1$ and two positive solutions for $R_{g}<1$
(appearing a sub-threshold $R^{c}<1$, and at $R_{g}=R^{c}$, the solutions $%
C_{+}^{\ast }$ and $C_{-}^{\ast }$ collapse in one solution $C^{s}$, and
only trivial equilibrium is found for $R_{g}<R^{c}$). Therefore, the complex
equation (\ref{equil2}), a $5^{th}$ degree polynomial in $C$, must present
backward or forward bifurcation (see Figure \ref{for_back} in Appendix \ref%
{part1}) depending on $\varepsilon _{1}$ and $\varepsilon _{2}$.

In the preceding section, we showed that the trivial-equilibrium point $%
P^{0} $ is locally stable when $R_{g}<1$, but globally stable only if
collaborator-dependent rates are $\varepsilon _{1}=\varepsilon _{2}=0$ (see
Appendix \ref{global}). Considering the first particular case of the model
in Appendix \ref{part1}, we showed that when collaborators adhere to
whistleblowing program\ ($\varepsilon _{1}>0$ and/or $\varepsilon _{2}>0$),
a sub-threshold $R^{c}$ appears. Hence, the parameters $\varepsilon _{1}$
and $\varepsilon _{2}$ increase the number of law-offenders caught by the
justice system, even they do not affect on $R_{0}$ and $Q$. Another
extraordinary result in the appearance of backward bifurcation is the
incarceration of law offenders even when $R_{g}<1$. Notice that when $%
R^{c}<R_{g}<1$, the backward bifurcation indicates that the corruption is
controlled if a sufficient number of criminals are caught and incarcerated,
which task is fulfilled by an effective whistleblowing policy, showing that
the quasi-ideal society harbors more untouchable corruptors (the highest top
in the hierarchy). However, when $R_{g}<R^{c}$, there is only the trivial
equilibrium point $P^{0}$ and an ideal corruption-free society is achieved.

Next, we present numerical simulations of the dynamic system (\ref{system})
to corroborate our analytical results -- the cost-benefit analysis by crime
susceptible individuals before committing a crime and the adherence to the
whistleblowing program by those caught by the justice system.

\section{Results}

We simulate equation (\ref{system}) using $4^{th}$ order Runge-Kutta method,
and the initial conditions are given by equation (\ref{init}) unless
explicitly cited. The values for the model parameters are fixed and given in
Table \ref{Tab_param}, except when explicitly cited. In the table, the
reciprocal of the parameters' values given as quotient is the average time
spent in the corresponding compartment, except $\phi $ (we considered $\phi
=\mu $ to have a constant population). For instance, $1/\mu =78$ $years$\ is
the population's life expectancy. We assess qualitatively two dilemmas --
cost-benefit balance by crime-susceptible individuals and the rewards in
adhering to the whistleblowing program by individuals caught by police
investigation.

The crime-prevention rate $\eta $ appears in $R_{0}$ (see equation (\ref%
{btrpart}) for thresholds) through $S^{0}$ (see equation (\ref{equil0}),
with $S^{0}$ following sigmoid-shape from $0$ to $q_{0}$, when $\eta $
varies from $0$ to $\infty $). This parameter is unchanged, and we study the
corruption behavior by varying the per-capita crime-influencing rates $\beta
_{1}$ and $\beta _{2}$, the inhibition coefficients $\kappa _{1}$ and $%
\kappa _{2}$, and the collaborator-depending whistleblowing rates $%
\varepsilon _{1}$ and $\varepsilon _{2}$.

We assume $\beta _{2}=z_{\beta }\beta _{1}$, $\kappa _{2}=z_{\kappa }\kappa
_{1}$ and $\varepsilon _{2}=z_{\varepsilon }\varepsilon _{1}$, where $%
z_{i}\geq 1$ for $i=\beta ,\kappa ,\varepsilon $. The reason behind it is
the enhanced influence to commit a crime (and the effectiveness of the
whistleblowing program) by those caught by police investigation, but waiting
for court trial in freedom ($F$) than the uncaught individuals ($U$).
However, incarcerated individuals ($I$) inhibit crime more than those caught
by police individuals ($C$). We assumed two principal actors in preventing
and combating criminals: police agents who investigate to catch offenders
and judges who carry on trials in a tribunal court. Hence, to be imprisoned,
an offender must be caught by investigators and sentenced by a judge.

\subsection{Crime-susceptible individual's dilemma -- Cost-benefit balance
to commit a crime (corruption)}

Initially, let us assess the cost-benefit balance to commit crime
considering the absence of collaborator-depending whistleblowing program ($%
\varepsilon _{1}=\varepsilon _{2}=0$). The dilemma of crime-susceptible
individuals increases as law offenders are caught and sentenced by the
justice system. Hence, we evaluate how such dilemma ($\kappa _{1}>0$ and $%
\kappa _{2}>0$) affects criminality compared with the absence of crime
inhibition ($\kappa _{1}=\kappa _{2}=0$).

\subsubsection{Case $\protect\kappa _{1}=\protect\kappa _{2}=0$ -- Without
inhibition}

The crime-susceptible individuals are attracted to offend law without any
inhibition by letting the coefficients zero ($\kappa _{1}=\kappa _{2}=0$).
The effectiveness of the justice system (defined by $C^{\ast }$ and $I^{\ast
}$) is evaluated by varying $\beta _{1}$, assuming that $\beta _{2}=z_{\beta
}\beta _{1}$ and letting $z_{\beta }=1.5$ arbitrarily.

\begin{figure}[h]
\centering                                                                   
\subfloat[]{
\includegraphics[scale=0.35]{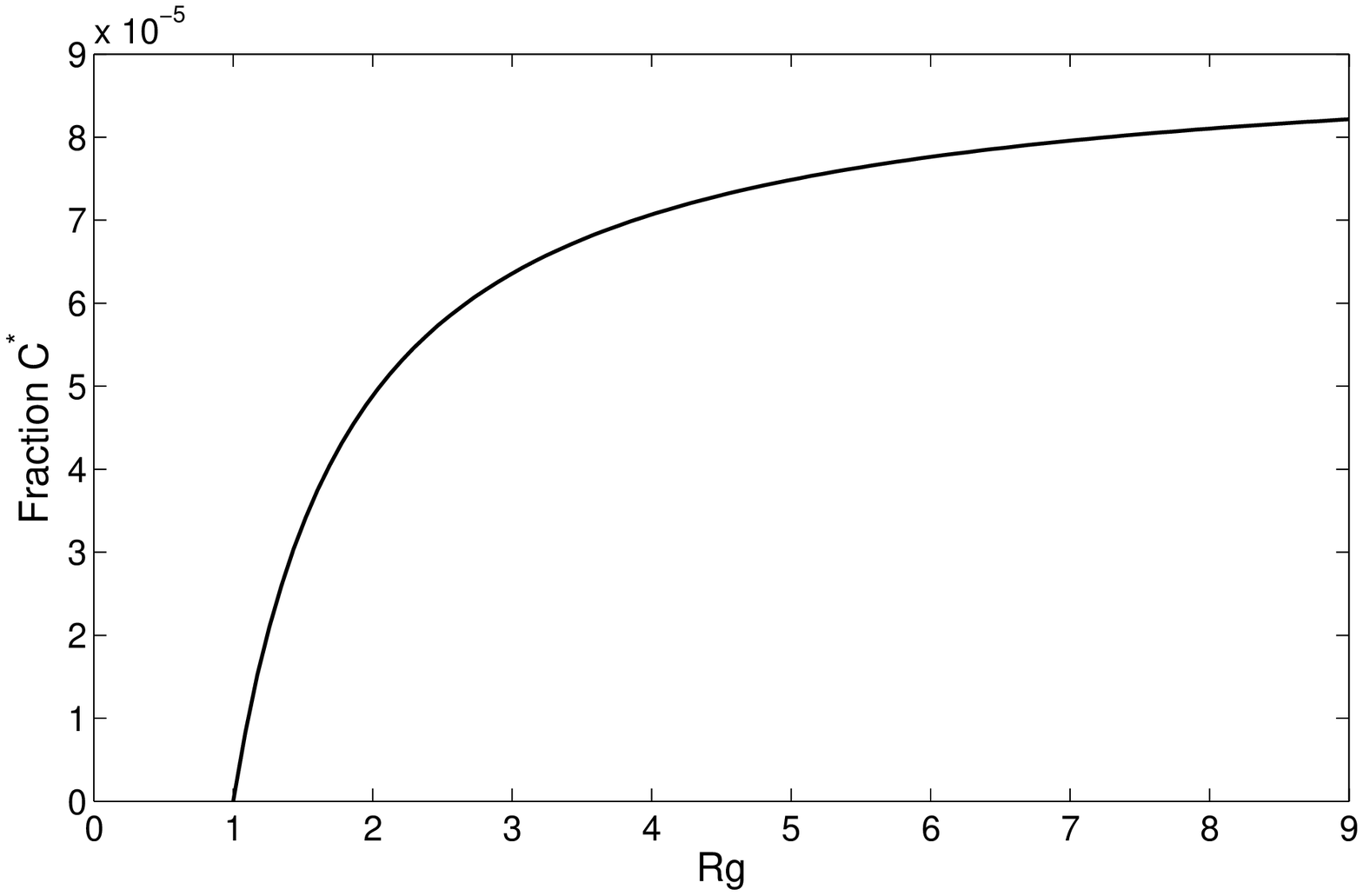}
} 
\subfloat[]{
\includegraphics[scale=0.35]{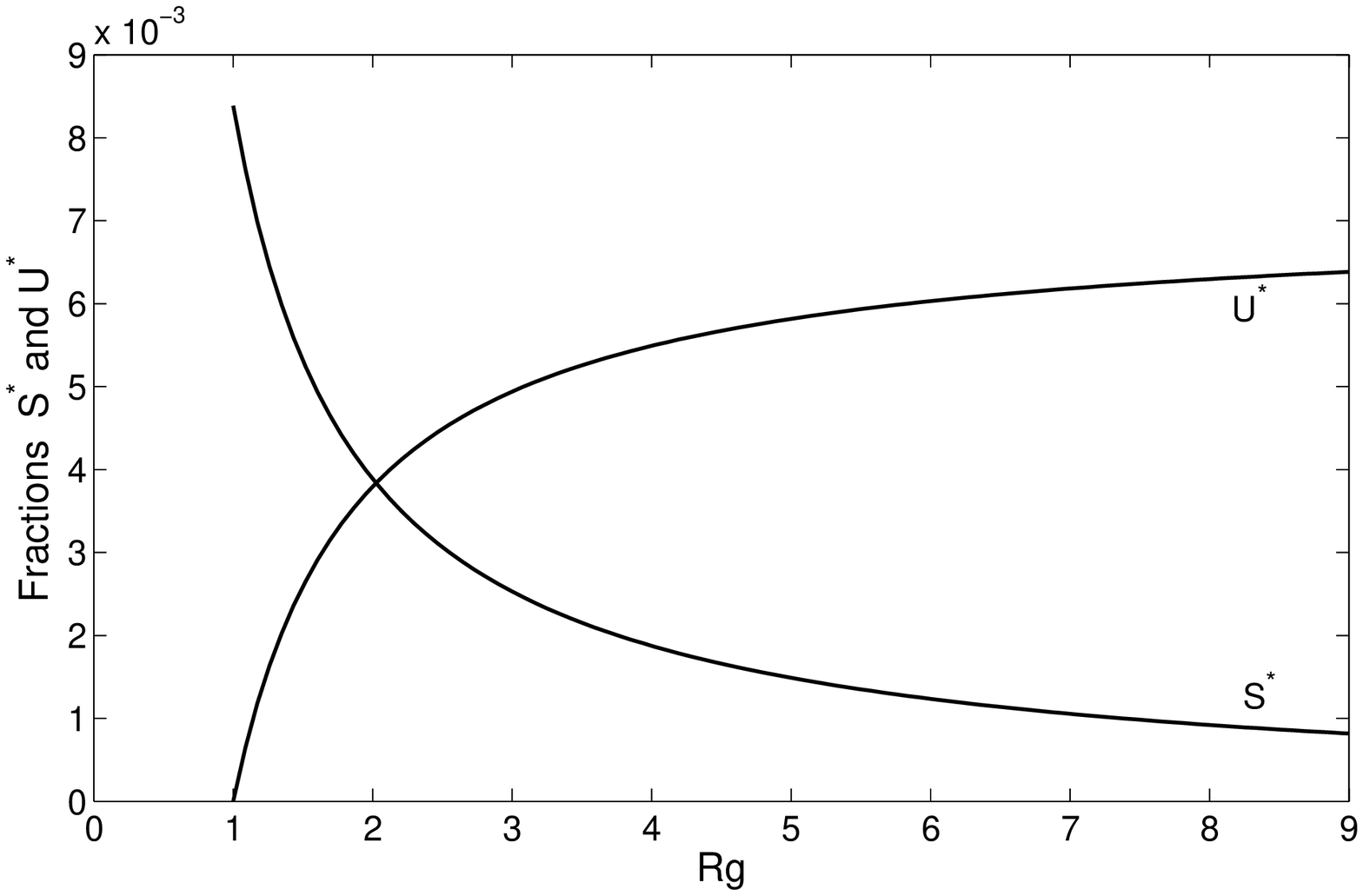}
}\newline
\subfloat[]{
\includegraphics[scale=0.35]{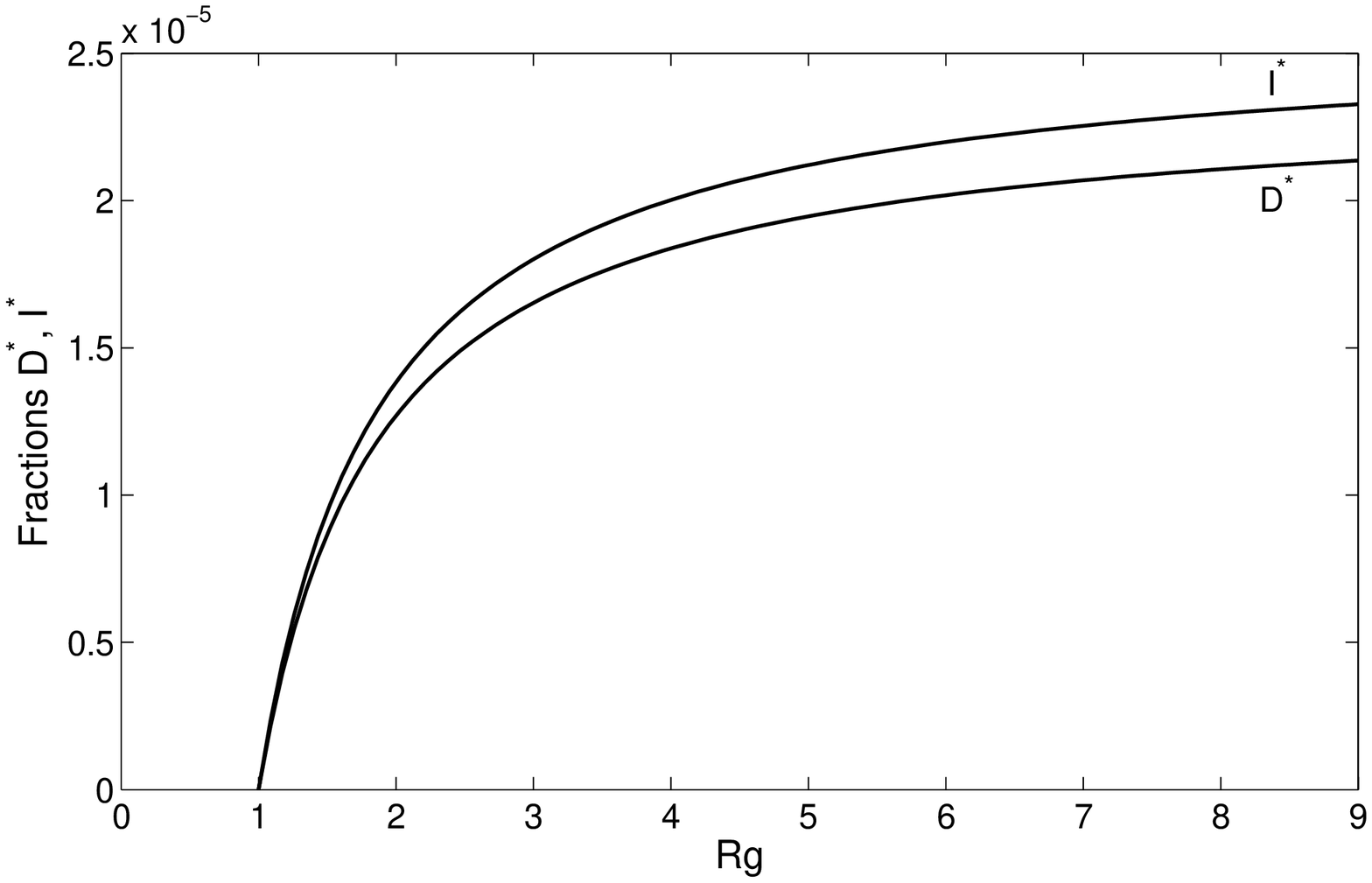}
} 
\subfloat[]{
\includegraphics[scale=0.35]{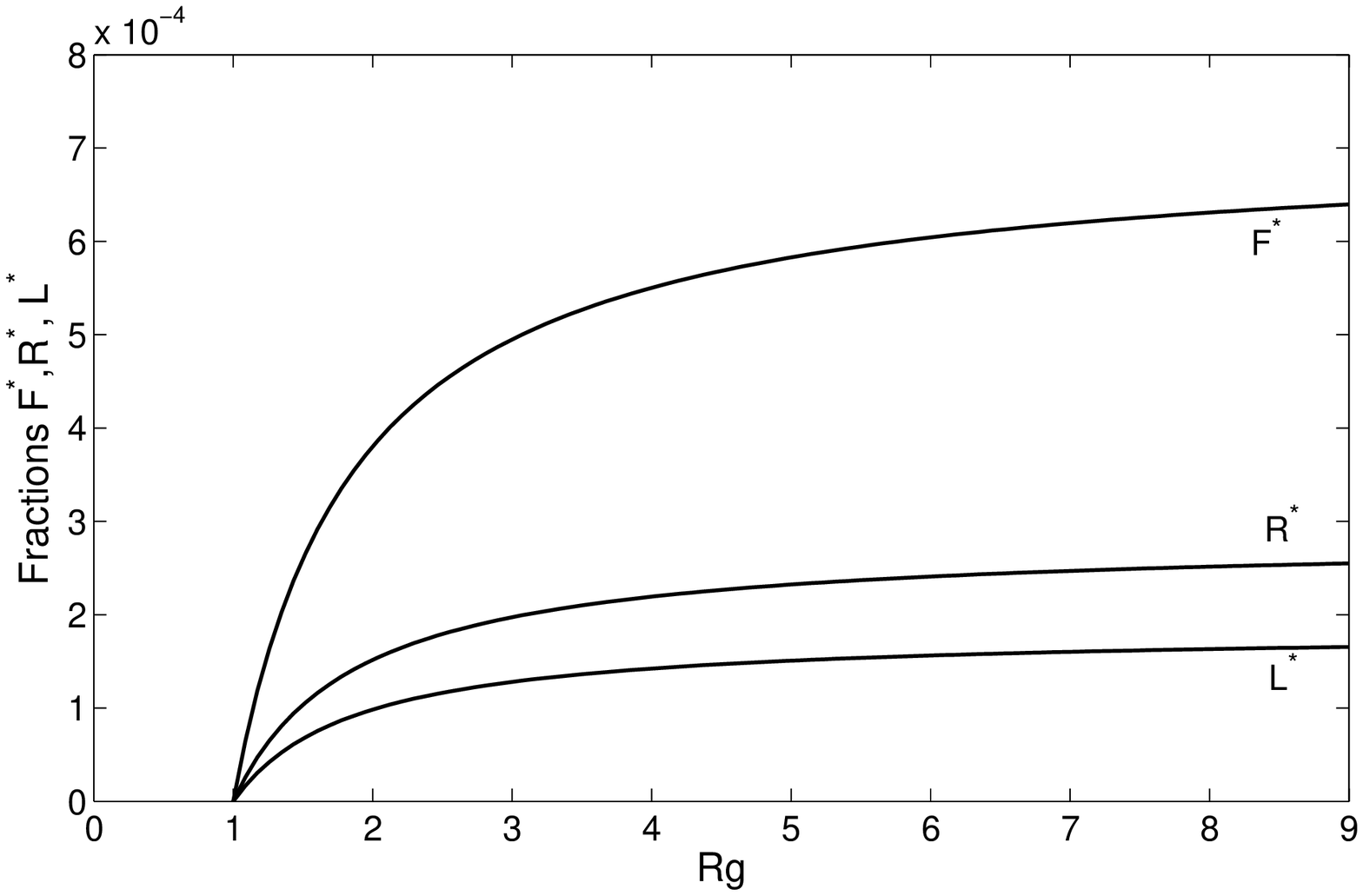}
}
\caption{The equilibrium value $C^{\ast }$ varying the reproduction number $%
R_{g}$ (a), and $S^{\ast }$ and $U^{\ast }$ (b), $D^{\ast }$ and $I^{\ast }$
(c), and $L^{\ast }$, $F^{\ast }$ and $R^{\ast }$ (d) calculated
substituting the solution of $C^{\ast }$ in equation (\protect\ref{equil2}).}
\label{Fig_C2}
\end{figure}

Figure \ref{Fig_C2} shows the equilibrium values varying the reproduction
number $R_{g}$. Figure \ref{Fig_C2}(a) shows the forward bifurcation
diagram: the individuals caught by police investigation $C^{\ast }$ by
varying the reproduction number $R_{g}$. Notice that $C^{\ast }$ is solution
of equation (\ref{equil2}) and $R_{g}$ is given by equation (\ref{Rg}). From
the solution of $C^{\ast }$, Figure \ref{Fig_C2} illustrates other
coordinates of the non-trivial equilibrium point $P^{\ast }$ calculated
using equation (\ref{equil1}): $S^{\ast }$ and $U^{\ast }$ (b), $D^{\ast }$
and $I^{\ast }$ (c), and $L^{\ast }$, $F^{\ast }$ and $R^{\ast }$ (d).

In Appendix \ref{part2}, we showed a unique positive solution given by
equation (\ref{Csol}) appearing for $R_{g}>1$ when $\varepsilon
_{1}=\varepsilon _{2}=0$. Figure \ref{Fig_C2}(a) showed the forward
bifurcation: when $R_{g}<1$, the trivial equilibrium $P^{0}$ ($C^{\ast }=0$)
is globally stable, and for $R_{g}>1$, the non-trivial equilibrium $P^{\ast
} $ ($C^{\ast }>0$)\ is stable (see Figure \ref{for_back}(a) in Appendix \ref%
{part1}). The plateaux (asymptotic value for $R_{g}\rightarrow \infty $) is
approached near $R_{g}=9$.

\subsubsection{Case $\protect\kappa _{1}>0$ and $\protect\kappa _{2}>0$ --
With inhibition}

To assess the dilemma of crime-susceptible individuals to offend the law, we
vary $\kappa _{1}$ assuming that $\kappa _{2}=z_{\kappa }\kappa _{1}$,
letting $z_{\kappa }=1.5$ arbitrarily. From equation (\ref{Csol}) in
Appendix \ref{part2}, when $\kappa _{1}=\kappa _{2}=0$, the inhibition does
not exist at all and $C^{\ast }$ is the maximum for each $R_{g}$, while for $%
\left( \kappa _{1},\kappa _{2}\right) \rightarrow \infty $, the inhibition
is perfect and $C^{\ast }\rightarrow 0$.

\begin{figure}[h]
\centering \subfloat[]{
\includegraphics[scale=0.35]{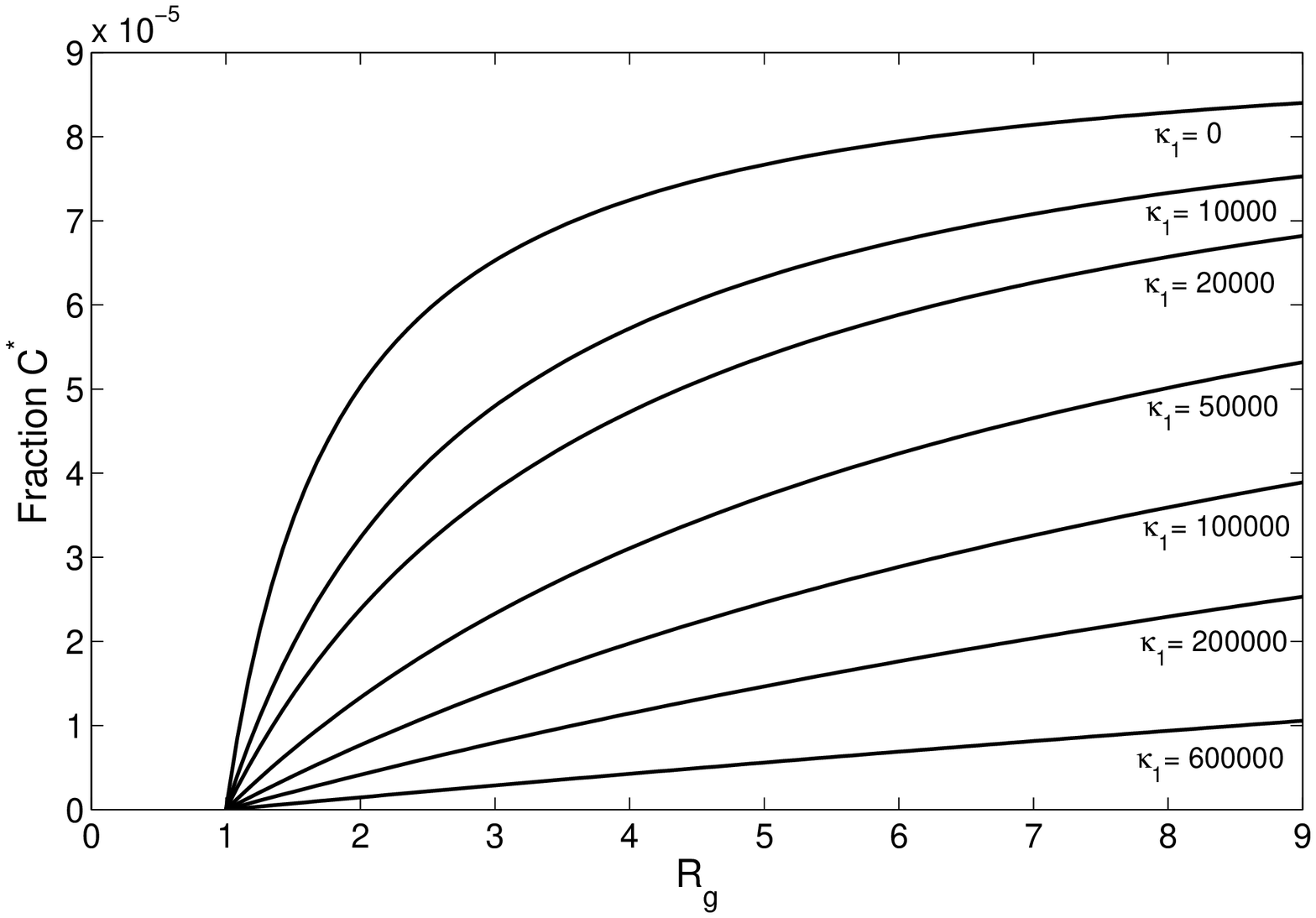}
} 
\subfloat[]{
\includegraphics[scale=0.35]{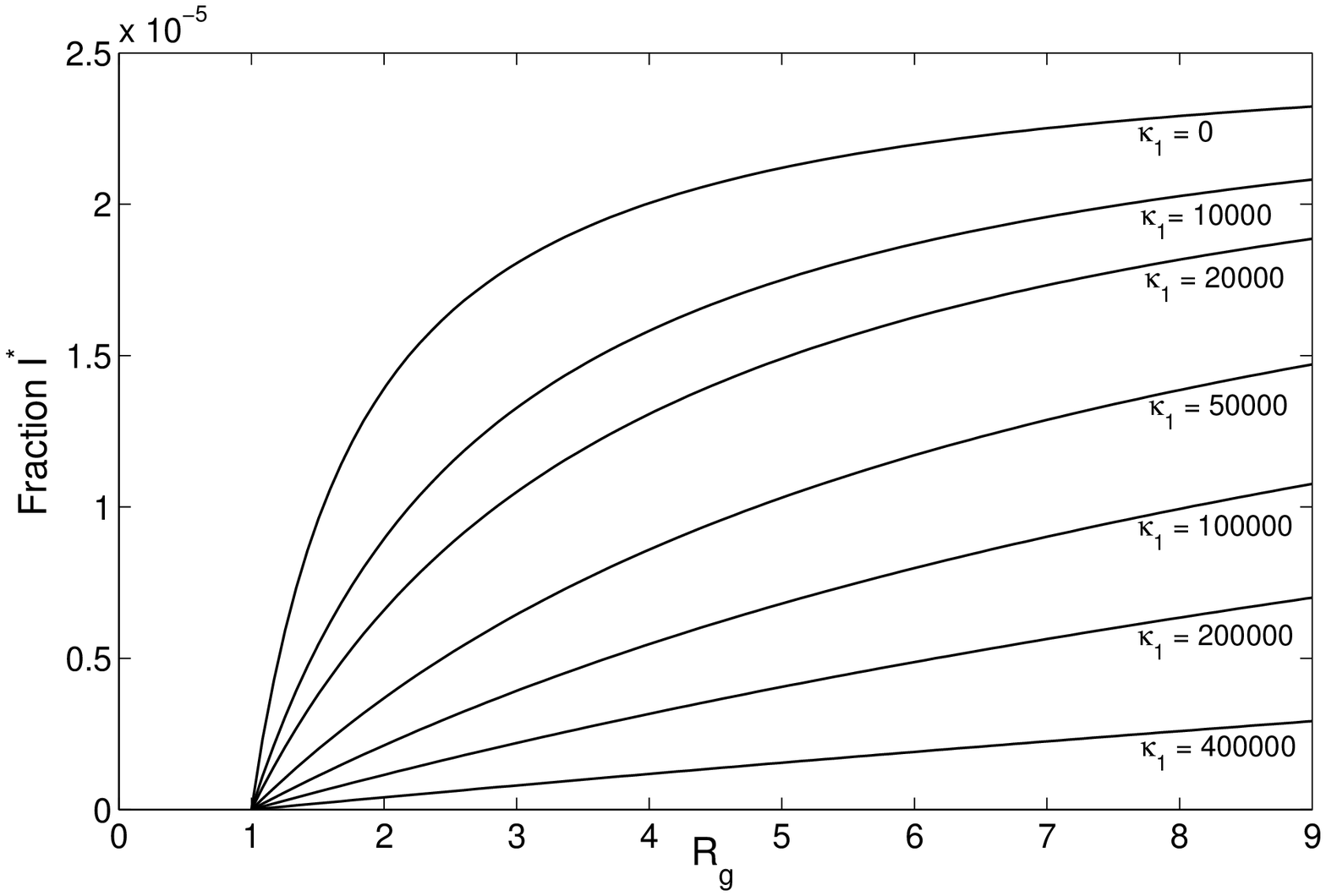}
}
\caption{The equilibrium values $C^{\ast }$ (a) and $I^{\ast }$ (b) varying
the reproduction number $R_{g}$ for values of $\protect\kappa _{1}$ (with $%
\protect\kappa _{2}=1.5\protect\kappa _{1}$) from $0$ to $60\times 10^{4}$.
Both $C^{\ast }$ and $I^{\ast }$ decrease as inhibition coefficients $%
\protect\kappa _{1}$ and $\protect\kappa _{2}$ increase.}
\label{Fig_k1k2}
\end{figure}

Figure \ref{Fig_k1k2} shows the equilibrium values $C^{\ast }$ (a) and $%
I^{\ast }$ (b) varying the reproduction number $R_{g}$ for values of $\kappa
_{1}$ (with $\kappa _{2}=1.5\kappa _{1}$) from $0$ to $60\times 10^{4}$.
Notice that the coefficients $\kappa _{1}$ and $\kappa _{2}$ effectively
inhibit the law offending when $\kappa _{1}C+\kappa _{2}I$ is higher (for
instance, when $\kappa _{1}C+\kappa _{2}I=1$, the force of law offending,
given by equation (\ref{lambda1}), is reduced by half). The model is
structured as fractions in the compartments; hence $\kappa _{1}$ and $\kappa
_{2}$ must assume higher values (order of $10^{4}$ to reduce crime
prevalence significantly).

Table \ref{Tab_k1k2} shows the equilibrium values for $\beta _{1}=7.051$ and 
$\beta _{2}=10.577$ (both in $years^{-1}$) resulting in $R_{g}=4$, for $%
\kappa _{1}=0$, $\kappa _{1}=2\times 10^{4}$, $\kappa _{1}=5\times 10^{4}$, $%
\kappa _{1}=10\times 10^{4}$, and $\kappa _{1}=40\times 10^{4}$ (with $%
\kappa _{2}=1.5\kappa _{1}$). In the absence of criminality, the coordinates
of the trivial equilibrium point $P^{0}$ are $S_{1}^{0}=0.1613$, $%
E^{0}=0.8303$, and $S^{0}=0.0084$. The sum of all crime-related coordinates
(each column) of the non-trivial equilibrium point $P^{\ast }$ is equal to $%
S^{0}$. When $\kappa _{1}=0$, the crime prevalence is $C^{\ast }+I^{\ast
}=9.06\times 10^{-5}$.

\begin{table}[h]
\caption{The equilibrium values for $\protect\beta _{1}=7.051$ and $\protect%
\beta _{2}=10.577$ (both in $years^{-1}$) resulting in $R_{g}=4$, for $%
\protect\kappa _{1}=0$, $\protect\kappa _{1}=2\times 10^{4}$, $\protect%
\kappa _{1}=5\times 10^{4}$, $\protect\kappa _{1}=10\times 10^{4}$, and $%
\protect\kappa _{1}=40\times 10^{4}$ (with $\protect\kappa _{2}=1.5\protect%
\kappa _{1}$).}
\label{Tab_k1k2}\centering                                                   
\begin{tabular}{lllllll}
\hline
$\kappa _{1}$ & $0$ & $2\times 10^{4}$ & $5 \times 10^{4}$ & $10\times
10^{4} $ & $40\times 10^{4}$ &  \\ \hline
$S^{\ast }$ & $1.87\times 10^{-3}$ & $4.6\times 10^{-3}$ & $5.57\times
10^{-3}$ & $6.72\times 10^{-3}$ & $7.87\times 10^{-3}$ &  \\ 
$L^{\ast }$ & $1.42\times 10^{-4}$ & $0.9\times 10^{-4}$ & $0.58\times
10^{-4}$ & $0.36\times 10^{-4}$ & $0.11\times 10^{-4}$ &  \\ 
$U^{\ast }$ & $5.49\times 10^{-3}$ & $3.47\times 10^{-3}$ & $2.24\times
10^{-3}$ & $1.41\times 10^{-3}$ & $0.44\times 10^{-3}$ &  \\ 
$C^{\ast }$ & $7.06\times 10^{-5}$ & $4.47\times 10^{-5}$ & $2.89\times
10^{-5}$ & $1.81\times 10^{-5}$ & $0.56\times 10^{-5}$ &  \\ 
$I^{\ast }$ & $2.00\times 10^{-5}$ & $1.27\times 10^{-5}$ & $0.82\times
10^{-5}$ & $0.51\times 10^{-5}$ & $0.16\times 10^{-5}$ &  \\ 
$F^{\ast }$ & $5.55\times 10^{-4}$ & $3.48\times 10^{-4}$ & $2.25\times
10^{-4}$ & $1.41\times 10^{-4}$ & $0.44\times 10^{-4}$ &  \\ 
$D^{\ast }$ & $1.83\times 10^{-5}$ & $1.16\times 10^{-5}$ & $0.75\times
10^{-5}$ & $0.47\times 10^{-5}$ & $0.15\times 10^{-5}$ &  \\ 
$R^{\ast }$ & $2.19\times 10^{-4}$ & $1.39\times 10^{-4}$ & $0.90\times
10^{-4}$ & $0.56\times 10^{-4}$ & $0.17\times 10^{-4}$ &  \\ \hline
\end{tabular}%
\end{table}

As $\kappa _{1}$ and $\kappa _{2}$ increase, the criminals caught by police
investigation ($C^{\ast }$) and the incarcerated individuals ($I^{\ast }$)
decrease. The crime prevalence $C^{\ast }+I^{\ast }$, in comparison with $%
\kappa _{1}=0$, is reduced to $63.4\%$ ($\kappa _{1}=2\times 10^{4}$), $%
40.9\%$ ($\kappa _{1}=5\times 10^{4}$), $20.5\%$ ($\kappa _{1}=10\times
10^{4}$), and $7.9\%$ ($\kappa _{1}=40\times 10^{4}$). In other words, the
effects of the offenders caught by police and incarceration of criminals on
the naive individuals increase the dilemma to commit a crime.

\begin{figure}[h]
\centering                                                                   
\subfloat[]{
\includegraphics[scale=0.35]{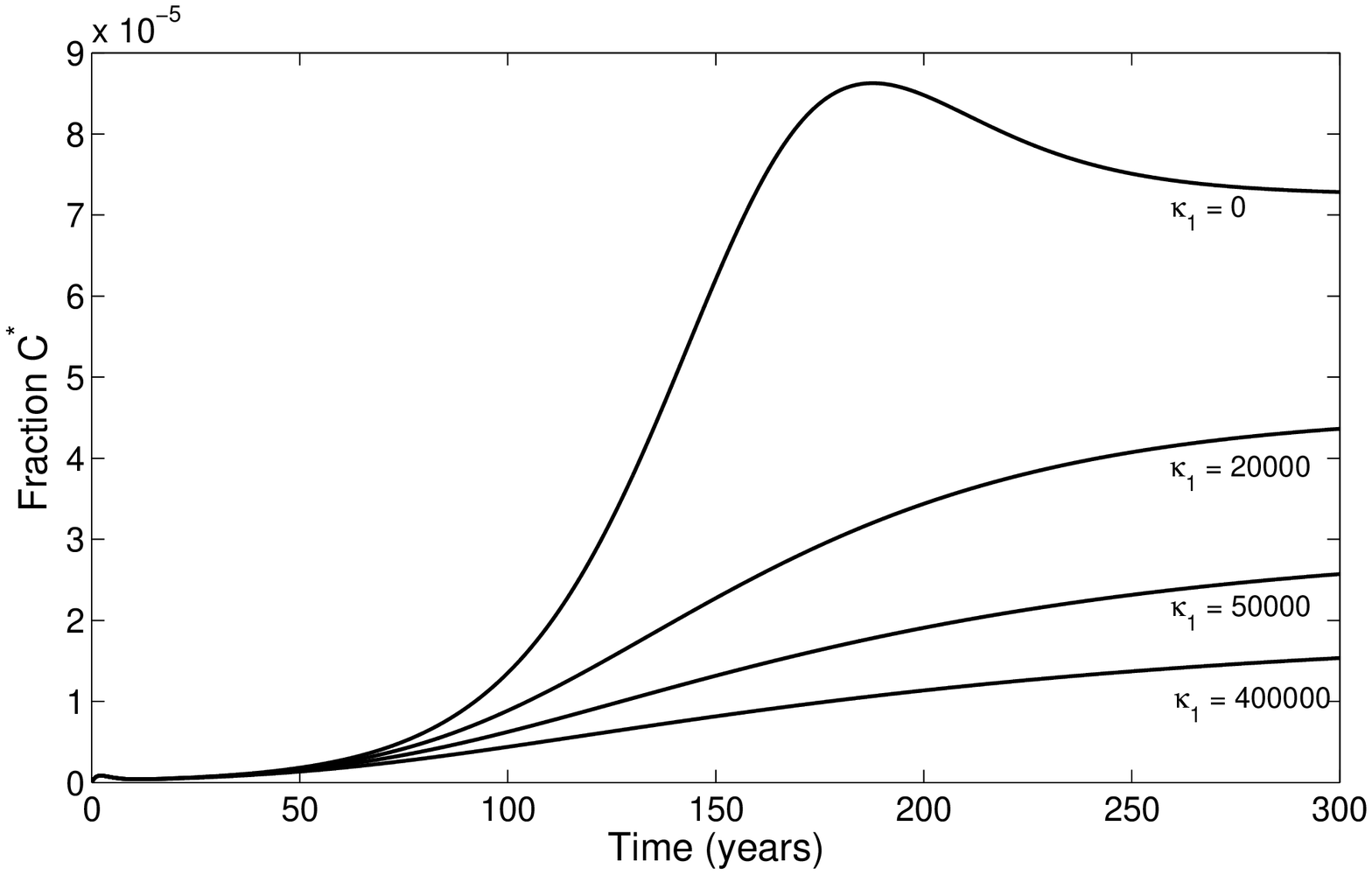}
} 
\subfloat[]{
\includegraphics[scale=0.35]{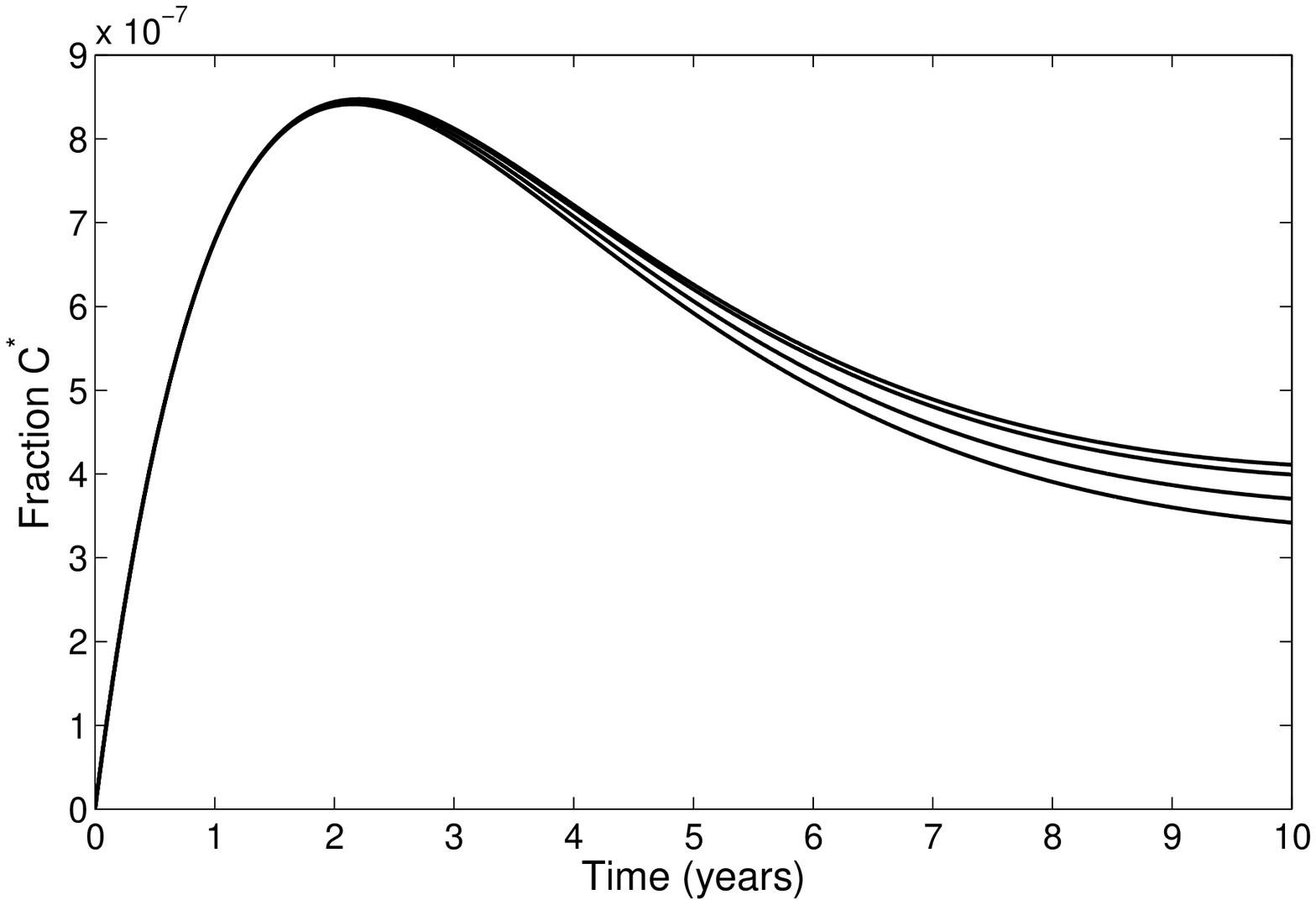}
}
\caption{The long-term (a) and short-term (b) trajectories of $C$, for $%
\protect\kappa _{1}=0$, $\protect\kappa _{1}=2\times 10^{4}$, $\protect%
\kappa _{1}=5\times 10^{4}$, and $\protect\kappa _{1}=40\times 10^{4}$ (with 
$\protect\kappa _{2}=1.5\protect\kappa _{1}$). In Figure 4(b), the curves
with increasing values of $\protect\kappa _{1}$ are from top to bottom.}
\label{Fig_rk1}
\end{figure}

Figure \ref{Fig_rk1} shows the long-term (a) and short-term (b) trajectories
of $C$, for $\kappa _{1}=0$, $\kappa _{1}=2\times 10^{4}$, $\kappa
_{1}=5\times 10^{4}$, and $\kappa _{1}=40\times 10^{4}$ (with $\kappa
_{2}=1.5\kappa _{1}$). In Figure \ref{Fig_rk1}(b), the curves with
increasing values of $\kappa _{1}$ are from top to bottom. The dynamic
trajectories are obtained by solving equation (\ref{system}) with $\beta
_{1}=7.235$ and $\beta _{2}=10.853$ (both in $years^{-1}$) resulting in $%
R_{g}=4$, and using the initial conditions given by equation (\ref{init})
letting arbitrarily $f=0.001S^{0}$.

From Figure \ref{Fig_rk1}(b), the short-term (zoom near $t=0$) trajectories
of $C$ showed a similar increasing phase in the first wave for all values of 
$\kappa _{1}$ and $\kappa _{2}$, with the curves separating after
approximately $2$ $years$. The reason behind this behavior is the fact that $%
R_{g}$ is the same and does not depend on $\kappa _{1}$ and $\kappa _{2}$.
(We recall that $R_{0}$ in epidemiology measures how fast the infection
spreads out initially, while the effects of $Q$ appear later on the
long-term epidemic.)

In Appendix \ref{local}, we showed that the fraction of crime-susceptible
individuals at the steady-state $\chi ^{-1}$ is given by equation (\ref%
{suscpart}) for $\rho _{1}=0$. However, we supposed that $\chi ^{-1}$ is
given by equation (\ref{suscgeral}) for $\rho _{1}>0$. Indeed, the
asymptotic value $S^{\ast }/S^{0}$, where $S^{\ast }=\lim_{t\rightarrow
\infty }S$, calculated from the dynamic trajectories $S$ (Figure \ref%
{Fig_rk1}(a) showed only the dynamic trajectories of $C$) are equal to $%
(1-Q)/R_{0}$ given by equation (\ref{suscgeral}). Therefore, we have two
thresholds $R_{g}=R_{0}+Q$ and $\chi ^{-1}=(1-Q)/R_{0}$.

The justice system can handle the crime inhibition coefficients $\kappa _{1}$
and $\kappa _{2}$ to avoid crime practice by naive individuals. For
instance, the effective police investigation and efficient court trial
spread by media, and the use of handcuffs, coercive conduction, pre-trial
detention, and incarceration after sentence confirmed by a lower federal
court (second instance) could increase $\kappa _{1}$ and $\kappa _{2}$.

\subsection{Prisoner's dilemma -- Adhere or not to the whistleblowing program%
}

Let us consider a systemic (or epidemic) corruption in a country
characterized by the crime reproduction number $R_{g}=4$ ($\beta _{1}=7.051$
and $\beta _{2}=10.577$, both in $years^{-1}$), and by the crime inhibition $%
\kappa _{1}=5\times 10^{4}$ and $\kappa _{2}=7.5\times 10^{4}$. This
case-study corresponds to curves labeled $\kappa _{1}=50000$ in Figures \ref%
{Fig_k1k2} and \ref{Fig_rk1}(a), with values $C^{\ast }=3.11\times 10^{-5}$
and $I^{\ast }=0.86\times 10^{-5}$ from Table \ref{Tab_k1k2}. We assess the
effectiveness of the whistleblowing program to increase the justice system ($%
C^{\ast }$ and $I^{\ast }$) by varying $\varepsilon _{1}$ and $\varepsilon
_{2}$ (rewards offered to the law offenders to participate in the plea
bargain).

\begin{figure}[h]
\centering                                                                   
\subfloat[]{
\includegraphics[scale=0.35]{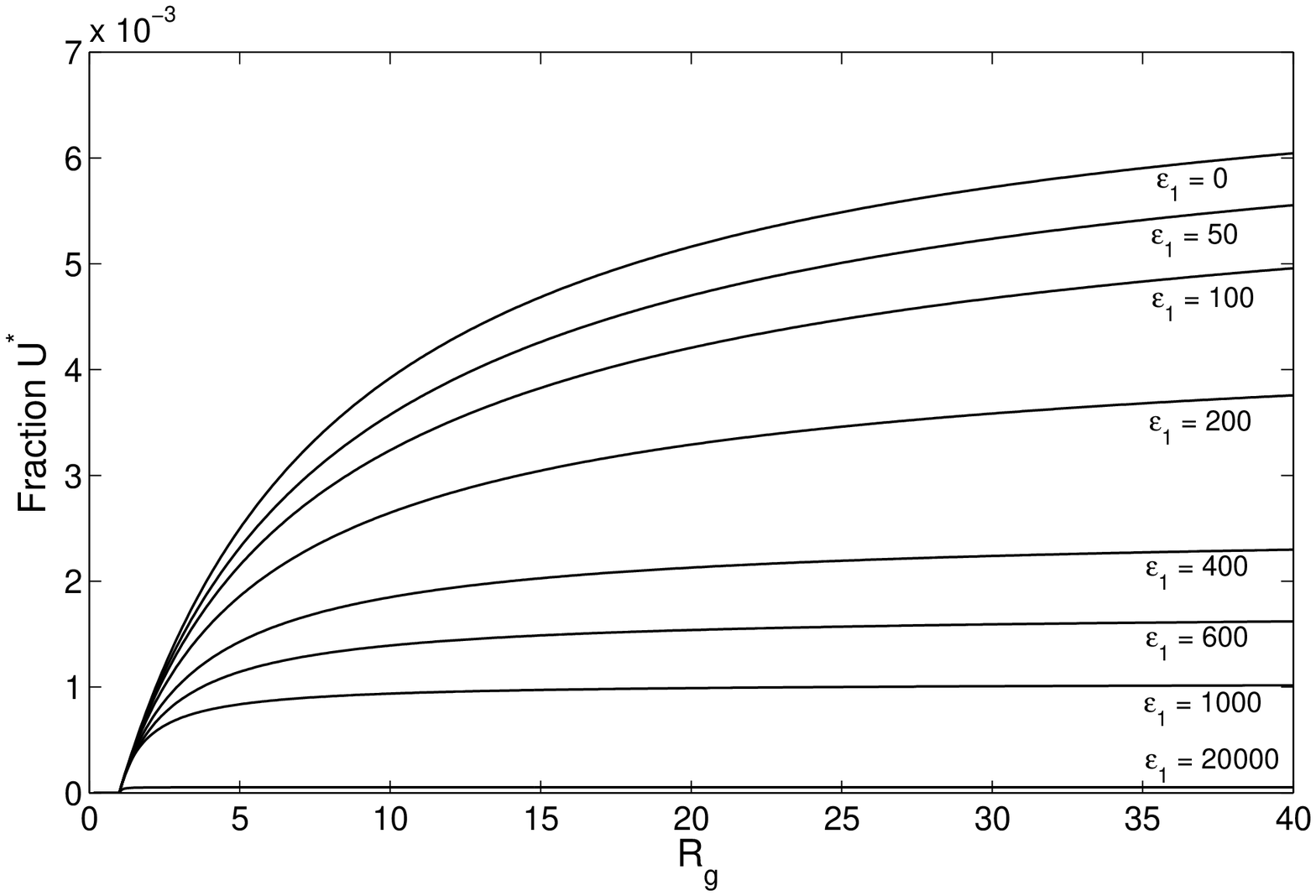}
} 
\subfloat[]{
\includegraphics[scale=0.35]{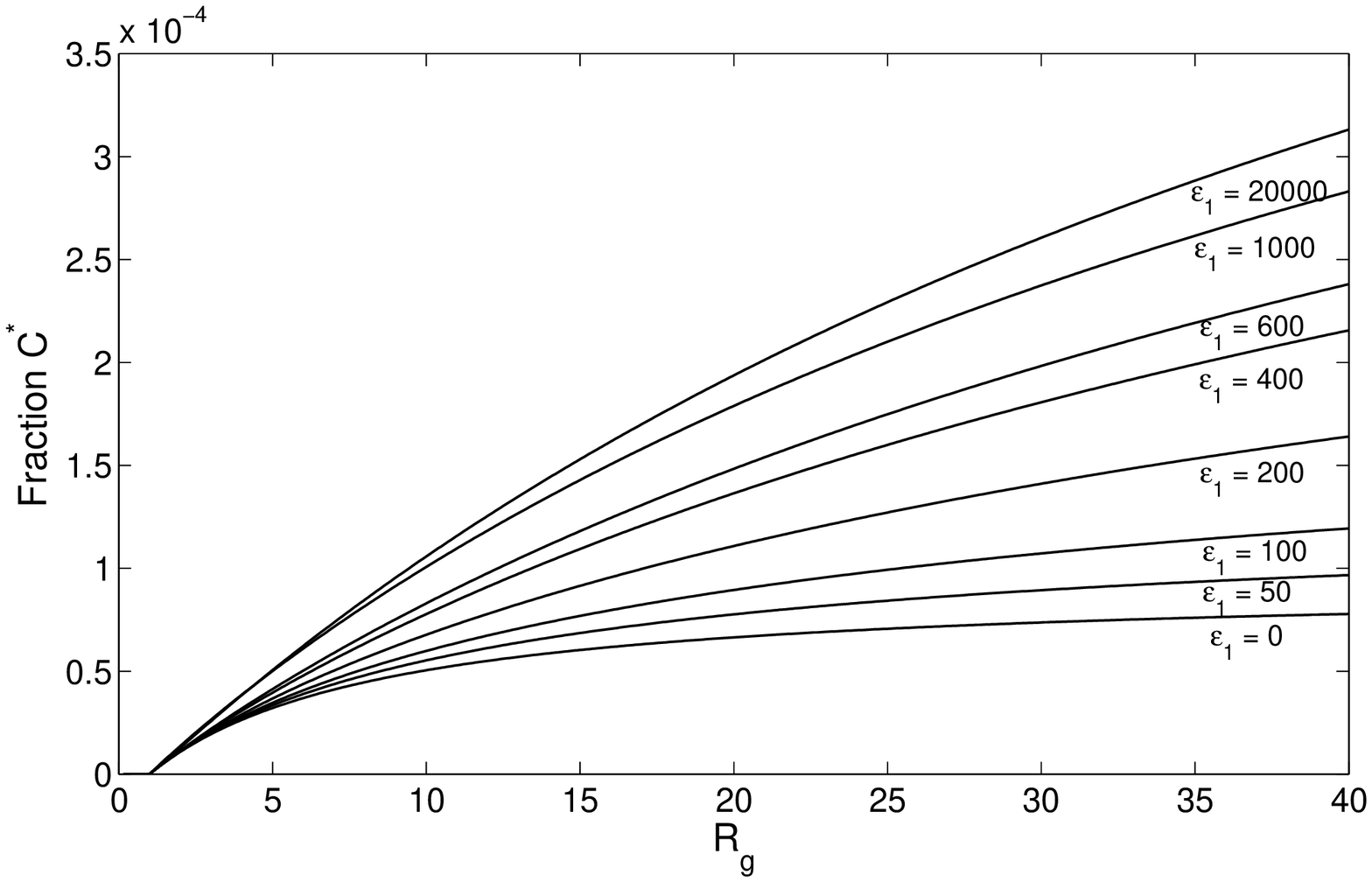}
}
\caption{The equilibrium values $U^{\ast }$ (a) and $C^{\ast }$ (b) letting $%
\protect\varepsilon _{2}=0$ and varying the reproduction number $R_{g}$ for $%
\protect\varepsilon _{1}$ from $0$ to $2\times 10^{4}$ $years^{-1}$.}
\label{Fig_uc}
\end{figure}

Figure \ref{Fig_uc} shows the equilibrium values $U^{\ast }$ (a) and $%
C^{\ast }$ (b) letting $\varepsilon _{2}=0$ and varying the reproduction
number $R_{g}$ for $\varepsilon _{1}$ from $0$ to $2\times 10^{4}$ $%
years^{-1}$. In Figure \ref{Fig_uc}(b), the curve labeled $\varepsilon
_{1}=0 $ corresponds to the curve labeled $\kappa _{1}=50000$ in Figure \ref%
{Fig_k1k2}(a). As $\varepsilon _{1}$ increases, the number of detentions of
the uncaught by police investigation $C^{\ast }$ increases.

\begin{figure}[h]
\centering                                                                   
\subfloat[]{
\includegraphics[scale=0.35]{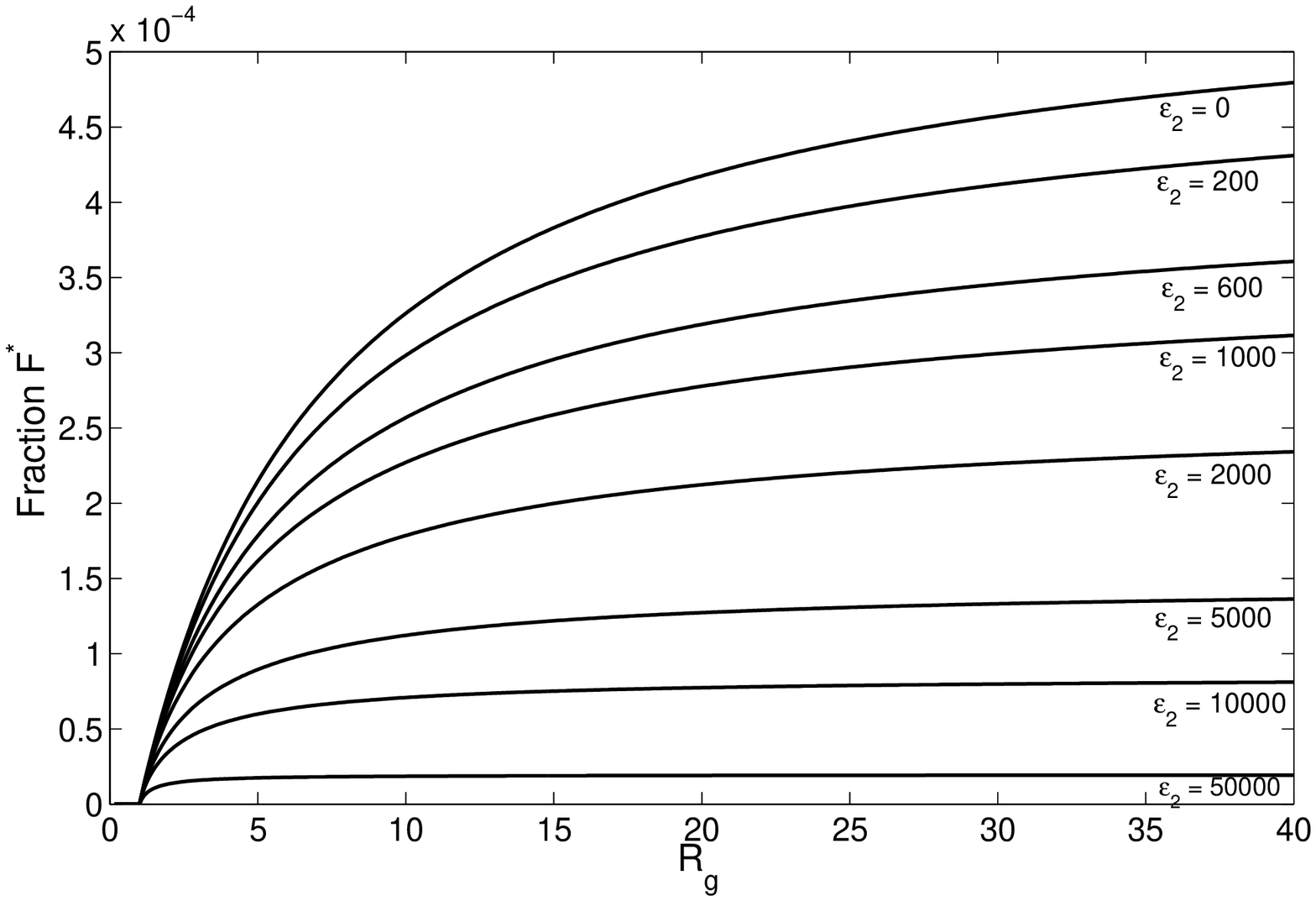}
} 
\subfloat[]{
\includegraphics[scale=0.35]{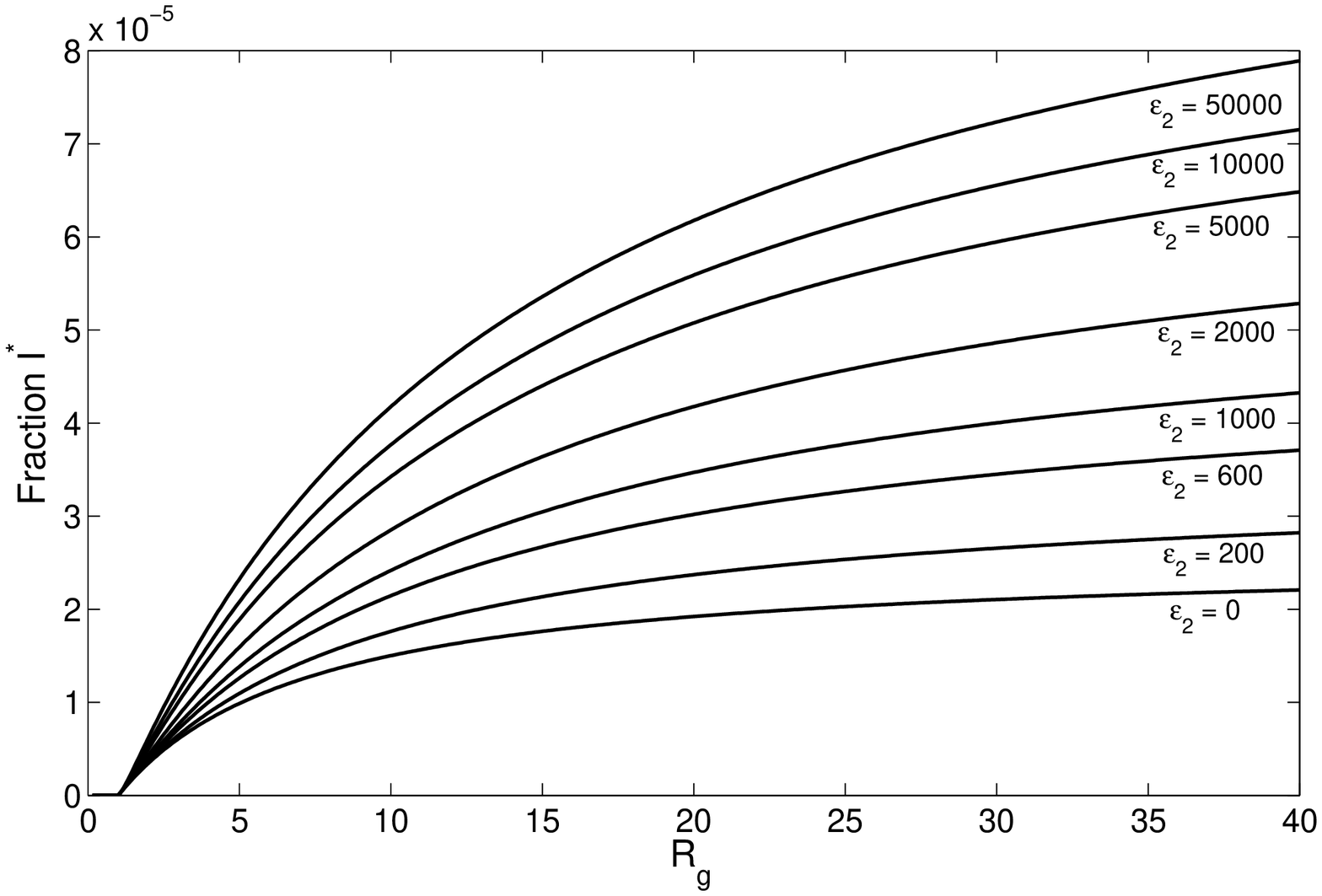}
}
\caption{The equilibrium values $F^{\ast }$ (a) and $I^{\ast }$ (b) letting $%
\protect\varepsilon _{1}=0$ and varying the reproduction number $R_{g}$ for $%
\protect\varepsilon _{2}$ from $0$ to $5\times 10^{4}$ $years^{-1}$.}
\label{Fig_fi}
\end{figure}

Figure \ref{Fig_fi} shows the equilibrium values $F^{\ast }$ (a) and $%
I^{\ast }$ (b) letting $\varepsilon _{1}=0$ and varying the reproduction
number $R_{g}$ for $\varepsilon _{2}$ from $0$ to $5\times 10^{4}$ $%
years^{-1}$. In Figure \ref{Fig_fi}(b), the curve labeled $\varepsilon
_{2}=0 $ corresponds to the curve labeled $\kappa _{1}=50000$ in Figure \ref%
{Fig_k1k2}(b). As $\varepsilon _{2}$ increases, the number of detentions of
the offenders waiting trial in freedom $I^{\ast }$ increases.

Let us illustrate the simultaneous variation of $\varepsilon _{1}$ and $%
\varepsilon _{2}$. Letting $\varepsilon _{2}=z_{\varepsilon }\varepsilon
_{1} $, with $z_{\varepsilon }=1.5$, and varying the reproduction number $%
R_{g}$ for $\varepsilon _{1}$ from $0$ to $20000$ $years^{-1}$, Figure \ref%
{Fig_eps12} shows $C^{\ast }$ for small (a) and high (b) values of $%
\varepsilon _{1}$. The joint action enhances the effectiveness of the
justice system by increasing both $C^{\ast }$ and $I^{\ast }$ (not shown).

\begin{figure}[h]
\centering                                                                   
\subfloat[]{
\includegraphics[scale=0.35]{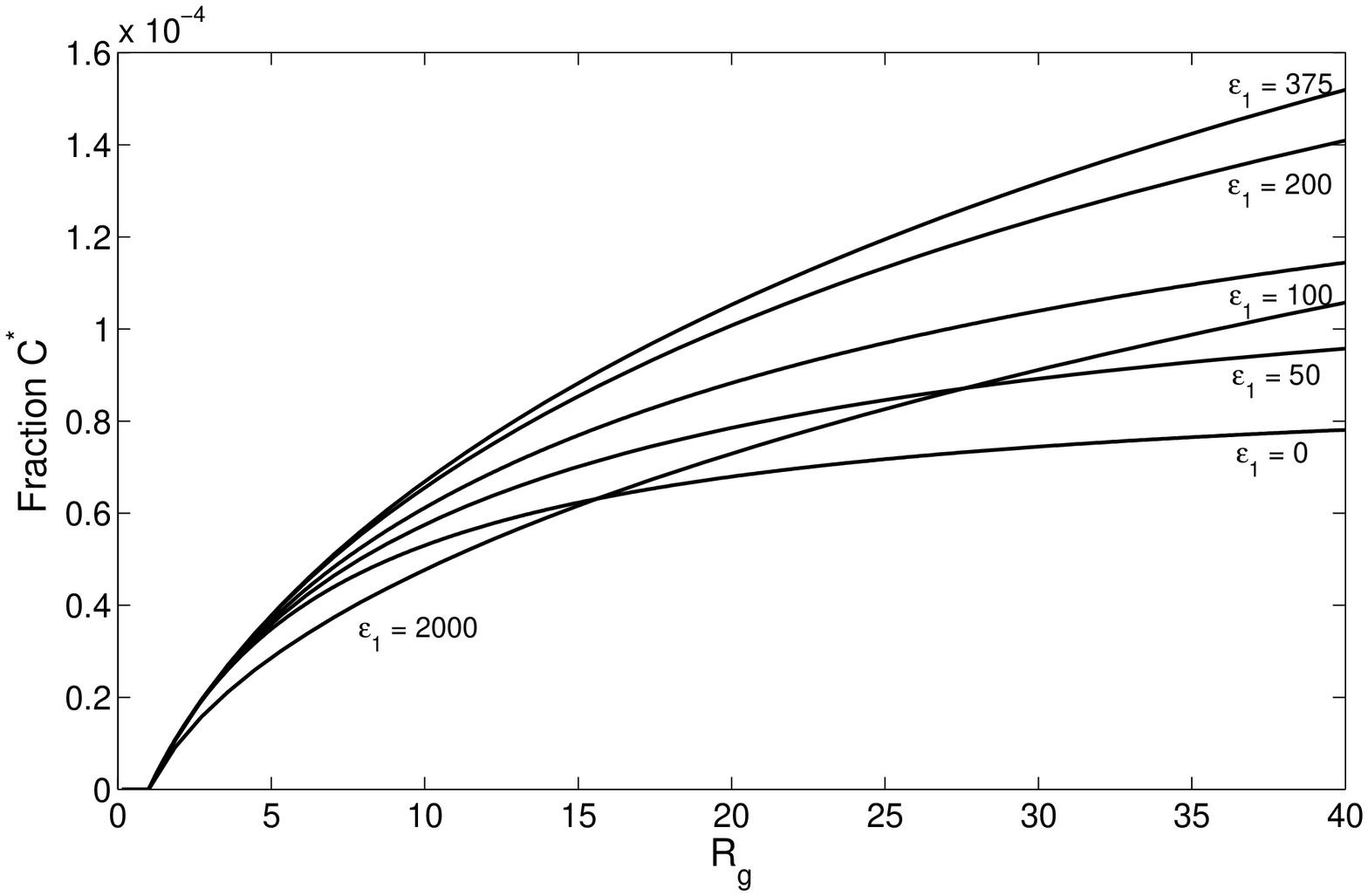}
} 
\subfloat[]{
\includegraphics[scale=0.35]{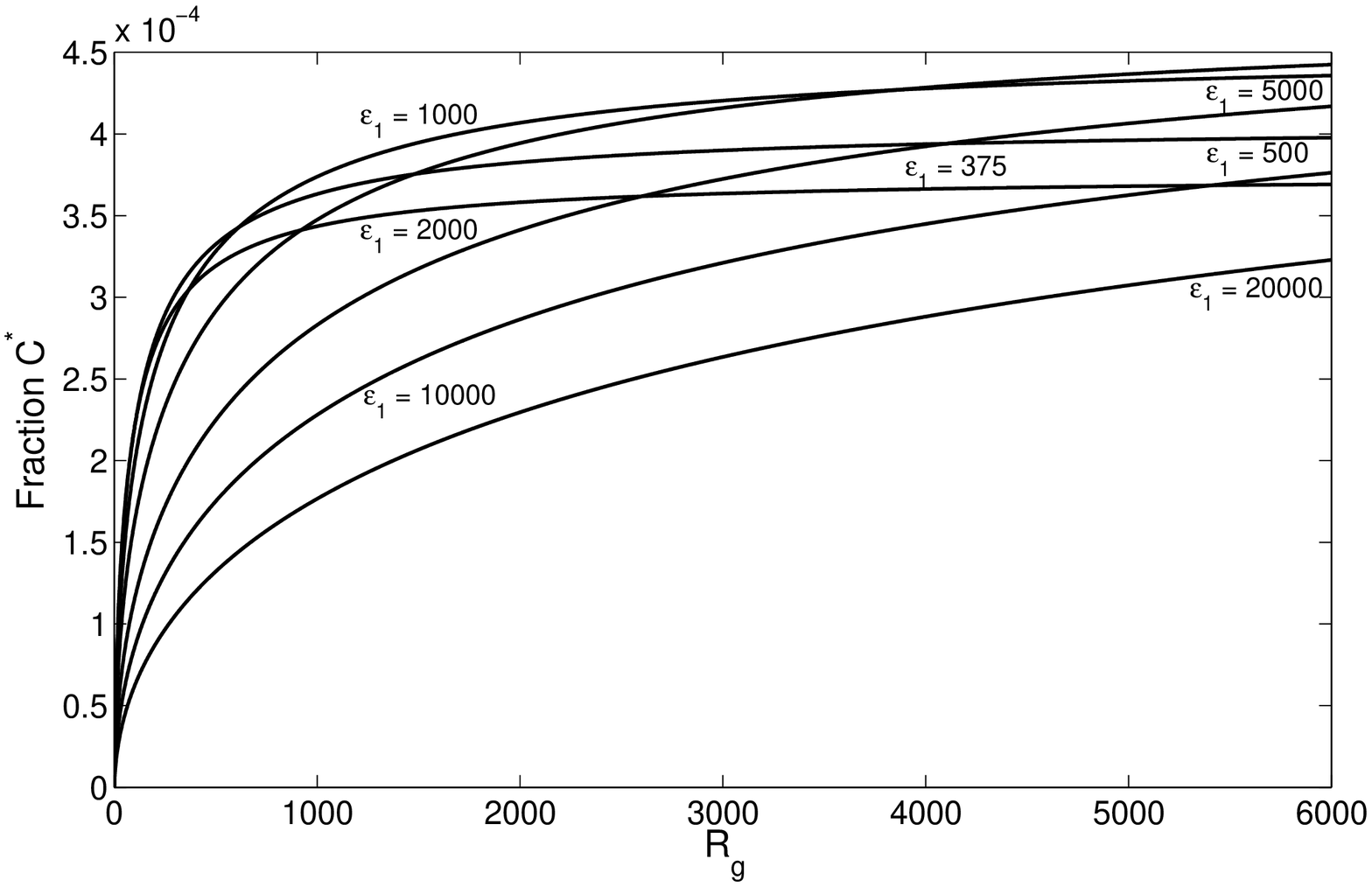}
}
\caption{The equilibrium values $C^{\ast }$ (a) and $I^{\ast }$ (b) varying
the crime reproduction number $R_{g}$ for $\protect\varepsilon _{1}$ from $0$
to $20000$ $years^{-1}$, with $\protect\varepsilon _{2}=1.5\protect%
\varepsilon _{1}$.}
\label{Fig_eps12}
\end{figure}

In Figures \ref{Fig_uc} and \ref{Fig_fi}, we let either $\varepsilon _{2}=0$
or $\varepsilon _{1}=0$ and varied the other parameter. When we vary $%
\varepsilon _{1}$ (or $\varepsilon _{2}$), individuals from class $U$ (or $F$%
) are transferred to $C$ (or $I$); still, the number of individuals in class 
$F$ (or $U$) is not affected directly by the whistleblower's collaboration.
For this reason, as $\varepsilon _{1}$ (or $\varepsilon _{2}$) increases,
the number of individuals in class $U$ (or $F$) diminishes; still, the other
class $F$ (or $U$) is not affected, resulting in smooth changes in the
dynamic (see equation (\ref{lambda1}) for the force of law offending $%
\lambda $). However, the simultaneous variation of $\varepsilon _{1}$ and $%
\varepsilon _{2}$ illustrated in Figure \ref{Fig_eps12} decreased both $U$
and $F$. Consequently, $\lambda $ decreases rapidly, and the dynamic changes
qualitatively at around $\varepsilon _{1}=375$ $years^{-1}$. When $%
\varepsilon _{1}$ and $\varepsilon _{2}$ assume higher values but $R_{g}$ is
small, the force of law offending decreases due to highly reduced numbers of
individuals in $U$ and $F$, and this behavior was shown in Figure \ref%
{Fig_eps12}(a) labeled $\varepsilon _{1}=2000$. However, as $R_{g}$
increases, the numbers of individuals in $U$ and $F$ raise, and the behavior
shown in Figure \ref{Fig_eps12}(b) followed that shown in Figures \ref%
{Fig_uc} and \ref{Fig_fi} -- as $\varepsilon _{1}$ and $\varepsilon _{2}$
increase, $C^{\ast }$ increases.

Table \ref{Tab_eps1eps2} summarizes Figures \ref{Fig_uc}-\ref{Fig_eps12}
showing the equilibrium values varying $\varepsilon _{1}$ and $\varepsilon
_{2}$. We fixed $\beta _{1}=7.051$ and $\beta _{2}=10.577$ (both in $%
years^{-1}$) resulting in $R_{g}=4$, and $\kappa _{1}=50000$ and $\kappa
_{2}=75000$. When $\varepsilon _{1}=\varepsilon _{2}=0$, we have $U^{\ast
}=2.24\times 10^{-3}$, $C^{\ast }=2.89\times 10^{-5}$, $F^{\ast }=2.25\times
10^{-4}$, and $I^{\ast }=0.82\times 10^{-5}$.

\begin{table}[h]
\caption{Summary of Figures \protect\ref{Fig_uc}-\protect\ref{Fig_eps12}
showing the equilibrium values varying $\protect\varepsilon _{1}$ and $%
\protect\varepsilon _{2}$. The parameters $\protect\beta _{1}=7.051$ and $%
\protect\beta _{2}=10.577$ (both in $years^{-1}$) resulting in $R_{g}=4$,
and $\protect\kappa _{1}=50000$ and $\protect\kappa _{2}=75000$ are fixed.}
\label{Tab_eps1eps2}\centering                                               
\begin{tabular}{lllllllll}
\hline
$%
\begin{array}{l}
\varepsilon _{1} \\ 
\varepsilon _{2}=0%
\end{array}%
$ & $%
\begin{array}{l}
U^{\ast } \\ 
\times 10^{-3}%
\end{array}%
$ & $%
\begin{array}{l}
C^{\ast } \\ 
\times 10^{-5}%
\end{array}%
$ & $%
\begin{array}{l}
\varepsilon _{2} \\ 
\varepsilon _{1}=0%
\end{array}%
$ & $%
\begin{array}{l}
F^{\ast } \\ 
\times 10^{-4}%
\end{array}%
$ & $%
\begin{array}{l}
I^{\ast } \\ 
\times 10^{-5}%
\end{array}%
$ & $%
\begin{array}{l}
\varepsilon _{1} \\ 
\varepsilon _{2}=1.5\varepsilon _{1}%
\end{array}%
$ & $%
\begin{array}{l}
C^{\ast } \\ 
\times 10^{-5}%
\end{array}%
$ & $%
\begin{array}{l}
I^{\ast } \\ 
\times 10^{-5}%
\end{array}%
$ \\ \hline
$0$ & $2.24$ & $2.89$ & $0$ & $2.25$ & $0.82$ & $0$ & $2.89$ & $0.82$ \\ 
$100$ & $1.95$ & $3.08$ & $600$ & $1.84$ & $1.03$ & $50$ & $2.94$ & $0.88$
\\ 
$400$ & $1.34$ & $3.50$ & $2000$ & $1.35$ & $1.29$ & $100$ & $2.99$ & $0.94$
\\ 
$1000$ & $0.78$ & $3.80$ & $5000$ & $0.90$ & $1.50$ & $200$ & $3.03$ & $1.05$
\\ 
$20000$ & $0.05$ & $4.07$ & $50000$ & $0.17$ & $1.79$ & $375$ & $3.03$ & $%
1.20$ \\ \hline
\end{tabular}%
\end{table}

As $\varepsilon _{1}$ and $\varepsilon _{2}$ increase, the criminals caught
by police investigation ($C^{\ast }$) and the incarcerated individuals ($%
I^{\ast }$) increase, showing the effectiveness of the plea bargain. When $%
\varepsilon _{2}=0$, the variation in $\varepsilon _{1}$ up to $20000$ $%
years^{-1}$ reduced $U^{\ast }$ up to $2.23\%$ and increased $C^{\ast }$ up
to $141\%$. When $\varepsilon _{1}=0$, the variation in $\varepsilon _{2}$
up to $50000$ $years^{-1}$ reduced $F^{\ast }$ up to $7.6\%$ and increased $%
I^{\ast }$ up to $218\%$. In other words, the dilemma to participate in the
plea bargain must be increased by valuable rewards to decrease the
individuals influencing positively ($U^{\ast }$ and $F^{\ast }$) and
increase those impacting negatively ($C^{\ast }$ and $I^{\ast }$) to commit
a crime. Notice that the overall crime is also decreased -- $U^{\ast
}+C^{\ast }$ up to $4\%$, and $F^{\ast }+I^{\ast }$ up to $15\%$.

\begin{figure}[h]
\centering                                                                   
\subfloat[]{
\includegraphics[scale=0.31]{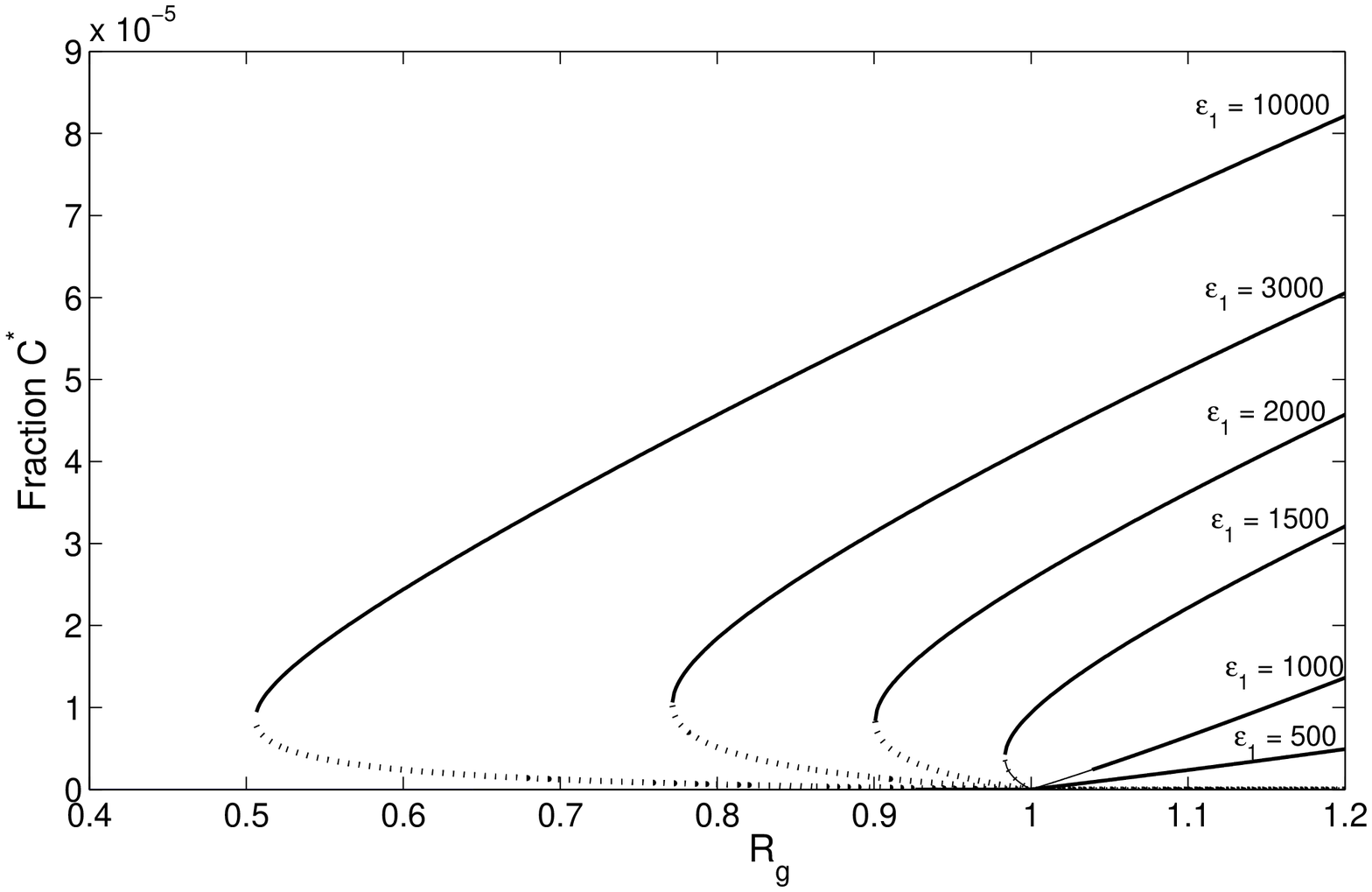}
} 
\subfloat[]{
\includegraphics[scale=0.31]{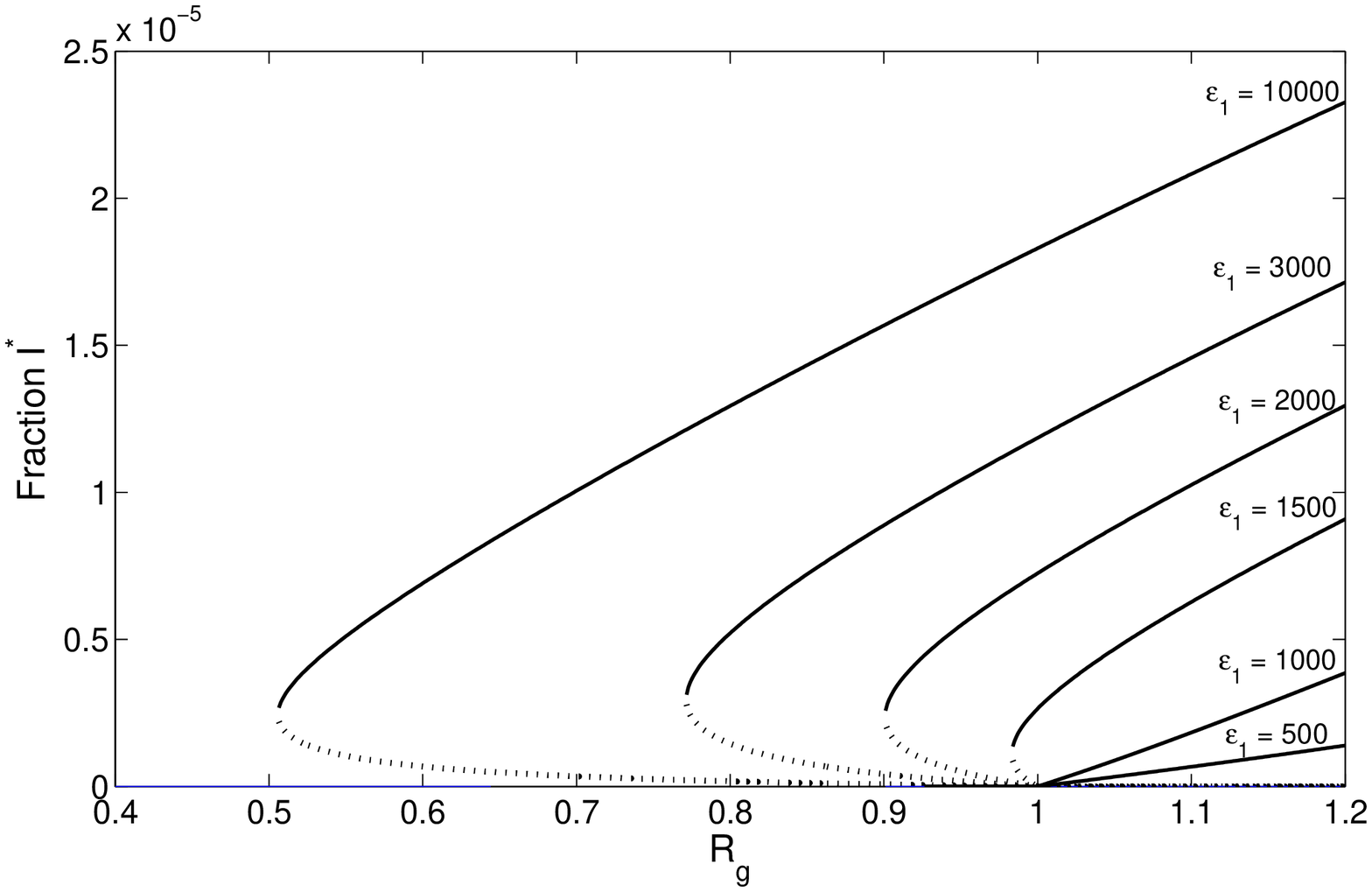}
}
\caption{The equilibrium values $C^{\ast }$ (a) and $I^{\ast }$ (b) near
threshold $R_{g}=1$ and subthreshold $R^{c}$, with the critical value $%
\protect\varepsilon _{1}^{c}=1311.8$ $years^{-1}$.}
\label{Fig_ci_near}
\end{figure}

The backward bifurcation occurs depending on the non-linear
collaborator-dependent rate $\varepsilon _{1}$. In Appendix \ref{part1}, we
showed the occurrence of backward bifurcation when $\beta _{1}=0$, $\kappa
_{2}=0$, and $\rho _{1}=0$. Notice that $\beta _{1}=0$ means that new
offenders not caught by police investigation remain \textquotedblleft
invisible\textquotedblright\ to the society and do not influence
crime-susceptible individuals to commit a crime. Using values in Table \ref%
{Tab_param}, and letting $\varepsilon _{2}=0$, $\beta _{1}=0$, $\beta
_{2}=18.090$ $years^{-1}$ (resulting in $R_{g}=1$), $\kappa _{1}=50000$ and $%
\kappa _{2}=1.5\kappa _{1}$, we have $C^{\ast }=0$ when $\varepsilon
_{1}<\varepsilon _{1}^{c}$, and $C^{\ast }>0$ for $\varepsilon _{1}\geq
\varepsilon _{1}^{c}$, with $\varepsilon _{1}^{c}=1311.8$ $years^{-1}$.
Figure \ref{Fig_ci_near} shows the equilibrium values $C^{\ast }$ (a) and $%
I^{\ast }$ (b) near threshold $R_{g}=1$ and subthreshold $R^{c}\left(
\varepsilon _{1}\right) $, for $\varepsilon _{1}$ from $0$ to $10000$ $%
years^{-1}$ -- the backward bifurcation occurs when $\varepsilon _{1}\geq
\varepsilon _{1}^{c}$. For $\varepsilon _{1}=1500$, $2000$, $3000$, and $%
10000$ (all $years^{-1}$), we have $R^{c}=0.984$, $0.902$, $0.772$, and $%
0.507$. In the backward bifurcation, the dotted curve (lower branch formed
with small solutions $C_{<}^{\ast }$) assumes $C_{<}^{\ast }=0$ at $R_{g}=1$%
. Forward bifurcation is illustrated with $\varepsilon _{1}=500$ and $1000$
(all in $years^{-1}$).

\begin{figure}[h]
\centering                                                                   
\subfloat[]{
\includegraphics[scale=0.31]{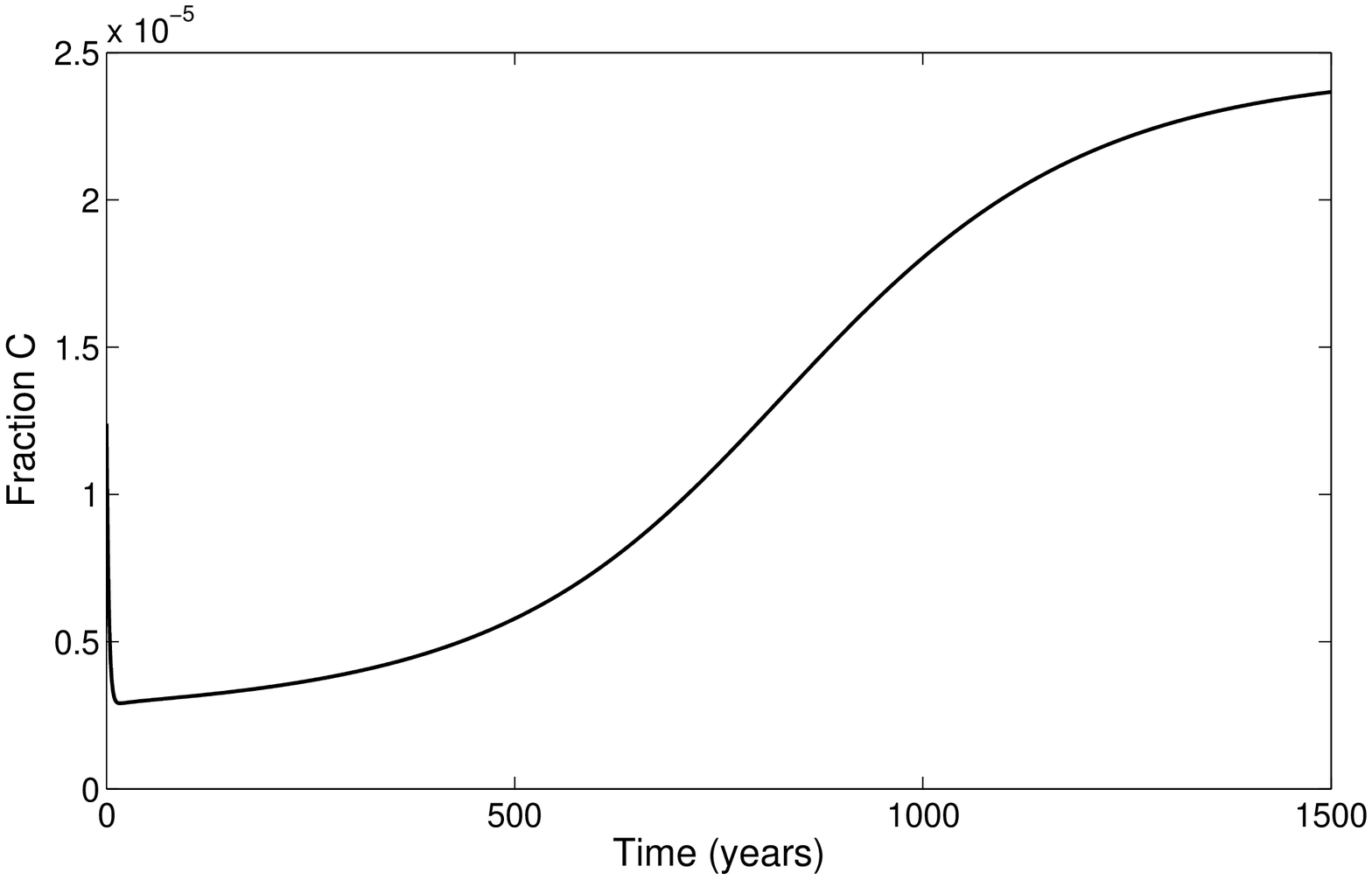}
} 
\subfloat[]{
\includegraphics[scale=0.31]{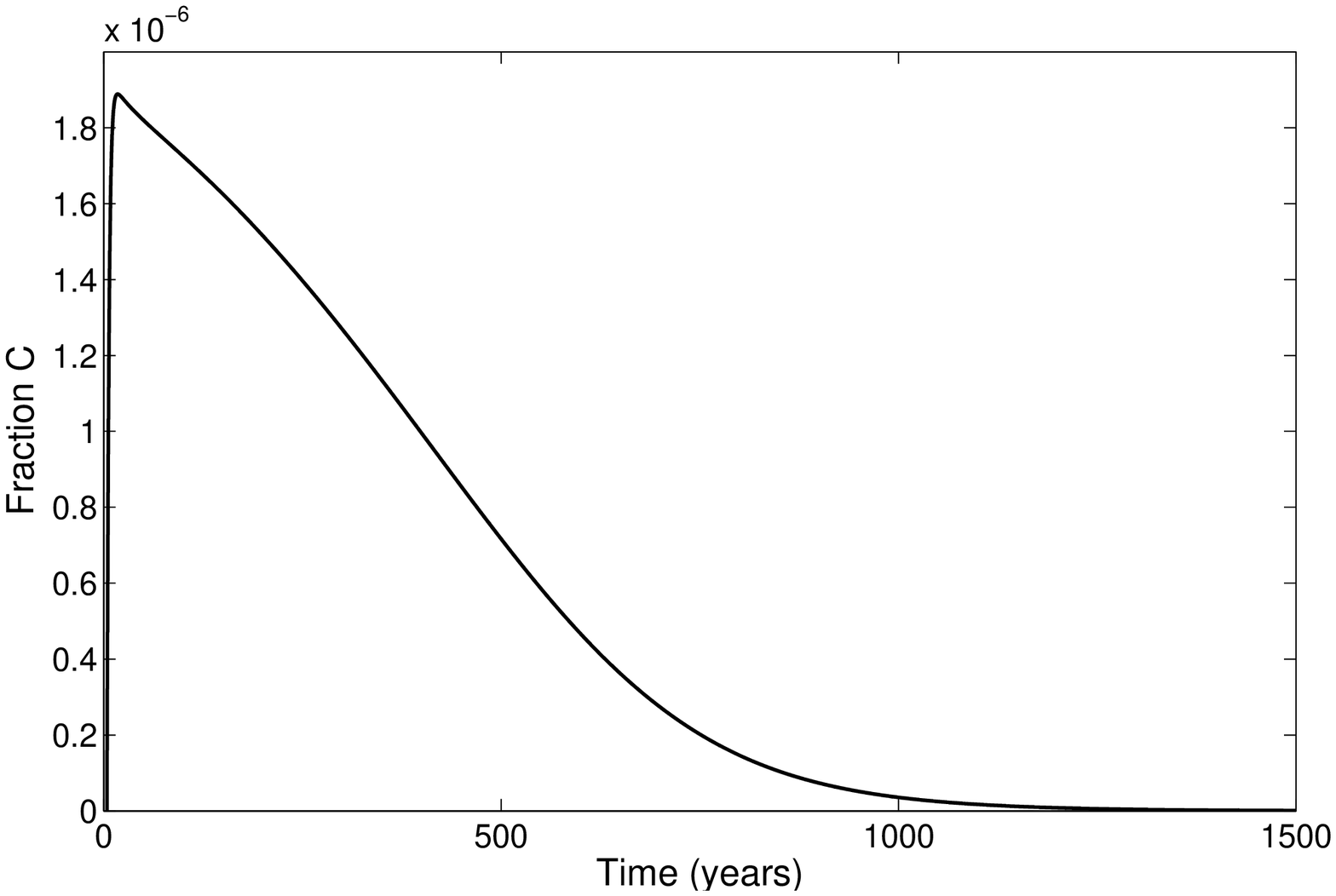}
}
\caption{Illustration of two attracting basins. If $C_{0}=C_{<}^{\ast }-%
\protect\zeta $, the attracting point is the trivial equilibrium point $%
P^{0} $ (a), while for $C_{0}=C_{<}^{\ast }+\protect\zeta $, the attracting
point is the non-trivial equilibrium point $P_{>}^{\ast }$ (b), with $%
\protect\zeta =0.00001C_{<}^{\ast }$.}
\label{Fig_rk2}
\end{figure}

From Figure \ref{Fig_ci_near}, when $\varepsilon \geq \varepsilon _{1}^{c}$,
we have (A) a unique $C^{\ast }>0$ for $R_{g}\geq 1$, corresponding to the
non-trivial equilibrium $P^{\ast }$; (B) two equilibrium values $C_{<}^{\ast
}$\ (unstable equilibrium $P_{<}^{\ast }$) and $C_{>}^{\ast }$ (stable
equilibrium $P_{>}^{\ast }$) for $R^{c}<R_{g}<1$; (C) at $R_{g}=R^{c}$ the
two values collapse to one $C_{<}^{\ast }=C_{>}^{\ast }$; and (D) only $%
C^{\ast }=0$ for $R_{g}<R^{c}$, corresponding to the trivial equilibrium $%
P^{0}$. The backward bifurcation is characterized by the unstable branch
formed by the coordinates of the equilibrium $P_{<}^{\ast }$ separating two
attracting basins -- letting the initial conditions at the coordinates of $%
P_{<}^{\ast }$, except $C(0)=C_{<}^{\ast }-\zeta $, where $\zeta >0$, the
dynamic system approaches the trivial equilibrium $P^{0}$; however, for $%
C(0)=C_{<}^{\ast }+\zeta $, the trajectories approach to $P_{>}^{\ast }$
(see Appendix \ref{part1} and Figure \ref{for_back}(b)) \cite{yang0}.
Letting $\kappa _{1}=50000$, $\kappa _{2}=75000$, $\varepsilon _{1}=10000$ $%
years^{-1}$, $\varepsilon _{2}=0$, $\beta _{1}=0$, $\beta _{2}=9.711$ $%
years^{-1}$ (resulting in $R_{g}=0.6$), Figure \ref{Fig_rk2} illustrates the
role of the unstable equilibrium point $P_{<}^{\ast }$ in the dynamic
system; hence, the initial conditions are%
\begin{equation*}
\left\{ 
\begin{array}{l}
S_{1}(0)=S_{1<}^{\ast },E(0)=E_{<}^{\ast },S(0)=S_{<}^{\ast
},L(0)=L_{<}^{\ast },U(0)=U_{<}^{\ast }, \\ 
C(0)=C_{0},F(0)=F_{<}^{\ast },I(0)=I_{<}^{\ast },D(0)=D_{<}^{\ast
},R(0)=R_{<}^{\ast },%
\end{array}%
\right.
\end{equation*}%
where the coordinates of $P_{<}^{\ast }$ are given by equation (\ref{equil1}%
) substituting the small solution $C_{<}^{\ast }$. Letting a tiny $\zeta
=0.00001C_{<}^{\ast }$, if $C_{0}=C_{<}^{\ast }-\zeta $, the attracting
point is the trivial equilibrium point $P^{0}$, represented by the
trajectory of $C$ in Figure \ref{Fig_rk2}(a). However, for $%
C_{0}=C_{<}^{\ast }+\zeta $, the attracting point is the non-trivial
equilibrium point $P_{>}^{\ast }$, represented by the trajectory of $C$ in
Figure \ref{Fig_rk2}(b).

The plea bargain plays an essential role in criminality. The whistleblowing
program benefits the justice system by helping both police investigations
and court trials. Additionally, suppose that efficient measures adopted by
the justice system decreased the crime reproduction number $R_{g}$ lower
than one. In the absence of, or in a weak whistleblowing program ($%
\varepsilon <\varepsilon _{1}^{c}$), no criminals are caught ($C^{\ast }=0$%
). However, in an enhanced whistleblowing program, more offenders will be
arrested if a critical number of criminals are convinced to collaborate with
the justice system. This case can portray the imprisonments of corruptors at
the highest top of the hierarchy.

\section{Discussion}

We developed a simplified model to describe criminality, a highly complex
social phenomenon. The preceding criminality results are compared with the
Covid-19 pandemic controlling efforts. The reason behind it is the common
aspects of the systemic white-collar crime faced by the \textquotedblleft
car wash operation\textquotedblright\ and the Covid-19 epidemic in Brazil.

The first characteristic is the invisible agent. The struggle against
corruption targets `invisible enemies'\ that deviate budgets from education,
healthcare, etc. -- \textquotedblleft they are often difficult to discover,
to prove, and to punish. Such crimes are usually committed in secret, by
powerful people, and with some degree of sophistication\textquotedblright\ 
\cite{moro}. Like other coronaviruses, SARS-CoV-2 has four structural
proteins, known as the S (spike), E (envelope), M (membrane), and N
(nucleocapsid) proteins; the N protein holds the RNA genome, and the S, E,
and M proteins together create the viral envelope. The diameter of
SARS-CoV-2 virion is 50-200 nanometers (from $5.0\times 10^{-5}$ to $%
20.0\times 10^{-5}$ $mm$) \cite{chen}, and the image at the atomic level of
the spike was obtained by cryogenic electron microscopy \cite{wrapp}.

Another characteristic is novelty. The task force leading the
\textquotedblleft car wash operation\textquotedblright\ in Brazil faced a
novel situation -- the imprisonment of politicians and businessmen. Both
federal judges and prosecutors\ had to learn how to proceed to face
`powerful and invisible enemy'\ hiding as untouchable individuals --
\textquotedblleft police, prosecutors, and the judiciary are often not well
prepared for the investigation, prosecution, and judgment of these highly
sophisticated crimes\textquotedblright\ \cite{moro}. Similarly, SARS-Cov-2
was initially confounded with the SARS-CoV-1 transmission, which resulted in
many misguiding control efforts. The current Covid-19 pandemic showed
utterly unknown, which demanded tremendous researchers' efforts to
understand the transmission of the virus, the efficient treatment, and
effective vaccine \cite{yang4}.

We describe the fights against corruption in Brazil\ and associate them with
the control mechanisms to mitigate the Covid-19 epidemic in Brazil to shed
light on understanding and explaining the \textquotedblleft car wash
operation\textquotedblright .

\subsection{Crime-prevention education}

Figure \ref{Fig_C2} showed the equilibrium values corresponding to the
law-offenders' classes increasing as the crime reproduction number $R_{g}$
increases. The model allows to analyze the reduction in the criminality by
varying the crime-prevention parameters $\eta $ and $q_{0}$, which affect
the crime reproduction number $R_{g}$ given by equation (\ref{Rg}).
According to Dossetor \cite{dossetor}, \textquotedblleft a number of the
most significant `crime prevention' studies have not been done by criminal
justice or criminology field, but by the early childhood and/or health
fields, where crime prevention is but one of a number of effects derived
from early childhood or home visiting interventions\textquotedblright .

The rapid spread of the Covid-19 epidemic required the implementation of
quarantine and protective measures (use of face masks, sanitization of
hands, and social distancing) \cite{yang6}, which showed effectiveness in
controlling the pandemic. Further, the containment of the epidemic was
reinforced by the introduction of vaccines. It is worth stressing the
adherence to these control measures by the majority of the population. The
quarantine, protective measures, and further mass vaccination to avoid
infection can be compared to the crime-prevention parameters $\eta $ and $%
q_{0}$.

In the epidemiology model, higher values for the basic reproduction number $%
R_{0}$ imply a fast propagation of the disease, and more efforts are needed
to eradicate the epidemic. For instance, the attempt to eliminate infection
by a vaccine is achieved when the effective reproduction number $R_{ef}=1$;
in other words, we must vaccinate at least a proportion $p=1-1/R_{0}$ (see
Anderson and May \cite{anderson}). In the case of Covid-19, it is necessary
vaccinating at least $89.2\%$ (for $R_{0}=9.24$ \cite{yang6}) of the
population to eradicate the disease.

Using values for $q_{1}$, $q_{2}$, $q_{3}$, and $\mu $ given in Table \ref%
{Tab_param}, $\beta _{1}=7.051$ and $\beta _{2}=10.577$ (both in $years^{-1}$%
), the threshold $q_{0}^{th}=1.75\times 10^{-3}$ was calculated from
equation (\ref{q0lim}) obtained letting $\left( \eta ,\sigma ,\gamma \right)
\rightarrow \infty $ and $\rho _{1}=\rho _{2}=0$. (Notice that $%
R_{0}=q_{0}/q_{0}^{th}=5.7$, for $q_{0}=0.01$.) Therefore, it is necessary
to crime-protecting at least $99.8\%$ of the population to eliminate the
corruption. Interestingly, considering $R_{0}=7.63$, the Covid-19 epidemic's
first wave fades out when $99.4\%$ of individuals are immunized \cite{yang7}.

Another similarity with the epidemiological modeling is the role of the
crime reproduction number $R_{g}=R_{0}+Q$. Figure \ref{Fig_rk1} showed that
the short-term dynamic was driven by the crime reproduction number $R_{g}$,
while the inhibition parameters $\kappa _{1}$ and $\kappa _{2}$ affected the
long-term dynamic. (We recall that $R_{0}$ in epidemiology measures how fast
the infection spreads out initially, while the effects of $Q$ appear later
on the long-term epidemic \cite{yang2}.)

\subsection{Inhibition of the corruption by\ justice system arresting and
sentencing law-offenders}

Our model was formulated similar to the epidemiological modeling, where the
infection is propagated depending on the contact between susceptible and
infectious individuals -- the force of infection increases proportionally to
the number of infectious individuals. In the Covid-19 epidemic modeling, the
asymptomatic individuals and fraction of mild Covid-19 cases are
transmitting infection, while isolated severe Covid-19 patients in treatment
are not transmitting infection; on the contrary, they influence people to
adhere to control mechanisms. The criminal modeling follows these ideas: (1)
the individuals uncaught ($U$) by police investigation and those caught by
police but not condemned individuals (waiting for a court trial in freedom, $%
F$) influence and encourage crime-susceptible individuals to commit a crime
(benefiting the products of crime); (2) however, the law offenders caught by
police ($C$) and those sentenced by court trial ($I$) inhibit the spreading
of criminality (the cost is being incarcerated).\ Therefore, the force of
law offending increases proportionally to the law offenders evading the
justice system but decreases inversely proportional to the efficacy of the
justice system by catching and incarcerating law offenders. Hence, the force
of law offending $\lambda $ given by equation (\ref{lambda}) mimics the
cost-benefit associated with the decision to participate in corruption.

When the \textquotedblleft car wash operation\textquotedblright\ began
(2014), the justice system was a vital ally to combat crime -- a critical
aspect to be mentioned is that offenders were imprisoned after condemnation
sentences pronounced by lower justice courts (first and second instances).
This interpretation of law resulted in the condemnation and further
imprisonment of politicians and businessmen who committed corruption once
the individual rights of corrupts were temporarily suspended. Moreover,
media covered the combat against the corruption, showing the use of
handcuffs, coercive conduction, and pre-trial detention of those under
investigation. All these measures approved by the Supremo Tribunal Federal
(STF, Supreme Court of Brazil) inhibited somehow the corruption.

Let us assess the decrease of corruption in an enhanced crime inhibiting
society. When the media appropriately covers the corruption and the laws to
combat white-collar crimes are enhanced, from Table \ref{Tab_k1k2}, the
crime prevalence $C^{\ast }+I^{\ast }$, when $\kappa _{1}$ increases from $0$
to $400000$, decreases up to $7.9\%$. The decrease in the corruption when
the inhibition coefficients $\kappa _{1}$ and $\kappa _{2}$ increase was
shown in Figure \ref{Fig_k1k2}. From equation (\ref{Rpartesp}), the increase
in the effectiveness of police investigation ($q_{1}$) and tribunal court ($%
q_{2}$) decreases the basic crime reproduction number $R_{0}$, decreasing
the recruitment of new law offenders. Additionally, the increase in $C$ and $%
I$ decreases the force of law offending $\lambda $.

The Covid-19 pandemic shed light to show how an invisible enemy, the
SARS-CoV-2, must be faced: individuals' rights were suppressed not for days,
but months \cite{yang7}. Indeed, the quarantine associated with individual
(sanitization of hands, use of face masks) and collective (social
distancing) measures contributed to controlling the pandemic. Despite these
control measures, hundreds of thousands of individuals died by lethal
Covid-19 in Brazil. For this reason, the justice system correctly endorsed
the suppression of individual rights for almost two years to face the
struggle against an invisible enemy. In other words, individual rights were
suppressed to result in collective welfare.

Summarizing, the lethality of Covid-19 resulted in the populational
adherence to accept the control measures, which decreased the transmission
of the virus. Roughly compared, the judicial measures suppressing
white-collar criminal's rights temporarily (use of handcuffs, coercive
conduction, pre-trial detentions, for instance) and appropriate coverage of
deleterious effects of corruption by media helped the \textquotedblleft car
wash operation\textquotedblright\ to combat the systemic corruption.

\subsection{Plea bargain}

We showed that the law offenders decreased when the inhibition coefficients $%
\kappa _{1}$ and $\kappa _{2}$ increased. However, the combat against
corruption can be improved by the whistleblowing programs, which arrest and
incarcerate `invisible' white-collar criminals. In organized criminality,
the identification of further law offenders is possible by convincing those
caught by police investigation to adhere to the whistleblowing program. The
collaboration of whistleblowers increases the efficiency of the justice
system, not only catching corrupts but also inhibiting crime-susceptible
individuals from committing a crime. Hence, the police-collaboration rate $%
\epsilon _{1}$ and judge-collaboration rate $\epsilon _{2}$ given by
equation (\ref{epsilon}) mimic the reward associated with the decision to
participate in the plea bargain.

As `invisible enemy', the corrupts are identified by a pool of indirect
shreds of evidence (no one signs a receipt of a bribe). Many judicial
measures (use of handcuffs, coercive conduction, pre-trial detentions, for
instance) must be used with parsimony to combat systemic corruption. Those
suspicious corrupts are sentenced by court tribune after further
investigations confirming their practice. These individuals are convinced to
adhere to the whistleblowing program. How effective the whistleblowing
program is implemented, higher becomes the justice system's effectiveness
measured by the increased incarceration of corrupts which inhibits the
practice of corruption. From Table \ref{Tab_eps1eps2}, the adoption of a
plea bargain by the justice system, when $\varepsilon _{2}=0$ and $%
\varepsilon _{1}$ increases from $0$ to $20000$ $years^{-1}$, $U^{\ast }$
decreases up to $2.23\%$ and $C^{\ast }$ increases up to $141\%$, while when 
$\varepsilon _{1}=0$ and $\varepsilon _{2}$ increases from $0$ to $50000$ $%
years^{-1}$, $F^{\ast }$ decreases up to $7.6\%$ and $I^{\ast }$ increases
up to $218\%$. Figures \ref{Fig_uc} and \ref{Fig_fi} showed that as the
effectiveness of whistleblowing program increases, individuals uncaught by
police investigation ($U$) and those waiting in freedom the court trial ($F$%
) are transferred to the classes of caught by police ($C$) and sentenced by
court trial ($I$). In practice, the increased individuals in the classes $C$
and $I$ increase the inhibition of crime-susceptible individuals to
participate in the corruption, while the decreasing in the classes $U$ and $%
F $ decreases the force of law-offending by crime-susceptible individuals.

As invisible enemy, Covid-19 is detected by symptoms, and those presenting
symptoms are hospitalized and treated if the Real-Time Polymerase Chain
Reaction (RT-PCR) test is positive. For this reason, the identification of
the asymptomatic individuals by mass test and isolating those positively
tested individuals is an efficient mechanism to control the SARS-CoV-2
transmission. At the early phase of the Covid-19 outbreak, many countries
applied this mass screening to catch and isolate the asymptomatic
individuals, which was not implemented in Brazil due to the lack of RT-PCR
and serological tests. As many asymptomatic individuals should be isolated,
the SARS-CoV-2 transmission must fade out quickly.

The initial investigations carried on by the \textquotedblleft car wash
operation\textquotedblright\ were able to catch a small number of corrupts.
However, a well-functioning whistleblowing program helped the
\textquotedblleft car wash operation\textquotedblright\ to identify and
investigate `invisible' agents, sentencing, even so, a former president of
Brazil. Roughly compared, a fraction of SARS-CoV-2 infection manifests
Covid-19 symptoms, but mass tests identified asymptomatic individuals
helping to control the epidemic.

As in the epidemiological modeling, the appearance of backward bifurcation
due to the well-functioning whistleblowing policies showed the existence of
more untouchable corrupts than usual (see Figures \ref{Fig_ci_near} and \ref%
{Fig_rk2}). In other words, to catch the highest hierarchy leaders in
organized crime (corruption), the plea bargain must be very effective allied
to a critical number of whistleblowers. In the disease transmission
modeling, the appearance of backward bifurcation is an additional challenge
to eradicate the infection because we must decrease the effective
reproduction number below the sub-threshold $R_{ef}=R^{c}$ \cite{yang0}.

\subsection{Remarks about \textquotedblleft car wash
operation\textquotedblright\ and Covid-19 epidemic's control in Brazil}

We showed the crucial role of crime inhibition and plea bargain in the
preceding sections. The flowchart shown in Figure \ref{Fig_diagram} and the
crime reproduction number $R_{g}$ established that the police investigation
(parameter $q_{1}$) is the foundation of the justice system. Based on these
results, the court trial (parameter $q_{2}$ and time of court trial $%
1/\gamma $) must be efficient in sentencing criminals. In the
\textquotedblleft car wash operation\textquotedblright\ during 6 years, the
efficient investigation of federal police (parameter $q_{1}$ higher) allied
to further court trial (parameter $q_{2}$ higher and time of court trial $%
1/\gamma $ lower) convicted 165 individuals in both first and second
instances, 49 individuals signed collaboration agreements, and 14 companies
signed leniency agreements \cite{bechara}. In comparison, the STF convicted
only 4 individuals during this period \cite{sardinha}. Additionally, federal
judge Moro convicted the first criminal after 12 months, while STF took 39
months \cite{sardinha}. This discrepancy can be naively explained by
equation (\ref{Rpartesp}) -- the crime reproduction number $R_{g}$ increases
if $q_{1}$ and $q_{2}$ decrease, showing that the STF should not be
productive unprovided of investigators.

It is worth stressing that, since 2018, the STF changed the jurisprudence
and prohibited the incarceration of criminals after condemnation sentences
pronounced by lower justice courts (first and second instances).
Additionally, other justice measures were suppressed or weakened (use of
handcuffs, coercive conduction, pre-trial detention, and plea bargain). To
avoid condemnation, powerful and wealthy offenders caught by police
investigation were defended and well-succeeded by famous lawyers' offices by
manipulating the flaws in law (court procedures, not the innocency of their
customers) and coopting media to spread their point of view. Additionally,
of greater gravity, one member of STF not only accepted their argument but
declared the federal judge Moro suspicious based on records obtained by
hackers. Moreover, citing a non-peer-reviewed chapter of a book \cite%
{marques}, one member of STF released many confessed corrupts sentenced by
lower instances' federal judges stating that \textquotedblleft the `car wash
operation'\ promoted unemployment and economic crisis in
Brazil\textquotedblright .

Let us assess the increase of corruption in a society when crime control
measures fail. When the media does not cover the corruption appropriately,
and the laws to combat white-collar crimes are weakened, from Table \ref%
{Tab_k1k2}, the crime prevalence $C^{\ast }+I^{\ast }$, in comparison with $%
\kappa _{1}=40\times 10^{4}$, is increased to $1258\%$ ($\kappa _{1}=0$), $%
797\%$ ($\kappa _{1}=2\times 10^{4}$), $515\%$ ($\kappa _{1}=5\times 10^{4}$%
), and $322\%$ ($\kappa _{1}=10\times 10^{4}$). On the other hand, from
Table \ref{Tab_eps1eps2}, the relaxation of the plea bargain by the justice
system, when $\varepsilon _{2}=0$ and $\varepsilon _{1}$ is decreased from $%
20000$ $years^{-1}$ to $0$, $U^{\ast }$ is increased up to $4480\%$ and $%
C^{\ast }$ is decreased up to $141\%$, while when $\varepsilon _{1}=0$ and $%
\varepsilon _{2}$ is decreased from $50000$ $years^{-1}$ to $0$, $F^{\ast }$
is increased up to $1324\%$ and $I^{\ast }$ is decreased up to $218\%$. The
overall effect is the encouragement of organized crime by increasing both $%
R_{g}$ and $\lambda $.

Unfortunately, we have a parallel with the Covid-19 epidemic control in
Brazil. The hugely increased number of deaths due to the Covid-19 in Brazil
may have several explanations -- the spread of fake news (for instance, the
broad use of non-scientifically proved treatments such as chloroquine,
hydroxychloroquine, and ivermectin) \cite{singh}, the opposition to
vaccination, and depreciating the public health-threatening by opposing to
the quarantine under economic argumentation (long-lasting closure of their
business will affect the economy and increase unemployment). The continuous
and insistently propagation of the use of non-efficient drugs as early
treatment (encouraging anti-quarantine movement and non-use of individual
and collective protections by a relatively significant proportion of the
population), resulted in a delay in the mass immunization. As a consequence,
a more virulent mutation of SARS-CoV-2 appeared \cite{yang7}.

Corruption does not kill as Covid-19 in a short period; still, the question
is: how many and how long will the corruption practiced by the
\textquotedblleft invisible enemy\textquotedblright\ deviating national
budgets result in deaths? The pandemic will last for some years, but
corruption influences generations. It is worth stressing that Covid-19
fatalities increased in Brazil due to the deviation of resources allocated
to the public healthcare system by politicians and businessmen. How many
people died due to the lack of ICUs, healthcare workers, and vaccines? The
harmful effects of systemic corruption are observed in several areas of
society, for instance, the healthcare system, public education, and security.

Linking Covid-19 and corruption, we summarize that the federal authorities
of Brazil opposed effective combat against an invisible enemy, letting them
as a secondary public health threat. Similarly, the STF relaxed laws to
fight against \textquotedblleft secret organized\textquotedblright\
corruption. As pointed out, the plea bargain programs are not viable in weak
institution environments, where protection is imperfect and court precision
is low \cite{buccirossi}. Additionally, the changes of jurisprudence by STF
will signal to the Legislative (Congress) and the Executive to weaken laws
to combat criminality.

The model developed to describe criminality was applied to corruption.
However, this model considering crime-susceptible's and prisoner's dilemmas
can also be used to organized crime. The justice system's effectiveness can
enhance both dilemmas to discourage the entrance into organized crime and
offer rewards to disestablish the organization (by incarceration of leaders
and cutting the flow of money).

\section{Conclusions}

We developed a deterministic model to describe organized crime considering
dilemmas related to crime commitment and adherence to plea bargain. The
model was applied to describe the \textquotedblleft car wash
operation\textquotedblright\ against corruption in Brazil. To better
understand the rise and fall of the \textquotedblleft car wash
operation\textquotedblright , we compared the combat to corruption with the
actions of STF and the President of Brazil to control the Covid-19 pandemic.

When \textquotedblleft car wash operation\textquotedblright\ initiated the
fight against corruption, the STF endorsed all judicial measures -- (1)
pre-trial detention, incarceration after sentenced by the Second Federal
Court Instance (isolation of criminals by restricting individual rights),
(2) uncover of hidden white-collar criminals by plea bargain (whistleblowing
program), and (3) visible judicial measures (use of handcuffs and coercive
conduction accompanied by appropriate media coverage). Recently, to face the
Covid-19 pandemic, restrictive measures were adopted worldwide -- (1) the
quarantine together with restriction on the economic activities (suppressing
individual rights), (2) an active search of asymptomatic and mild Covid-19
cases by mass tests (uncovering infectious individuals and isolating them),
and (3) use of individual and collective protective measures appropriately
propagated by the media. All these measures to control a lethal SARS-CoV-2
transmission were correctly endorsed by the STF to gain collective benefit.

The STF, however, abandoned the initial support to \textquotedblleft car
wash operation\textquotedblright\ changing the jurisprudence and restricting
or even prohibiting the use of handcuffs, coercive conduction, pre-trial
detention, incarceration after sentenced by the lower Federal Court,
accepting fake-argumentations by lawyers, among others. It is expected that
the relaxing in the combat to the corruption by justice system may increase
the activities of the white-collar criminals, which misconducting Covid-19
epidemic control can shed lights. In Brazil, the President of Brazil and his
staff relaxed and even opposed to the Covid-19 controlling efforts,
encouraging anti-quarantine and anti-vaccine movements, use of
non-scientifically proved treatments, and, also, spreading fake-news. All
these actions somehow could explain the hugely increased number of deaths
due to the Covid-19.

The fight against the invisible Covid-19 enemy showed that in extreme
situations, extreme measures must be demanded. Indeed, our model showed that
drastic measures to restrict individual rights -- using handcuffs, coercive
conduction, pre-trial detention, plea bargain, and other judicial actions --
must be implemented with cautions to inhibit the corruption by powerful and
wealthy \textquotedblleft invisible enemies\textquotedblright . If the
justice system is complacent with them, limiting police investigation, plea
bargain, incarceration after being sentenced by the second federal court,
and other harmful measures, the corruption will increase.



\section*{Conflict of interest}

The authors declare there is no conflict of interest.

\appendix

\renewcommand{\theequation}{\Alph{section}.\arabic{equation}} %
\renewcommand{\thefigure}{\Alph{section}.\arabic{figure}} %
\renewcommand{\thetable}{\Alph{section}.\arabic{table}} %
\setcounter{equation}{0} \setcounter{figure}{0} \setcounter{table}{0}

\section{The trivial equilibrium point $P^{0}$ \label{p_trivial}}

The trivial equilibrium point $P^{0}$ of the system of equations (\ref%
{system}), a perfect society without corruption, is given by%
\begin{equation*}
P^{0}=\left( \bar{S}_{1}=S_{1}^{0},\bar{E}=E^{0},\bar{S}=S^{0},\bar{L}=0,%
\bar{U}=0,\bar{C}=0,\bar{F}=0,\bar{I}=0,\bar{D}=0,\bar{R}=0\right) ,
\end{equation*}%
where%
\begin{equation}
\begin{array}{ccccccc}
S_{1}^{0}=\frac{\mu }{\mu +\eta }, &  & E^{0}=\left( 1-q_{0}\right) \frac{%
\eta }{\mu +\eta }, &  & \mathrm{and} &  & S^{0}=q_{0}\frac{\eta }{\mu +\eta 
},%
\end{array}
\label{equil0}
\end{equation}%
with $S_{1}^{0}+E^{0}+S^{0}=1$, and $S^{0}$ is the fraction of population's
size at the risk of trespassing the law. This equilibrium point, considering 
$\eta \rightarrow \infty $ (well-elaborated crime-prevention policy),
reduces to%
\begin{equation}
\begin{array}{ccccccc}
S_{1}^{0}=0, &  & E^{0}=1-q_{0}, &  & \mathrm{and} &  & S^{0}=q_{0},%
\end{array}
\label{equil0esp}
\end{equation}%
that is, the entire population is divided into two crime-protected ($E$) and
crime-susceptible ($S$) classes depending only on the proportion of
crime-prevention failure $q_{0}$.

The stability of $P^{0}$ is assessed by applying the next generation matrix
theory considering the vector of variables $x=\left( L,U,C,F,I,R\right) $ 
\cite{diekman}. We apply the method proposed in \cite{yang} and proved in 
\cite{yang1}. The global stability is presented in a particular case.

\subsection{The local stability \label{local}}

We present the existence of two thresholds\ -- The gross reproduction number 
$R_{g}$ and the steady-state fraction of the crime-susceptible individuals $%
\chi ^{-1}$ \cite{yang5}.

The Jacobian matrix $\bar{J}$ corresponding to the complementary vector $%
\bar{x}=\left( S_{1},E,S,D\right) $ evaluated at the trivial equilibrium
point $P^{0}$ has negative eigen-values $\xi _{1}=-\left( \mu +\eta \right) $
and $\xi _{2}=\xi _{3}=\xi _{4}=-\mu $ (see equation (\ref{jacobian})
below). Therefore, the stability of the trivial equilibrium point $P^{0}$ is
determined by the next generation matrix theory considering the vector of
variables $x=\left( L,U,C,F,I,R\right) $.

\subsubsection{The gross crime reproduction number $R_{g}$}

Let us define the crime vector $f$ and the transition vector $v$ associated
with the sub-system of equations (\ref{system}) restricted to the variables $%
L$, $U$, $C$, $F$, $I$ and $R$ as 
\begin{equation}
\begin{array}{lll}
f^{T}=\left( 
\begin{array}{c}
\frac{\beta _{1}U+\beta _{2}F}{1+\kappa _{1}C+\kappa _{2}I}S+\rho _{1}R+\rho
_{2}F \\ 
\left( 1-q_{1}\right) \sigma L \\ 
q_{1}\sigma L+\left( \varepsilon _{10}+\varepsilon _{1}D\right) C \\ 
\left( 1-q_{2}-q_{3}\right) \gamma C \\ 
q_{2}\gamma C+\left( \varepsilon _{20}+\varepsilon _{2}D\right) F \\ 
\theta I%
\end{array}%
\right) & \mathrm{and} & v^{T}=\left( 
\begin{array}{c}
\left( \mu +\sigma \right) L \\ 
\left( \mu +\varepsilon _{10}+\varepsilon _{1}D\right) U \\ 
\left( \mu +\gamma \right) C \\ 
\left( \mu +\rho _{2}+\varepsilon _{20}+\varepsilon _{2}D\right) F \\ 
\left( \mu +\theta \right) I \\ 
\left( \mu +\rho _{1}\right) R%
\end{array}%
\right) ,%
\end{array}
\label{fv}
\end{equation}%
where $T$ stands for transposition of a matrix.

The derivatives of $f$ and $v$ with respect to $\left( L,U,C,F,I,R\right) $
evaluated at the trivial equilibrium point are denoted by the matrices $F$
and $V$, where $F$ is%
\begin{equation}
F=\left( 
\begin{array}{cccccc}
0 & \beta _{1}S^{0} & 0 & \beta _{2}S^{0}+\rho _{2} & 0 & \rho _{1} \\ 
\left( 1-q_{1}\right) \sigma & 0 & 0 & 0 & 0 & 0 \\ 
q_{1}\sigma & \varepsilon _{10} & 0 & 0 & 0 & 0 \\ 
0 & 0 & \left( 1-q_{2}-q_{3}\right) \gamma & 0 & 0 & 0 \\ 
0 & 0 & q_{2}\gamma & \varepsilon _{20} & 0 & 0 \\ 
0 & 0 & 0 & 0 & \theta & 0%
\end{array}%
\right) ,  \label{f0}
\end{equation}%
and $V$ is%
\begin{equation}
V=\left( 
\begin{array}{cccccc}
\mu +\sigma & 0 & 0 & 0 & 0 & 0 \\ 
0 & \mu +\varepsilon _{10} & 0 & 0 & 0 & 0 \\ 
0 & 0 & \mu +\gamma & 0 & 0 & 0 \\ 
0 & 0 & 0 & \mu +\rho _{2}+\varepsilon _{20} & 0 & 0 \\ 
0 & 0 & 0 & 0 & \mu +\theta & 0 \\ 
0 & 0 & 0 & 0 & 0 & \mu +\rho _{1}%
\end{array}%
\right) .  \label{v}
\end{equation}

The next generation matrix, denoted by $FV^{-1}$, is given by

\begin{equation*}
FV^{-1}=\left( 
\begin{array}{cccccc}
0 & \frac{\beta _{1}S^{0}}{\mu +\varepsilon _{10}} & 0 & \frac{\beta
_{2}S^{0}+\rho _{2}}{\mu +\rho _{2}+\varepsilon _{20}} & 0 & \frac{\rho _{1}%
}{\mu +\rho _{1}} \\ 
\frac{\left( 1-q_{1}\right) \sigma }{\mu +\sigma } & 0 & 0 & 0 & 0 & 0 \\ 
\frac{q_{1}\sigma }{\mu +\sigma } & \frac{\varepsilon _{10}}{\mu
+\varepsilon _{10}} & 0 & 0 & 0 & 0 \\ 
0 & 0 & \frac{\left( 1-q_{2}-q_{3}\right) \gamma }{\mu +\gamma } & 0 & 0 & 0
\\ 
0 & 0 & \frac{q_{2}\gamma }{\mu +\gamma } & \frac{\varepsilon _{20}}{\mu
+\rho _{2}+\varepsilon _{20}} & 0 & 0 \\ 
0 & 0 & 0 & 0 & \frac{\theta }{\mu +\theta } & 0%
\end{array}%
\right) ,
\end{equation*}%
with the characteristic equation corresponding to $FV^{-1}$ being given by%
\begin{equation}
\xi ^{6}-R_{01}\xi ^{4}-\left( R_{02}+Q_{F}^{1}\right) \xi ^{3}-\left(
R_{03}+Q_{F}^{2}+Q_{R}^{1}\right) \xi ^{2}-\left( Q_{R}^{2}+Q_{R}^{3}\right)
\xi -Q_{R}^{4}=0,  \label{eigen1}
\end{equation}%
where the basic crime reproduction number $R_{0}$ is%
\begin{equation}
R_{0}=R_{01}+R_{02}+R_{03},  \label{R0}
\end{equation}%
with%
\begin{equation}
\begin{array}{lllllll}
R_{01}=\beta _{1}^{th}\beta _{1}, &  & R_{02}=\beta _{2}^{th}\beta _{2}, & 
& \mathrm{and} &  & R_{03}=\beta _{3}^{th}\beta _{2},%
\end{array}
\label{R0part}
\end{equation}%
where $\beta _{1}^{th}$, $\beta _{2}^{th}$, and $\beta _{3}^{th}$ are the
thresholds of parameters $\beta _{1}$ and $\beta _{2}$ given by%
\begin{equation}
\left\{ 
\begin{array}{lll}
\beta _{1}^{th} & = & \displaystyle\frac{\sigma }{\mu +\sigma }\left(
1-q_{1}\right) \frac{S^{0}}{\mu +\varepsilon _{10}} \\ 
\beta _{2}^{th} & = & \displaystyle\frac{\sigma }{\mu +\sigma }q_{1}\frac{%
\gamma }{\mu +\gamma }\left( 1-q_{2}-q_{3}\right) \frac{S^{0}}{\mu +\rho
_{2}+\varepsilon _{20}} \\ 
\beta _{3}^{th} & = & \displaystyle\frac{\sigma }{\mu +\sigma }\left(
1-q_{1}\right) \frac{\varepsilon _{10}}{\mu +\varepsilon _{10}}\frac{\gamma 
}{\mu +\gamma }\left( 1-q_{2}-q_{3}\right) \frac{S^{0}}{\mu +\rho
_{2}+\varepsilon _{20}};%
\end{array}%
\right.  \label{btrpart}
\end{equation}%
and the additional crime reproduction number $Q$ is%
\begin{equation}
Q=Q_{F}^{1}+Q_{F}^{2}+Q_{R}^{1}+Q_{R}^{2}+Q_{R}^{3}+Q_{R}^{4},  \label{Q}
\end{equation}%
with%
\begin{equation}
\left\{ 
\begin{array}{lll}
Q_{F}^{1} & = & \displaystyle\frac{\rho _{2}}{\mu +\rho _{2}+\varepsilon
_{20}}\frac{\sigma }{\mu +\sigma }q_{1}\frac{\gamma }{\mu +\gamma }\left(
1-q_{2}-q_{3}\right) \\ 
Q_{F}^{2} & = & \displaystyle\frac{\rho _{2}}{\mu +\rho _{2}+\varepsilon
_{20}}\frac{\sigma }{\mu +\sigma }\left( 1-q_{1}\right) \frac{\varepsilon
_{10}}{\mu +\varepsilon _{10}}\frac{\gamma }{\mu +\gamma }\left(
1-q_{2}-q_{3}\right) \\ 
Q_{R}^{1} & = & \displaystyle\frac{\rho _{1}}{\mu +\rho _{1}}\frac{\sigma }{%
\mu +\sigma }q_{1}\frac{\gamma }{\mu +\gamma }q_{2}\frac{\theta }{\mu
+\theta } \\ 
Q_{R}^{2} & = & \displaystyle\frac{\rho _{1}}{\mu +\rho _{1}}\frac{\sigma }{%
\mu +\sigma }q_{1}\frac{\gamma }{\mu +\gamma }\left( 1-q_{2}-q_{3}\right) 
\frac{\varepsilon _{20}}{\mu +\rho _{2}+\varepsilon _{20}}\frac{\theta }{\mu
+\theta } \\ 
Q_{R}^{3} & = & \displaystyle\frac{\rho _{1}}{\mu +\rho _{1}}\frac{\sigma }{%
\mu +\sigma }\left( 1-q_{1}\right) \frac{\varepsilon _{10}}{\mu +\varepsilon
_{10}}\frac{\gamma }{\mu +\gamma }q_{2}\frac{\theta }{\mu +\theta } \\ 
Q_{R}^{4} & = & \displaystyle\frac{\rho _{1}}{\mu +\rho _{1}}\frac{\sigma }{%
\mu +\sigma }\left( 1-q_{1}\right) \frac{\varepsilon _{10}}{\mu +\varepsilon
_{10}}\frac{\gamma }{\mu +\gamma }\left( 1-q_{2}-q_{3}\right) \frac{%
\varepsilon _{20}}{\mu +\rho _{2}+\varepsilon _{20}}\frac{\theta }{\mu
+\theta }.%
\end{array}%
\right.  \label{Qpart}
\end{equation}%
We interpret below $R_{0i}$, $i=1,2,3$, and $Q_{j}^{i}$, $j=F,R$ and $%
i=1,2,3,4$.

Equation (\ref{eigen1}) does not have an analytical expression for the
spectral radius. However, by applying the conjecture proposed in \cite{yang}
and proved in \cite{yang1}, we concluded that the gross crime reproduction
number $R_{g}$, defined by%
\begin{equation}
R_{g}=R_{0}+Q,  \label{Rg}
\end{equation}%
is one of the thresholds \cite{yang2}. $R_{0}$ is the basic crime
reproduction number, and $Q$ is an additional crime reproduction number due
to the crime relapse of released ($R$) and court waiting in freedom ($F$)
individuals. In epidemiological modeling, $R_{0}$ dictates the beginning of
the epidemic, while the effects of $Q$ appear on the long-term dynamic (see 
\cite{yang2}). Hence, $P^{0}$ is locally asymptotically stable if $R_{g}<1$.

The partial reproduction numbers $R_{02}$ and $R_{03}$ depending on $\beta
_{2}$ obey%
\begin{equation*}
\begin{array}{lllll}
R_{02}\geq R_{03} &  & \mathrm{if} &  & q_{1}\geq \frac{\varepsilon _{10}}{%
\mu +2\varepsilon _{10}}.%
\end{array}%
\end{equation*}%
If $\varepsilon _{10}=0$, then $q_{1}\geq 0$\ (for all values of $q_{1}$, $%
R_{02}\geq R_{03}$); and for $\varepsilon _{10}\rightarrow \infty $, $%
R_{02}\geq R_{03}$ for $q_{1}\geq 1/2$.

Notice that $Q_{j}^{i}$ given by equation (\ref{Qpart}), with $j=F,R$ and $%
i=1,2,3,4$, depends on $\left( \varepsilon _{10},\varepsilon _{20},\rho
_{1},\rho _{2}\right) $. When all parameters are zero, $\left( \varepsilon
_{10},\varepsilon _{20},\rho _{1},\rho _{2}\right) =0$, then $Q_{j}^{i}=0$.
However, when $\left( \varepsilon _{10},\varepsilon _{20},\rho _{1},\rho
_{2}\right) \rightarrow \infty $, using that%
\begin{equation*}
\begin{array}{lll}
\lim\limits_{\left( \rho _{2},\varepsilon _{20}\right) \rightarrow \infty
}\rho _{2}/\left( \mu +\rho _{2}+\varepsilon _{20}\right) =\frac{1}{2} & 
\mathrm{and} & \lim\limits_{\left( \rho _{2},\varepsilon _{20}\right)
\rightarrow \infty }\varepsilon _{20}/\left( \mu +\rho _{2}+\varepsilon
_{20}\right) =\frac{1}{2},%
\end{array}%
\end{equation*}%
$Q_{j}^{i}$ depend only on $q_{1}$, $q_{2}$, and $q_{3}$ resulting, from
equation (\ref{Q}), in $Q=1-q_{3}<1$. Hence, we have%
\begin{equation}
0\leq Q=Q_{F}^{1}+Q_{F}^{2}+Q_{R}^{1}+Q_{R}^{2}+Q_{R}^{3}+Q_{R}^{4}<1
\label{Qrange}
\end{equation}%
when all parameters vary from $0$ to $\infty $.

Let us analyze particular cases for the basic crime reproduction number $%
R_{0}$ and additional crime reproduction number $Q$. We assume the absence
of self-whistleblowers, that is, $\varepsilon _{10}=0$ and $\varepsilon
_{20}=0$.

\begin{enumerate}
\item $\sigma \rightarrow \infty $, that is, the cover-up period is zero --
in this case, $\sigma /\left( \mu +\sigma \right) \rightarrow 1$ and $%
R_{g}=R_{0}+Q$, with%
\begin{equation*}
\left\{ 
\begin{array}{lll}
R_{0} & = & \displaystyle\left( 1-q_{1}\right) \frac{\beta _{1}S^{0}}{\mu }%
+q_{1}\frac{\gamma }{\mu +\gamma }\left( 1-q_{2}-q_{3}\right) \frac{\beta
_{2}S^{0}}{\mu +\rho _{2}} \\ 
Q & = & \displaystyle\frac{\rho _{2}}{\mu +\rho _{2}}q_{1}\frac{\gamma }{\mu
+\gamma }\left( 1-q_{2}-q_{3}\right) +\frac{\rho _{1}}{\mu +\rho _{1}}q_{1}%
\frac{\gamma }{\mu +\gamma }q_{2}\frac{\theta }{\mu +\theta }.%
\end{array}%
\right.
\end{equation*}

\item $\sigma \rightarrow \infty $ plus $\gamma \rightarrow \infty $, that
is, the court trial period is zero -- in this case, $\sigma /\left( \mu
+\sigma \right) \rightarrow 1$ and $\gamma /\left( \mu +\gamma \right)
\rightarrow 1$, and $R_{g}=R_{0}+Q$, with%
\begin{equation*}
\left\{ 
\begin{array}{lll}
R_{0} & = & \displaystyle\left( 1-q_{1}\right) \frac{\beta _{1}S^{0}}{\mu }%
+q_{1}\left( 1-q_{2}-q_{3}\right) \frac{\beta _{2}S^{0}}{\mu +\rho _{2}} \\ 
Q & = & \displaystyle\frac{\rho _{2}}{\mu +\rho _{2}}q_{1}\left(
1-q_{2}-q_{3}\right) +\frac{\rho _{1}}{\mu +\rho _{1}}q_{1}q_{2}\frac{\theta 
}{\mu +\theta }.%
\end{array}%
\right.
\end{equation*}

\item Besides $\sigma \rightarrow \infty $ and $\gamma \rightarrow \infty $,
letting $\rho _{1}=0$ and $\rho _{2}=0$ -- in this case, $Q=0$ and%
\begin{equation}
R_{g}=R_{0}=\displaystyle\left( 1-q_{1}\right) \frac{\beta _{1}S^{0}}{\mu }%
+q_{1}\left( 1-q_{2}-q_{3}\right) \frac{\beta _{2}S^{0}}{\mu }.
\label{Rpart}
\end{equation}%
Using $S^{0}=q_{0}$ from equation (\ref{equil0esp}), whenever $%
q_{0}^{th}<q_{0}$, where $q_{0}^{th}$ is the threshold of $q_{0}$ (obtained
imposing $R_{0}=1$) given by%
\begin{equation}
q_{0}^{th}=\displaystyle\frac{\mu }{\left( 1-q_{1}\right) \beta
_{1}+q_{1}\left( 1-q_{2}-q_{3}\right) \beta _{2}},  \label{q0lim}
\end{equation}%
we have $R_{0}<1$, resulting in the stability of the trivial equilibrium
point $P^{0}$.
\end{enumerate}

To understand better the role of the fractions $q_{1}$, $q_{2}$,and $q_{3}$
in equation (\ref{Rpart}), let us consider $\beta _{1}=\beta _{2}=\beta $,
resulting in%
\begin{equation}
R_{g}=R_{0}=\displaystyle\left[ 1-q_{1}\left( q_{2}+q_{3}\right) \right] 
\frac{\beta S^{0}}{\mu }.  \label{Rpartesp}
\end{equation}%
The basic crime reproduction number $R_{0}$ assumes the highest value when $%
q_{1}=0$ (the police investigation is completely inhibited or absent);
however, for $q_{1}>0$, $R_{0}$ decreases proportionally to the increase in
the incerceration by the court trial ($q_{2}$) and/or the participation of
individuals caught by the police investigation in the whistleblowing program
($q_{3}$). It is worth stressing that the police innvestigation ($q_{1}$) is
crucial to inhibit criminal activities.

\subsubsection{The fraction of susceptible individuals $\protect\chi ^{-1}$}

To calculate the second threshold, we consider a particular case, letting $%
\rho _{1}=0$. In this case, we remove $R$ and, using $x=\left(
L,U,C,F,I\right) $, we construct the vectors $f$ and $v$ as%
\begin{equation*}
\begin{array}{lll}
f^{T}=\left( 
\begin{array}{c}
\frac{\beta _{0}+\beta _{1}U+\beta _{2}F}{1+\kappa _{1}C+\kappa _{2}I}S \\ 
0 \\ 
\varepsilon _{1}DC \\ 
0 \\ 
\varepsilon _{2}DF%
\end{array}%
\right) & \mathrm{and} & v^{T}=\left( 
\begin{array}{c}
\rho _{1}R-\rho _{2}F-\left( \mu +\sigma \right) L \\ 
-\left( 1-q_{1}\right) \sigma L+\left( \mu +\varepsilon _{10}+\varepsilon
_{1}D\right) U \\ 
-q_{1}\sigma L-\varepsilon _{10}U+\left( \mu +\gamma \right) C \\ 
-\left( 1-q_{2}-q_{3}\right) \gamma C+\left( \mu +\rho _{2}+\varepsilon
_{20}+\varepsilon _{2}D\right) F \\ 
-q_{2}\gamma C-\varepsilon _{20}F+\left( \mu +\theta \right) I%
\end{array}%
\right) ,%
\end{array}%
\end{equation*}%
and the corresponding matrices $F$ and $V$ are%
\begin{equation*}
F=\left( 
\begin{array}{ccccc}
0 & \beta _{1}S^{0} & 0 & \beta _{2}S^{0} & 0 \\ 
0 & 0 & 0 & 0 & 0 \\ 
0 & 0 & 0 & 0 & 0 \\ 
0 & 0 & 0 & 0 & 0 \\ 
0 & 0 & 0 & 0 & 0%
\end{array}%
\right)
\end{equation*}

\noindent and%
\begin{equation*}
V=\left( 
\begin{array}{ccccc}
\mu +\sigma & 0 & 0 & -\rho _{2} & 0 \\ 
-\left( 1-q_{1}\right) \sigma & \mu +\varepsilon _{10} & 0 & 0 & 0 \\ 
-q_{1}\sigma & -\varepsilon _{10} & \mu +\gamma & 0 & 0 \\ 
0 & 0 & -\left( 1-q_{2}-q_{3}\right) \gamma & \mu +\rho _{2}+\varepsilon
_{20} & 0 \\ 
0 & 0 & -q_{2}\gamma & -\varepsilon _{20} & \mu +\theta%
\end{array}%
\right) .
\end{equation*}%
The next generation matrix $FV^{-1}$ is given by%
\begin{equation*}
FV^{-1}=\left( 
\begin{array}{ccccc}
\frac{R_{01}+R_{02}+R_{03}}{1-\left( Q_{F}^{1}+Q_{F}^{2}\right) } & A & B & C
& 0 \\ 
0 & 0 & 0 & 0 & 0 \\ 
0 & 0 & 0 & 0 & 0 \\ 
0 & 0 & 0 & 0 & 0 \\ 
0 & 0 & 0 & 0 & 0%
\end{array}%
\right) ,
\end{equation*}%
where $A$, $B$, and $C$ are omitted ($V^{-1}$ is obtained after a tedious
calculation), and the corresponding characteristic equation is given by%
\begin{equation}
\xi ^{5}-\frac{R_{01}+R_{02}+R_{03}}{1-\left( Q_{F}^{1}+Q_{F}^{2}\right) }%
\xi ^{4}=0,  \label{eigen2}
\end{equation}%
where $R_{01}$, $R_{02}$, $R_{03}$, $Q_{F}^{1}$, and $Q_{F}^{2}$\ are given
by equations (\ref{R0part}) and (\ref{Qpart}). Defining the second threshold
as%
\begin{equation*}
\chi =\frac{R_{01}+R_{02}+R_{03}}{1-\left( Q_{F}^{1}+Q_{F}^{2}\right) },
\end{equation*}%
$P^{0}$ is locally asymptotically stable if $\chi <1$.

Notice that $\chi \neq 1/R_{g}$, thus the second threshold is the
multiplicative inverse of the fraction of susceptible individuals given by%
\begin{equation}
\chi ^{-1}=\frac{S^{\ast }}{S^{0}}=\frac{1-\left( Q_{F}^{1}+Q_{F}^{2}\right) 
}{R_{01}+R_{02}+R_{03}}=\frac{1}{R_{0}}-\frac{Q_{F}^{1}+Q_{F}^{2}}{R_{0}},
\label{suscpart}
\end{equation}%
according to \cite{yang}. In the general case, $\rho _{1}>0$, the
calculation of $FV^{-1}$ is tough, but it is expected that%
\begin{equation}
\chi ^{-1}=\frac{S^{\ast }}{S^{0}}=\frac{1-\left(
Q_{F}^{1}+Q_{F}^{2}+Q_{R}^{1}+Q_{R}^{2}+Q_{R}^{3}+Q_{R}^{4}\right) }{%
R_{01}+R_{02}+R_{03}}=\frac{1}{R_{0}}-\frac{Q}{R_{0}},  \label{suscgeral}
\end{equation}%
with $Q<1$ according to equation (\ref{Qrange}).

\subsection{The global stability \label{global}}

The global stability of $P^{0}$ follows the method proposed in \cite{shuai}.
Let the vector of variables be $x=\left( L,U,C,F,I,R\right) $, vectors $f$
and $v$ given by equation (\ref{fv}), and matrices $F$ and $V$ given by
equations (\ref{f0}) and (\ref{v}). The vector $g$, constructed as%
\begin{equation*}
g^{T}=\left( F-V\right) x^{T}-f^{T}+v^{T},
\end{equation*}%
is%
\begin{equation*}
g^{T}=\left( 
\begin{array}{c}
\displaystyle\left( \beta _{1}U+\beta _{2}F\right) \left( S^{0}-\frac{S}{%
1+\kappa _{1}C+\kappa _{2}I}\right) \\ 
\varepsilon _{1}DC \\ 
-\varepsilon _{1}DC \\ 
\varepsilon _{2}DF \\ 
-\varepsilon _{2}DF \\ 
0%
\end{array}%
\right) ,
\end{equation*}%
where $g^{T}\geq 0$ if $S^{0}\geq S/\left( 1+\kappa _{1}C+\kappa
_{2}I\right) $ and $\varepsilon _{1}=\varepsilon _{2}=0$. Notice that $%
S^{0}\geq S$ is always true. Hence, we show the global stability for the
particular case $\varepsilon _{1}=\varepsilon _{2}=0$.

Let $v_{l}=\left( z_{1},z_{2},z_{3},z_{4},z_{5},z_{6}\right) $ be the left
eigenvector satisfying the equation $v_{l}V^{-1}F=\psi v_{l}$, where $\psi
=\psi \left( FV^{-1}\right) $ is the spectral radius of the characteristic
equation (\ref{eigen1}), and%
\begin{equation*}
V^{-1}F=\left[ 
\begin{array}{cccccc}
0 & \frac{\beta _{1}S^{0}}{\mu +\sigma } & 0 & \frac{\beta _{2}S^{0}+\rho
_{2}}{\mu +\sigma } & 0 & \frac{\rho _{1}}{\mu +\sigma } \\ 
\frac{\left( 1-q_{1}\right) \sigma }{\mu +\varepsilon _{10}} & 0 & 0 & 0 & 0
& 0 \\ 
\frac{q_{1}\sigma }{\mu +\gamma } & \frac{\varepsilon _{10}}{\mu +\gamma } & 
0 & 0 & 0 & 0 \\ 
0 & 0 & \frac{\left( 1-q_{2}-q_{3}\right) \gamma }{\mu +\rho
_{2}+\varepsilon _{20}} & 0 & 0 & 0 \\ 
0 & 0 & \frac{q_{2}\gamma }{\mu +\theta } & \frac{\varepsilon _{20}}{\mu
+\theta } & 0 & 0 \\ 
0 & 0 & 0 & 0 & \frac{\theta }{\mu +\rho _{1}} & 0%
\end{array}%
\right] .
\end{equation*}%
We must solve the system of equations%
\begin{equation*}
\left\{ 
\begin{array}{rll}
\displaystyle\frac{\left( 1-q_{1}\right) \sigma }{\mu +\varepsilon _{10}}%
z_{2}+\frac{q_{1}\sigma }{\mu +\gamma }z_{3} & = & \psi z_{1} \\ 
\displaystyle\frac{\beta _{1}S^{0}}{\mu +\sigma }z_{1}+\frac{\varepsilon
_{10}}{\mu +\gamma }z_{3} & = & \psi z_{2} \\ 
\displaystyle\frac{\left( 1-q_{2}-q_{3}\right) \gamma }{\mu +\rho
_{2}+\varepsilon _{20}}z_{4}+\frac{q_{2}\gamma }{\mu +\theta }z_{5} & = & 
\psi z_{3} \\ 
\displaystyle\frac{\beta _{2}S^{0}+\rho _{2}}{\mu +\sigma }z_{1}+\frac{%
\varepsilon _{20}}{\mu +\theta }z_{5} & = & \psi z_{4} \\ 
\displaystyle\frac{\theta }{\mu +\rho _{1}}z_{6} & = & \psi z_{5} \\ 
\displaystyle\frac{\rho _{1}}{\mu +\sigma }z_{1} & = & \psi z_{6},%
\end{array}%
\right.
\end{equation*}%
and the coordinates of the vector $v_{l}$ are given by%
\begin{equation*}
\left\{ 
\begin{array}{lll}
z_{1} & = & \displaystyle\psi ^{2}\frac{\left( \mu +\sigma \right) \left(
\mu +\rho _{1}\right) }{\theta \rho _{1}} \\ 
z_{2} & = & \displaystyle\frac{1}{\psi ^{3}}\left\{ \frac{\beta
_{1}S^{0}\left( \mu +\rho _{1}\right) }{\theta \rho _{1}}\psi ^{4}+\frac{%
\varepsilon _{10}}{\mu +\gamma }\right. \\ 
&  & \left. \times \displaystyle\left[ \frac{\left( 1-q_{2}-q_{3}\right)
\gamma }{\mu +\rho _{2}+\varepsilon _{20}}\left( \frac{\left( \mu +\rho
_{1}\right) \left( \beta _{2}S^{0}+\rho _{2}\right) }{\theta \rho _{1}}\psi
^{2}+\frac{\varepsilon _{20}}{\mu +\theta }\right) +\frac{q_{2}\gamma }{\mu
+\theta }\psi \right] \right\} \\ 
z_{3} & = & \displaystyle\frac{1}{\psi ^{2}}\left\{ \frac{\left(
1-q_{2}-q_{3}\right) \gamma }{\mu +\rho _{2}+\varepsilon _{20}}\left[ \frac{%
\left( \mu +\rho _{1}\right) \left( \beta _{2}S^{0}+\rho _{2}\right) }{%
\theta \rho _{1}}\psi ^{2}+\frac{\varepsilon _{20}}{\mu +\theta }\right] +%
\frac{q_{2}\gamma }{\mu +\theta }\psi \right\} \\ 
z_{4} & = & \displaystyle\frac{1}{\psi }\left[ \frac{\left( \mu +\rho
_{1}\right) \left( \beta _{2}S^{0}+\rho _{2}\right) }{\theta \rho _{1}}\psi
^{2}+\frac{\varepsilon _{20}}{\mu +\theta }\right] \\ 
z_{5} & = & 1 \\ 
z_{6} & = & \displaystyle\psi \frac{\mu +\rho _{1}}{\theta },%
\end{array}%
\right. .
\end{equation*}%
where we used the spectral radius $\psi $ as the solution of equation (\ref%
{eigen1}).

A Lyapunov function $L_{y}$ can be constructed as $L_{y}=v_{l}V^{-1}x^{T}$,
resulting in%
\begin{equation*}
\begin{array}{ccl}
L_{y} & = & \displaystyle\frac{z_{1}}{\mu +\sigma }L+\frac{z_{2}}{\mu
+\varepsilon _{10}}U+\frac{z_{3}}{\mu +\gamma }C+\frac{z_{4}}{\mu +\rho
_{2}+\varepsilon _{20}}F+\frac{1}{\mu +\theta }I+\frac{z_{6}}{\mu +\rho _{1}}%
R,%
\end{array}%
\end{equation*}%
which is always positive or zero ($L_{y}\geq 0$), and%
\begin{equation*}
\begin{array}{ccl}
\frac{d}{dt}L_{y} & = & -\displaystyle\frac{1}{\psi }\frac{z_{1}}{\mu
+\sigma }\left( \beta _{1}U+\beta _{2}F\right) \left( S^{0}-\psi \frac{S}{%
1+\kappa _{1}C+\kappa _{2}I}\right) -\frac{1}{\psi }\left( 1-\psi \right) \\ 
&  & \times \displaystyle\left[ \psi z_{1}L+z_{3}\left( \varepsilon
_{10}U+C\right) +\left( \frac{\rho _{2}}{\mu +\sigma }z_{1}+\frac{%
\varepsilon _{20}}{\mu +\theta }\right) F+\psi I+\frac{\mu +\rho _{1}}{%
\theta }R\right] ,%
\end{array}%
\end{equation*}%
which is negative or zero ($dL_{y}/dt\leq 0$) only if $\psi \leq 1$, and $%
S^{0}\geq S/\left( 1+\kappa _{1}C+\kappa _{2}I\right) $, the letter is the
condition to have $g^{T}\geq 0$.

Hence, the method proposed in \cite{shuai} is valid only for $\varepsilon
_{1}=\varepsilon _{2}=0$, in which case $P^{0}$ is globally stable if $\psi
\leq 1$, and $S^{0}\geq S/\left( 1+\kappa _{1}C+\kappa _{2}I\right) $. When $%
\varepsilon _{1}>0$ and/or $\varepsilon _{2}>0$, $P^{0}$ is locally
asymptotically stable, and two positive solutions can occur when $R_{g}<1$.

\subsection{Interpreting $R_{0}$ and $Q$ \label{interp}}

To interpret $R_{0}$ and $Q$, we define the following probabilities:

\begin{enumerate}
\item $\displaystyle p_{\sigma }=\frac{\sigma }{\mu +\sigma }$ --
Probability of surviving in the class $L$, and entering into the next class.

\item $\displaystyle p_{\gamma }=\frac{\gamma }{\mu +\gamma }$ --
Probability of surviving in the class $C$, and entering into the next class.

\item $\displaystyle p_{\varepsilon _{10}}=\frac{\varepsilon _{10}}{\mu
+\varepsilon _{10}}$ -- Probability of surviving in the class $U$, and
entering into the class $C$.

\item $\displaystyle p_{\varepsilon _{20}}=\frac{\varepsilon _{20}}{\mu
+\rho _{2}+\varepsilon _{20}}$ -- Probability of surviving and not relapsing
in the class $F$, and entering into the class $I$.

\item $\displaystyle p_{\rho _{2}}=\frac{\rho _{2}}{\mu +\rho
_{2}+\varepsilon _{20}}$ -- Probability of surviving and not being
incarcerated in the class $F$, and entering into the class $L$.

\item $\displaystyle p_{\rho _{1}}=\frac{\rho _{1}}{\mu +\rho _{1}}$ --
Probability of surviving in the class $R$, and entering into the class $L$.

\item $\displaystyle p_{\theta }=\frac{\theta }{\mu +\theta }$ --
Probability of surviving in the class $I$, and entering into the class $R$.
\end{enumerate}

\subsubsection{Understanding $R_{0}$}

The basic crime reproduction number $R_{0}=R_{01}+R_{02}+R_{03}$, given by
equation (\ref{R0}), measures the strength of the criminality outbreak
(beginning of the epidemic, short-term dynamic) influencing
crime-susceptible individuals. The force of law-offending $\lambda $, given
by equation (\ref{lambda}), depends on the individuals in classes $U$ and $F$%
. From $L$, there is a unique route to reach $U$, but two routes to reach $F$%
. Once in these classes, they influence crime-susceptible individuals to
offend the law.

\begin{enumerate}
\item Reaching the class $U$ from $L$ and influencing criminality -- $R_{01}$%
. One individual in class $L$ survives in this class with probability $%
p_{\sigma }$ and enters into class $U$ with probability $1-q_{1}$. During
the period $1/\left( \mu +\varepsilon _{10}\right) $ staying in class $U$,
he/she influences $\beta _{1}S^{0}/\left( \mu +\varepsilon _{10}\right) $
individuals to commit a crime.

\item Reaching the class $F$ from $L$ and influencing criminality.

\begin{description}
\item[2.a.] From $L$ to $C$ and reaching $F$ -- $R_{02}$. One individual in
class $L$ survives in this class with probability $p_{\sigma }$ and enters
into class $C$ with probability $q_{1}$; he/she survives in this class with
probability $p_{\gamma }$ and enters into class $F$ with probability $%
1-q_{2}-q_{3}$. During the period $1/\left( \mu +\rho _{2}+\varepsilon
_{20}\right) $ staying in class $F$, he/she influences $\beta
_{2}S^{0}/\left( \mu +\rho _{2}+\varepsilon _{20}\right) $ individuals to
commit a crime.

\item[2.b] From $L$ to $U$ to $C$\ and reaching $F$ -- $R_{03}$. One
individual in class $L$ survives in this class with probability $p_{\sigma }$
and enters into class $U$ with probability $1-q_{1}$; he/she survives in
this class with probability $p_{\varepsilon _{10}}$ and enters into class $C$%
, where survives with probability $p_{\gamma }$ and enters into class $F$
with probability $1-q_{2}-q_{3}$. During the period $1/\left( \mu +\rho
_{2}+\varepsilon _{20}\right) $ staying in class $F$, he/she influences $%
\beta _{2}S^{0}/\left( \mu +\rho _{2}+\varepsilon _{20}\right) $ individuals
to commit a crime.
\end{description}
\end{enumerate}

Therefore, $R_{0}$ is all secondary criminal co-optations produced by one
offender in a crime-free community. The basic crime reproduction number
depends on the non-linear terms in equation (\ref{system}).

\subsubsection{Understanding $Q$}

The additional crime reproduction number $Q$, given by equation (\ref{Q}),
measures the strength of the criminality epidemic (long-term dynamic)
influencing crime-susceptible individuals. $Q$ accounts for the additional
influence on the crime-susceptible individuals by the court waiting in
freedom ($F$) and released ($R$) individuals when they relapse and commit
crime again. Notice that an individual in $F$ returning to class $L$ has two
routes to reach $F$ again ($Q_{F}^{1}$ and $Q_{F}^{2}$). However, an
individual in $R$ returning to class $L$ has four routes to reach $R$ again (%
$Q_{R}^{1}$, $Q_{R}^{2}$, $Q_{R}^{3}$, and $Q_{R}^{4}$).

\begin{enumerate}
\item Leaving and returning to the same class $F$ and influencing
criminality again -- $Q_{F}^{1}$ and $Q_{F}^{2}$.

\begin{description}
\item[1.a.] From $F$ to $L$ to $C$ and reaching $F$ again -- $Q_{F}^{1}$.
One individual in class $F$ survives in this class with probability $p_{\rho
_{2}}$ and enters into class $L$, where he/she survives with probability $%
p_{\sigma }$ and enters into class $C$ probability $q_{1}$. He/she survives
this class with probability $p_{\gamma }$ and re-enters into class $F$ with
probability $1-q_{2}-q_{3}$.

\item[1.b.] From $F$ to $L$ to $U$ to $C$ and reaching $F$ again -- $%
Q_{F}^{2}$. One individual in class $F$ survives in this class with
probability $p_{\rho _{2}}$ and enters into class $L$, where he/she survives
with probability $p_{\sigma }$ and enters into class $U$ probability $%
1-q_{1} $. He/she survives this class with probability $p_{\varepsilon
_{10}} $ and enters into class $C$, where he/she survives with probability $%
p_{\gamma }$ and re-enters into class $F$ with probability $1-q_{2}-q_{3}$.
\end{description}

\item Leaving and returning to the same class $R$ and influencing
criminality again -- $Q_{R}^{1}$, $Q_{R}^{2}$, $Q_{R}^{3}$, and $Q_{R}^{4}$.

\begin{description}
\item[2.a.] From $R$ to $L$ to $C$ to $I$ and reaching $R$ again -- $%
Q_{R}^{1}$. One individual in class $R$ survives in this class with
probability $p_{\rho _{1}}$ and enters into class $L$, where he/she survives
with probability $p_{\sigma }$ and enters into class $C$ probability $q_{1}$%
. He/she survives this class with probability $p_{\gamma }$ and enters into
class $I$ with probability $q_{2}$, where he/she survives with probability $%
p_{\theta }$ and re-enters into class $R$.

\item[2.b] From $R$ to $L$ to $C$ to $F$ to $I$ and reaching $R$ again -- $%
Q_{R}^{2}$. One individual in class $R$ survives in this class with
probability $p_{\rho _{1}}$ and enters into class $L$, where he/she survives
with probability $p_{\sigma }$ and enters into class $C$ probability $q_{1}$%
. He/she survives this class with probability $p_{\gamma }$ and enters into
class $F$ with probability $1-q_{2}-q_{3}$, where he/she survives with
probability $p_{\varepsilon _{20}}$ and enters into class $I$, from which
survives with probability $p_{\theta }$ and re-enters into class $R$.

\item[2.c.] From $R$ to $L$ to $U$ to $C$ to $I$ and reaching $R$ again -- $%
Q_{R}^{3}$. One individual in class $R$ survives in this class with
probability $p_{\rho _{1}}$ and enters into class $L$, where he/she survives
with probability $p_{\sigma }$ and enters into class $U$ probability $%
1-q_{1} $. He/she survives this class with probability $p_{\varepsilon
_{10}} $ and enters into class $C$, where he/she survives with probability $%
p_{\gamma }$ and enters into class $I$ with probability $q_{2}$, from which
survives with probability $p_{\theta }$ and re-enters into class $R$.

\item[2.d.] From $R$ to $L$ to $U$ to $C$ to $F$ to $I$ and reaching $R$
again -- $Q_{R}^{4}$. One individual in class $R$ survives in this class
with probability $p_{\rho _{1}}$ and enters into class $L$, where he/she
survives with probability $p_{\sigma }$ and enters into class $U$
probability $1-q_{1}$. He/she survives this class with probability $%
p_{\varepsilon _{10}}$ and enters into class $C$, where he/she survives with
probability $p_{\gamma }$ and enters into class $F$ with probability $%
1-q_{2}-q_{3}$. He/she survives this class with probability $p_{\varepsilon
_{20}}$ and enters into class $I$, where he/she survives with probability $%
p_{\theta }$ and re-enters into class $R$.
\end{description}
\end{enumerate}

Therefore, $Q$ is all additional criminal co-optations produced by one
offender when relapsing and committing a crime again. The additional crime
reproduction number depends on the linear terms in equation (\ref{system}).

\renewcommand{\theequation}{\Alph{section}.\arabic{equation}} %
\renewcommand{\thefigure}{\Alph{section}.\arabic{figure}} %
\renewcommand{\thetable}{\Alph{section}.\arabic{table}} %
\setcounter{equation}{0} \setcounter{figure}{0} \setcounter{table}{0}

\section{The non-trivial equilibrium point $P^{\ast }$ \label{p_n_trivial}}

The non-trivial equilibrium point $P^{\ast }$ of the system of equations (%
\ref{system}), the law offenders' imprisonment, has the coordinates given by%
\begin{equation}
\begin{array}{lll}
P^{\ast } & = & \left( \bar{S}_{1}=S_{1}^{0},\bar{E}=E^{0},\bar{S}=S^{\ast },%
\bar{L}=L^{\ast },\bar{U}=U^{\ast },\right. \\ 
&  & \left. \bar{C}=C^{\ast },\bar{F}=F^{\ast },\bar{I}=I^{\ast },\bar{D}%
=D^{\ast },\bar{R}=R^{\ast }\right) ,%
\end{array}
\label{Pstar}
\end{equation}%
where $S^{0}=S^{\ast }+L^{\ast }+U^{\ast }+C^{\ast }+F^{\ast }+I^{\ast
}+D^{\ast }+R^{\ast }$, and $S^{0}$, $S_{1}^{0}$, and $E^{0}$ are given by
equation (\ref{equil0}). The coordinates of the non-trivial equilibrium
point $P^{\ast }$ given by equation (\ref{Pstar}) are%
\begin{equation}
\left\{ 
\begin{array}{lll}
S^{\ast } & = & \displaystyle\frac{S^{0}\left( 1+\kappa _{1}C^{\ast }+\kappa
_{2}I^{\ast }\right) }{1+\kappa _{1}C^{\ast }+\kappa _{2}I^{\ast }+\frac{%
\beta _{1}}{\mu }U^{\ast }+\frac{\beta _{2}}{\mu }F^{\ast }} \\ 
L^{\ast } & = & \displaystyle\frac{\mu +\gamma }{\mu }\times \frac{\mu
+\varepsilon _{10}+\frac{\varepsilon _{1}q_{3}\gamma }{\mu }C^{\ast }}{\frac{%
\sigma }{\mu }\left( \mu +\varepsilon _{10}+\frac{\varepsilon
_{1}q_{3}\gamma }{\mu }C^{\ast }\right) -\left( 1-q_{1}\right) \sigma }%
C^{\ast } \\ 
U^{\ast } & = & \displaystyle\frac{\left( 1-q_{1}\right) \sigma }{\mu
+\varepsilon _{10}+\frac{\varepsilon _{1}q_{3}\gamma }{\mu }C^{\ast }}\times 
\frac{\mu +\gamma }{\mu }\times \frac{\mu +\varepsilon _{10}+\frac{%
\varepsilon _{1}q_{3}\gamma }{\mu }C^{\ast }}{\frac{\sigma }{\mu }\left( \mu
+\varepsilon _{10}+\frac{\varepsilon _{1}q_{3}\gamma }{\mu }C^{\ast }\right)
-\left( 1-q_{1}\right) \sigma }C^{\ast } \\ 
F^{\ast } & = & \displaystyle\frac{\left( 1-q_{2}-q_{3}\right) \gamma }{\mu
+\rho _{2}+\varepsilon _{20}+\frac{\varepsilon _{2}q_{3}\gamma }{\mu }%
C^{\ast }}C^{\ast } \\ 
I^{\ast } & = & \displaystyle\frac{\left( 1-q_{3}\right) \gamma }{\mu
+\theta }C^{\ast }-\frac{\mu +\rho _{2}}{\mu +\theta }\times \frac{\left(
1-q_{2}-q_{3}\right) \gamma }{\mu +\rho _{2}+\varepsilon _{20}+\frac{%
\varepsilon _{2}q_{3}\gamma }{\mu }C^{\ast }}C^{\ast } \\ 
D^{\ast } & = & \displaystyle\frac{q_{3}\gamma }{\mu }C^{\ast } \\ 
R^{\ast } & = & \displaystyle\frac{\theta }{\mu +\rho _{1}}\left[ \frac{%
\left( 1-q_{3}\right) \gamma }{\mu +\theta }C^{\ast }-\frac{\mu +\rho _{2}}{%
\mu +\theta }\times \frac{\left( 1-q_{2}-q_{3}\right) \gamma }{\mu +\rho
_{2}+\varepsilon _{20}+\frac{\varepsilon _{2}q_{3}\gamma }{\mu }C^{\ast }}%
C^{\ast }\right] ,%
\end{array}%
\right.  \label{equil1}
\end{equation}%
where $C^{\ast }$\ is solution of $Pol_{5}(C)\times C=0$, with $Pol_{5}(C)$
being a $5^{th}$ degree polynomial given by%
\begin{equation}
Pol_{5}(C)=c_{5}C^{5}+c_{4}C^{4}+c_{3}C^{3}+c_{2}C^{2}+c_{1}C+c_{0},
\label{equil2}
\end{equation}%
with the coefficients $c_{i}$, $i=0,1,\cdots ,5$, being given by%
\begin{equation}
\left\{ 
\begin{array}{lll}
c_{5} & = & a_{6}^{2}a_{7}^{2}a_{9}a_{10}a_{12}\left( \kappa _{1}+\kappa _{2}%
\frac{a_{5}}{\mu +\theta }\right) \left[ 1-\frac{a_{1}a_{12}}{%
a_{1}a_{12}-a_{4}}\left( Q_{R}^{1}+Q_{R}^{3}+Q_{R}^{2th}+Q_{R}^{4th}\right) %
\right] \\ 
c_{4} & = & B_{2}D_{3}+B_{3}D_{2} \\ 
c_{3} & = & B_{1}D_{3}+B_{2}D_{2}+B_{3}D_{1}-A_{2}A_{5} \\ 
c_{2} & = & A_{3}D_{3}+B_{1}D_{2}+B_{2}D_{1}-\left(
A_{1}A_{5}+A_{2}A_{4}\right) \\ 
c_{1} & = & A_{3}D_{2}+B_{1}D_{1}-\left( A_{1}A_{4}+A_{2}A_{3}\right) \\ 
c_{0} & = & a_{1}a_{2}^{2}a_{9}a_{10}\left( a_{1}a_{12}-a_{4}\right) \left(
1-R_{g}\right) ,%
\end{array}%
\right.  \label{coefic}
\end{equation}%
where the auxiliary parameters $A_{i}$, $i=1,\cdots ,5$, are%
\begin{equation*}
\left\{ 
\begin{array}{l}
\begin{array}{ll}
A_{1}=\displaystyle\frac{S^{0}}{\mu }\left[ a_{2}a_{4}a_{9}\beta
_{1}+a_{3}\left( a_{1}a_{12}-a_{4}\right) \beta _{2}\right] , & A_{2}=%
\displaystyle\frac{S^{0}}{\mu }\left( a_{4}a_{7}a_{9}\beta
_{1}+a_{3}a_{6}a_{12}\beta _{2}\right) ,%
\end{array}
\\ 
\begin{array}{llll}
A_{3}=a_{2}\left( a_{1}a_{12}-a_{4}\right) , & A_{4}=a_{7}\left(
a_{1}a_{12}-a_{4}\right) +a_{2}a_{6}a_{12}, & \mathrm{and} & 
A_{5}=a_{6}a_{7}a_{12},%
\end{array}%
\end{array}%
\right.
\end{equation*}%
$B_{i}$, $i=1,2,3$, are%
\begin{equation*}
\left\{ 
\begin{array}{lll}
B_{1} & = & \displaystyle\left( a_{1}a_{12}-a_{4}\right) \left[
a_{7}+a_{2}\left( \kappa _{1}+\kappa _{2}\frac{a_{5}}{\mu +\theta }\right)
+a_{3}\left( \frac{\beta _{2}}{\mu }-a_{8}\kappa _{2}\right) \right] \\ 
&  & +\displaystyle a_{2}\left( a_{4}a_{9}\frac{\beta _{1}}{\mu }%
+a_{6}a_{12}\right) \\ 
B_{2} & = & \displaystyle a_{6}a_{12}\left[ a_{7}+a_{2}\left( \kappa
_{1}+\kappa _{2}\frac{a_{5}}{\mu +\theta }\right) +a_{3}\left( \frac{\beta
_{2}}{\mu }-a_{8}\kappa _{2}\right) \right] \\ 
&  & +\displaystyle\left( a_{1}a_{12}-a_{4}\right) a_{7}\left( \kappa
_{1}+\kappa _{2}\frac{a_{5}}{\mu +\theta }\right) +a_{4}a_{7}a_{9}\frac{%
\beta _{1}}{\mu } \\ 
B_{3} & = & \displaystyle a_{6}a_{7}a_{12}\left( \kappa _{1}+\kappa _{2}%
\frac{a_{5}}{\mu +\theta }\right) ,%
\end{array}%
\right.
\end{equation*}%
and $D_{i}$, $i=1,2,3$, are%
\begin{equation*}
\left\{ 
\begin{array}{lll}
D_{1} & = & a_{1}a_{2}a_{9}a_{10}\left( 1-Q\right) \\ 
D_{2} & = & a_{9}a_{10}\left\{ a_{1}a_{7}\left[ 1-\left(
Q_{R}^{1}+Q_{R}^{3}+Q_{R}^{2th}+Q_{R}^{4th}\right) \right] +a_{2}a_{6}\left(
1-\frac{a_{1}a_{12}}{a_{1}a_{12}-a_{4}}Q\right) \right\} \\ 
D_{3} & = & a_{6}a_{7}a_{9}a_{10}\left[ 1-\frac{a_{1}a_{12}}{%
a_{1}a_{12}-a_{4}}\left( Q_{R}^{1}+Q_{R}^{3}+Q_{R}^{2th}+Q_{R}^{4th}\right) %
\right] ,%
\end{array}%
\right.
\end{equation*}%
with $Q_{R}^{2th}$ and $Q_{R}^{4th}$ being the thresholds of $Q_{R}^{2}$ and 
$Q_{R}^{4}$,%
\begin{equation*}
\left\{ 
\begin{array}{lllll}
Q_{R}^{2th} & = & \lim\limits_{\varepsilon _{20}\rightarrow \infty }Q_{R}^{2}
& = & \displaystyle\frac{\rho _{1}}{\mu +\rho _{1}}\frac{\sigma }{\mu
+\sigma }q_{1}\frac{\gamma }{\mu +\gamma }\left( 1-q_{2}-q_{3}\right) \frac{%
\theta }{\mu +\theta } \\ 
Q_{R}^{4th} & = & \lim\limits_{\varepsilon _{20}\rightarrow \infty }Q_{R}^{4}
& = & \displaystyle\frac{\rho _{1}}{\mu +\rho _{1}}\frac{\sigma }{\mu
+\sigma }\left( 1-q_{1}\right) \frac{\varepsilon _{10}}{\mu +\varepsilon
_{10}}\frac{\gamma }{\mu +\gamma }\left( 1-q_{2}-q_{3}\right) \frac{\theta }{%
\mu +\theta },%
\end{array}%
\right.
\end{equation*}%
and $R_{g}$ and $Q_{R}^{i}$, $i=1,\cdots ,4$, are given by equations (\ref%
{Rg}) and (\ref{Qpart}). The parameters $a_{i}$, with $i=1,\cdots ,14$, are%
\begin{equation}
\left\{ 
\begin{array}{l}
\begin{array}{llll}
a_{1}=\mu +\varepsilon _{10}, & a_{2}=\mu +\rho _{2}+\varepsilon _{20}, & 
a_{3}=\left( 1-q_{2}-q_{3}\right) \gamma , & a_{4}=\left( 1-q_{1}\right)
\sigma ,%
\end{array}
\\ 
\begin{array}{lllll}
a_{5}=\left( 1-q_{3}\right) \gamma , & a_{6}=\displaystyle\frac{q_{3}\gamma 
}{\mu }\varepsilon _{1}, & a_{7}=\displaystyle\frac{q_{3}\gamma }{\mu }%
\varepsilon _{2}, & a_{8}=\displaystyle\frac{\mu +\rho _{2}}{\mu +\theta },
& a_{9}=\displaystyle\frac{\mu +\gamma }{\mu },%
\end{array}
\\ 
\begin{array}{llllll}
a_{10}=\displaystyle\frac{\mu +\sigma }{\mu }, & a_{11}=\displaystyle\frac{%
\rho _{1}}{\mu }, & a_{12}=\displaystyle\frac{\sigma }{\mu }, & a_{13}=%
\displaystyle\frac{\theta }{\mu }, & \mathrm{and} & a_{14}=\displaystyle%
\frac{\rho _{2}}{\mu }.%
\end{array}%
\end{array}%
\right.  \label{ais}
\end{equation}

Notice that $C^{\ast }=0$ is a solution, and for $C^{\ast }>0$, $C^{\ast }$\
is the positive solution(s) of $Pol_{5}(C)=0$. The denominator of $L^{\ast }$
is always positive; thus $C^{\ast }\geq 0$ is the unique restriction to be
populationally feasible. Letting $\varepsilon _{2}=0$ and varying $%
\varepsilon _{1}$, we are transferring individuals from compartment $U$
(evading police investigation) to $C$ (caught by police), which rate is
proportional to the product $U\times D$. However, the collaborator-dependent
rate $\varepsilon _{2}$ transfers individuals from compartment $F$ (waiting
for court trial) to $I$ (condemned and incarcerated).

\subsection{Particular cases}

Two particular cases of the model are presented.

\subsubsection{Case 1 -- $\protect\beta _{1}=0$, $\protect\kappa _{2}=0$,
and $\protect\rho _{1}=0$ \label{part1}}

The analytical assessment of the number of positive solutions for equation (%
\ref{equil2}) is not an easy task. For this reason, we consider a particular
case, letting $\beta _{1}=0$, $\kappa _{2}=0$ and $\rho _{1}=0$. This case
removes one influencing term (by $U$), one avoiding factor (by $I$), and one
relapsing to crime (by $R$). In other words, we have%
\begin{equation}
\lambda =\displaystyle\frac{\beta _{2}F}{1+\kappa _{1}C}.  \label{lamb}
\end{equation}%
In this case, $Pol_{5}(C)$ in equation (\ref{equil2}) becomes a $4^{th}$
degree polynomial $Pol_{4}(C)$ given by%
\begin{equation}
Pol_{4}\left( C\right) =c_{5}C^{4}+c_{4}C^{3}+c_{3}C^{2}+c_{2}C+c_{1},
\label{equil3}
\end{equation}%
with%
\begin{equation*}
\left\{ 
\begin{array}{lll}
c_{5} & = & \varepsilon _{1}\varepsilon _{2}^{2}a_{15}^{3}\kappa _{1} \\ 
c_{4} & = & \varepsilon _{1}\varepsilon _{2}a_{15}^{2}\left( \frac{\beta _{2}%
}{\mu }a_{3}+\varepsilon _{2}a_{15}+\kappa _{1}a_{2}\right) \\ 
&  & +\varepsilon _{2}a_{15}\kappa _{1}\left\{ \varepsilon _{1}a_{2}a_{15} 
\left[ 1-\left( Q_{F}^{1}+Q_{F}^{2th}\right) \right] +\varepsilon
_{2}a_{1}a_{15}\right\} \\ 
c_{3} & = & \varepsilon _{1}\varepsilon _{2}a_{2}a_{15}^{2}\left[ 1-\left(
R_{02}+R_{03}^{th}\right) \right] +\varepsilon _{2}a_{1}a_{2}a_{15}\kappa
_{1}\left( 1-Q\right) \\ 
&  & +\left\{ \varepsilon _{1}a_{2}a_{15}\left[ 1-\left(
Q_{F}^{1}+Q_{F}^{2th}\right) \right] +\varepsilon _{2}a_{1}a_{15}\right\}
\left( \frac{\beta _{2}}{\mu }a_{3}+\varepsilon _{2}a_{15}+\kappa
_{1}a_{2}\right) \\ 
c_{2} & = & \varepsilon _{1}a_{2}^{2}a_{15}\left[ 1-\left(
R_{02}+Q_{F}^{1}+R_{03}^{th}+Q_{F}^{2th}\right) \right] +\varepsilon
_{2}a_{1}a_{2}a_{15}\left( 1-R_{0}\right) \\ 
&  & +a_{1}a_{2}\left( \frac{\beta _{2}}{\mu }a_{3}+\varepsilon
_{2}a_{15}+\kappa _{1}a_{2}\right) \left( 1-Q\right) \\ 
c_{1} & = & a_{1}a_{2}^{2}\left( 1-R_{g}\right) ,%
\end{array}%
\right.
\end{equation*}%
where $Q=Q_{F}^{1}+Q_{F}^{2}$, $R_{g}=R_{0}+Q$ with $R_{0}=R_{02}+R_{03}$
(see equations (\ref{R0part}) and (\ref{Qpart})), $\left. a_{i}\right.
^{\prime }s$ are given by equation (\ref{ais}), and $R_{03}^{th}$ and $%
Q_{F}^{2th}$ are the thresholds of $R_{03}$ and $Q_{F}^{2}$ given by%
\begin{equation}
\left\{ 
\begin{array}{lllll}
R_{03}^{th} & = & \lim\limits_{\varepsilon _{10}\rightarrow \infty }R_{03} & 
= & \displaystyle\frac{\sigma }{\mu +\sigma }\left( 1-q_{1}\right) \frac{%
\gamma }{\mu +\gamma }\left( 1-q_{2}-q_{3}\right) \frac{\beta _{2}S^{0}}{\mu
+\rho _{2}+\varepsilon _{20}} \\ 
Q_{F}^{2th} & = & \lim\limits_{\varepsilon _{10}\rightarrow \infty }Q_{F}^{2}
& = & \displaystyle\frac{\sigma }{\mu +\sigma }\left( 1-q_{1}\right) \frac{%
\gamma }{\mu +\gamma }\left( 1-q_{2}-q_{3}\right) \frac{\rho _{2}}{\mu +\rho
_{2}+\varepsilon _{20}},%
\end{array}%
\right.  \label{RQthres}
\end{equation}%
with $R_{g}=R_{02}+R_{03}+Q_{F}^{1}+Q_{F}^{2}$. Here, we substituted $a_{6}$
and $a_{7}$ by%
\begin{equation*}
\begin{array}{lllll}
a_{6}=a_{15}\varepsilon _{1} & \mathrm{and} & a_{7}=a_{15}\varepsilon _{2},
& \mathrm{where} & a_{15}=\displaystyle\frac{q_{3}\gamma }{\mu }.%
\end{array}%
\end{equation*}

The coefficients of the $Pol_{4}\left( C\right) $ are $c_{4}>0$, $c_{3}>0$,
and $c_{0}<0$ if $R_{g}>1$. Hence, Descartes' signal rule assures the
existence of at least one positive solution if $R_{g}>1$. However, if $%
R_{g}<1$, we have $c_{0}>0$ and, depending on the signal of coefficients $%
c_{2}$ and $c_{1}$, zero or two positive solutions are feasible. Two
positive solutions for $R_{g}<1$ occur if at $R_{g}=1$ a positive solution
appears. When $R_{g}=1$, we have $c_{0}=0$, and for simplicity, let us
consider $\varepsilon _{10}=\varepsilon _{20}=0$ (these parameters do not
affect the qualitative behavior of the model). When $\varepsilon _{2}=0$, we
have $c_{5}=0$ and $c_{4}=0$, and%
\begin{equation*}
\left\{ 
\begin{array}{lll}
c_{3} & = & \varepsilon _{1}a_{2}a_{15}\left( \frac{\beta _{2}}{\mu }%
a_{3}+\kappa _{1}a_{2}\right) \left( R_{02}-Q_{F}^{2th}\right) \\ 
c_{2} & = & -a_{2}^{2}a_{15}\left( R_{03}^{th}+Q_{F}^{2th}\right) \left(
\varepsilon _{1}-\varepsilon _{1}^{c}\right) ,%
\end{array}%
\right.
\end{equation*}%
where $\varepsilon _{1}^{c}$ is the critical value for $\varepsilon _{1}$
given by%
\begin{equation*}
\displaystyle\varepsilon _{1}^{c}=\frac{a_{1}a_{2}\left( \frac{\beta _{2}}{%
\mu }a_{3}+\kappa _{1}a_{2}\right) \left( 1-Q\right) }{a_{2}^{2}a_{15}\left(
R_{03}^{th}+Q_{F}^{2th}\right) },
\end{equation*}%
and a positive $C^{\ast }=-c_{2}/c_{3}$ is possible if $c_{3}$ and $c_{2}$
have opposite signal. For simplicity, let us consider $\rho _{2}=0$, that
is, $Q_{F}^{2th}=0$, resulting in $c_{3}>0$. When $\varepsilon
_{1}>\varepsilon _{1}^{c}$, we have $c_{2}<0$ resulting in $C^{\ast }>0$. In
this case, we have a critical value (or sub-threshold) $R^{c}<1$ such that
we have only $C^{\ast }=0$ when $R_{g}\leq R^{c}$, two positive solutions $%
C_{<}^{\ast }$ (small) and $C_{>}^{\ast }$ (big) when $R^{c}<R_{g}<1$, and a
unique positive solution $C^{\ast }>0$ if $R_{g}>1$. (At $R_{g}=1$, the
small positive solution becomes zero, and assumes negative value since
after.) In this case we have backward bifurcation \cite{yang0}, and the
sub-threshold $R^{c}$ can be obtained numerically.

However, when $\varepsilon _{1}=0$, for all $\varepsilon _{2}>0$, we have $%
c_{5}=0$, and $c_{4}>0$, $c_{3}>0$, and $c_{2}>0$; hence, the unique
feasible solution is $C^{\ast }=0$. In this case, we have the forward
bifurcation \cite{yang0}. Additionally, if $\beta _{2}=0$, the force of law
offending is%
\begin{equation*}
\lambda =\displaystyle\frac{\beta _{1}U}{1+\kappa _{1}C+\kappa _{2}I},
\end{equation*}%
and it is not expected the occurrence of backward bifurcation \cite{yang0}
for all values of $\varepsilon _{1}$ and $\varepsilon _{2}$.

\begin{figure}[h]
\centering                                                                   
\subfloat[]{
\includegraphics[scale=0.40]{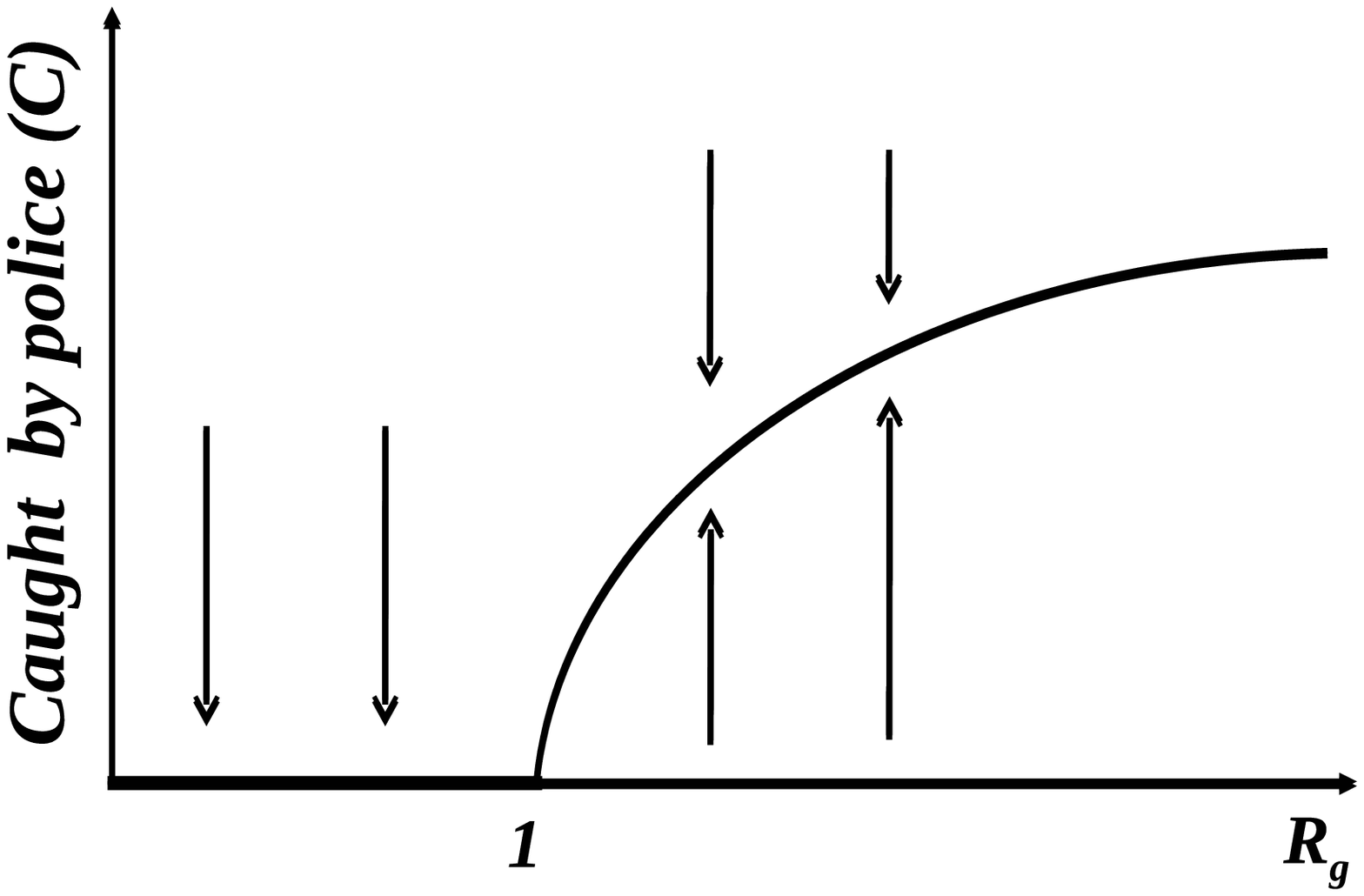}
} 
\subfloat[]{
\includegraphics[scale=0.30]{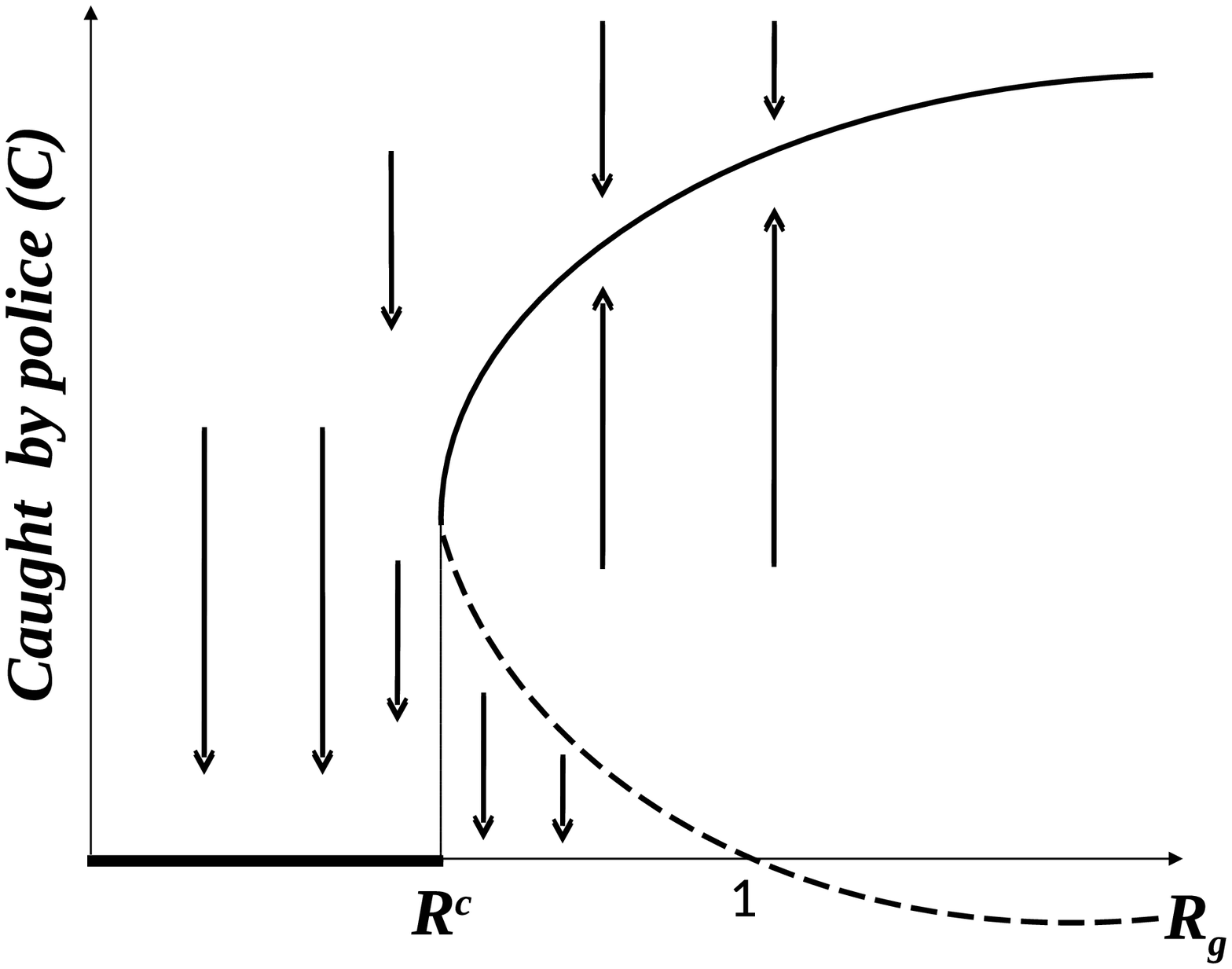}
}
\caption{Illustration of the forward (a) and backward (b) bifurcations, with
the arrows indicating the attracting equilibrium.}
\label{for_back}
\end{figure}

Summarizing, in Figure \ref{for_back} we illustrate the forward (a) and
backward (b) bifurcations, with the arrows indicating the attracting
equilibrium value. In the forward bifurcation, the trivial equilibrium ($%
C^{\ast }=0$) is stable up to $R_{g}=1$, which becomes unstable since after
and the unique non-trivial equilibrium ($C^{\ast }>0$) appears. In backward
bifurcation, there arises an interval $R^{c}<R_{g}<1$\ where the unstable
small equilibrium ($C_{-}^{\ast }$, dotted line) separates the attraction to
the stable trivial equilibrium ($C^{\ast }=0$) or to the stable big
equilibrium ($C_{+}^{\ast }$, continuous line). For $R_{g}>1$, the trivial
equilibrium ($C^{\ast }=0$) is unstable, and the unique non-trivial
equilibrium ($C_{+}^{\ast }$) is stable, which assumes $C_{+}^{\ast }>0$ at
the threshold $R_{g}=1$, but assumes $C_{+}^{\ast }=0$ at the sub-threshold $%
R^{c}<1$.

\subsubsection{Case 2 -- $\protect\varepsilon _{1}=\protect\varepsilon %
_{2}=0 $ \label{part2}}

Let us consider $\varepsilon _{1}=0$ and $\varepsilon _{2}=0$, that is, the
collaboration depending on the number of whistleblowers\ ($D$) rates are
removed. The outflows, depending on the whistleblowers, are absent from
compartments $U$ ($\varepsilon _{1}=0$, evading police investigation) and $F$
($\varepsilon _{2}=0$, postponing imprisonment by court trial), and
individuals caught by police investigation ($C$) and condemned by court
trial ($I$) assume the lowest values. Equation (\ref{equil2}) is simplified,
and $C^{\ast }$\ is the positive solution of equation $P_{1}\left( C\right)
\times C=0$, where the solution of the $1^{st}$ degree polynomial $%
P_{1}\left( C\right) $ is%
\begin{equation}
C^{\ast }=\frac{x_{1}}{x_{2}\left( 1-Q\right) }\left( R_{g}-1\right) ,
\label{Csol}
\end{equation}%
where $R_{g}=R_{0}+Q$ is given by equation (\ref{Rg}), $R_{0}$ and $Q$ are
given by equations (\ref{R0}) and (\ref{Q}), and 
\begin{equation*}
\left\{ 
\begin{array}{lll}
x_{1} & = & \displaystyle a_{1}a_{12}-a_{4}=\frac{q_{1}\sigma \left( \mu
+\varepsilon _{10}\right) +\left( 1-q_{1}\right) \sigma \varepsilon _{10}}{%
\mu }>0 \\ 
x_{2} & = & \displaystyle x_{1}\left[ \kappa _{1}+\kappa _{2}\left( \frac{%
a_{5}}{\mu +\theta }-\frac{a_{3}a_{8}}{a_{2}}\right) +\frac{\beta _{2}a_{3}}{%
\mu a_{2}}\right] +\frac{\beta _{1}a_{4}a_{9}}{\mu }>0.%
\end{array}%
\right.
\end{equation*}%
Therefore, if $R_{g}\geq 1$ we have $C^{\ast }\geq 0$ since $Q<1$.

\subsection{Local stability \label{stab}}

The local stability of $P^{\ast }$ is assessed numerically by calculating
the eigen-values of the Jacobian matrix evaluated at this point with
coordinates given by equation (\ref{Pstar}). We have two eigen-values $\xi
_{1}=-\left( \mu +\eta \right) $ and $\xi _{2}=-\mu $, corresponding to
equations for $S_{1}$ and $E$ in the system (\ref{system}), plus eight
eigen-values of the Jacobian matrix $J$ given by%
\begin{equation}
J=\left[ 
\begin{array}{cccccccc}
-d_{1} & 0 & -j_{1} & j_{4} & -j_{2} & j_{5} & 0 & 0 \\ 
j_{3} & -d_{2} & j_{1} & -j_{4} & \rho _{2}+j_{2} & -j_{5} & 0 & \rho _{1}
\\ 
0 & \left( 1-q_{1}\right) \sigma & -d_{3} & 0 & 0 & 0 & -\varepsilon
_{1}U^{\ast } & 0 \\ 
0 & q_{1}\sigma & \varepsilon _{10}+\varepsilon _{1}D^{\ast } & -d_{4} & 0 & 
0 & \varepsilon _{1}U^{\ast } & 0 \\ 
0 & 0 & 0 & \left( 1-q_{2}-q_{3}\right) \gamma & -d_{5} & 0 & -\varepsilon
_{2}F^{\ast } & 0 \\ 
0 & 0 & 0 & q_{2}\gamma & \varepsilon _{20}+\varepsilon _{2}D^{\ast } & 
-d_{6} & \varepsilon _{2}F^{\ast } & 0 \\ 
0 & 0 & 0 & q_{3}\gamma & 0 & 0 & -\mu & 0 \\ 
0 & 0 & 0 & 0 & 0 & \theta & 0 & -d_{7}%
\end{array}%
\right] ,  \label{jacobian}
\end{equation}%
where the positive diagonal elements are%
\begin{equation*}
\left\{ 
\begin{array}{l}
\begin{array}{cccc}
d_{1}=\mu +j_{3}, & d_{2}=\mu +\sigma , & d_{3}=\mu +\varepsilon
_{10}+\varepsilon _{1}D^{\ast }, & d_{4}=\mu +\gamma ,%
\end{array}
\\ 
\begin{array}{cccc}
d_{5}=\mu +\rho _{2}+\varepsilon _{20}+\varepsilon _{2}D^{\ast }, & 
d_{6}=\mu +\theta , & \mathrm{and} & d_{7}=\mu +\rho _{1},%
\end{array}%
\end{array}%
\right.
\end{equation*}%
and off diagonal elements are%
\begin{equation*}
\left\{ 
\begin{array}{l}
\begin{array}{lll}
j_{1}=\displaystyle\frac{\beta _{1}S^{\ast }}{1+\kappa _{1}C^{\ast }+\kappa
_{2}I^{\ast }}, & j_{2}=\displaystyle\frac{\beta _{2}S^{\ast }}{1+\kappa
_{1}C^{\ast }+\kappa _{2}I^{\ast }}, & j_{3}=\displaystyle\frac{\beta
_{1}U^{\ast }+\beta _{2}F^{\ast }}{1+\kappa _{1}C^{\ast }+\kappa _{2}I^{\ast
}},%
\end{array}
\\ 
\begin{array}{lll}
j_{4}=\displaystyle-\frac{\kappa _{1}\left( \beta _{1}U^{\ast }+\beta
_{2}F^{\ast }\right) S^{\ast }}{\left( 1+\kappa _{1}C^{\ast }+\kappa
_{2}I^{\ast }\right) ^{2}}, & \mathrm{and} & j_{5}=\displaystyle-\frac{%
\kappa _{2}\left( \beta _{1}U^{\ast }+\beta _{2}F^{\ast }\right) S^{\ast }}{%
\left( 1+\kappa _{1}C^{\ast }+\kappa _{2}I^{\ast }\right) ^{2}},%
\end{array}%
\end{array}%
\right.
\end{equation*}%
with the equilibrium values given by equations (\ref{equil1}) and (\ref%
{equil2}).

If all eigen-values of $J$ are negative or have negative real part if
complex, then the non-trivial equilibrium point $P^{\ast }$ is locally
asymptotically stable.

\end{document}